\documentclass[nofootinbib,aps,prd,groupedaddress,preprintnumbers,%
  showpacs,showkeys,floatfix,amssymb,amsfonts]{revtex4-1} 

\usepackage[top=.95in, bottom=.95in, left=.95in, right=.95in] {geometry}
\usepackage{empheq}
\usepackage{braket}
\usepackage{graphicx}
\usepackage{bm}
\usepackage{amsmath}
\usepackage{amssymb}
\usepackage{amsfonts}
\usepackage{bbold}
\usepackage{float}
\usepackage{dsfont}  
\usepackage{slashed}  
\usepackage{booktabs}
\usepackage{multirow}
\usepackage{subfigure}
\usepackage[sort&compress]{natbib}
\usepackage{ulem}

\newcommand{\be}{\begin{equation}}  
\newcommand{\ee}{\end{equation}}  
\newcommand{\beq}{\begin{eqnarray}} 
\newcommand{\eeq}{\end{eqnarray}}

\newcommand{\bea}{\begin{eqnarray}}
\newcommand{\eea}{\end{eqnarray}}

\newcommand{\MSb}{{\overline{\rm MS}}}
\newcommand{\MMSb}{{\rm{M}\overline{\rm MS}}}

\newcommand{\zmax}{{z_{\rm{max}}}}

\usepackage{xr-hyper}

\usepackage[svgnames,x11names,table]{xcolor}
\usepackage[colorlinks=true,linkcolor=MediumBlue,citecolor=MediumBlue,urlcolor=MediumBlue]{hyperref}
\usepackage{hyperref}

\begin{document}

\title{Chiral-even axial twist-3 GPDs of the proton from lattice QCD}

\author{
\vspace*{0.35cm}
  Shohini Bhattacharya$^{1,2}$,
 Krzysztof Cichy$^3$,
  Martha Constantinou$^{1}$, \\[0.75ex]
  Jack Dodson$^{1}$,
Andreas Metz$^{1}$,
Aurora Scapellato$^{1}$,
Fernanda Steffens$^{4}$ \\[3ex] \phantom{.}
}

\affiliation{
   $^1$ Department of Physics, Temple University, 1925 N. 12th Street, Philadelphia, PA 19122-1801, USA\\
     \vskip 0.05cm
   $^2$RIKEN BNL Research Center, Brookhaven National Laboratory, Upton, NY 11973, USA \\ 
  \vskip 0.05cm
  $^3$ {\it Faculty of Physics, Adam Mickiewicz University, Uniwersytetu Pozna\'{n}skiego 2, 61-614 Pozna\'{n}, Poland}\\
 \vskip 0.05cm
  $^4$ Institut f\"ur Strahlen- und Kernphysik, Rheinische
  Friedrich-Wilhelms-Universit\"at Bonn, \\ Nussallee 14-16, 53115 Bonn, Germany \\
  \vskip 0.05cm
  }

\begin{abstract}
This work presents the first lattice-QCD calculation of the twist-3 axial quark GPDs for the proton using the large-momentum effective theory approach. We calculate matrix elements with momentum-boosted proton states and a non-local axial operator. The calculation is performed using one ensemble of two degenerate light, a strange and a charm quark ($N_f=2+1+1$) of maximally twisted mass fermions with a clover term. The ensemble has a volume $32^3\times64$, lattice spacing 0.0934 fm, and corresponds to a pion mass of 260 MeV. The matrix elements are calculated for three values of the proton momentum, namely 0.83, 1.25, and 1.67 GeV. The light-cone GPDs are defined in the symmetric frame, which we implement here with a (negative) 4-momentum transfer squared of 0.69, 1.38, and 2.76 GeV$^2$, all at zero skewness. We also conduct several consistency checks, including assessing the local limit of the twist-3 GPDs and examining the Burkhardt-Cottingham-type as well as Efremov-Teryaev-Leader-type sum rules.
\end{abstract}


\maketitle

\section{Introduction}

Distribution functions of partonic content play a pivotal role in understanding proton structure, and they have been extensively studied worldwide in major laboratories, including BNL, JLab, Fermilab, DESY, SLAC, CERN, PSI, J-PARC, and MAMI. 
Key quantities for mapping proton structure encompass parton distribution functions (PDFs), generalized parton distributions (GPDs)~\cite{Mueller:1998fv, Ji:1996ek, Radyushkin:1996nd}, and transverse-momentum-dependent distributions (TMDs). 
These are inferred from experimental data from high-energy scattering processes, which is possible due to the asymptotic freedom of the strong coupling. 
In particular, asymptotic freedom enables the use of QCD factorization theorems~\cite{Collins:1989gx} to isolate the universal non-perturbative component of the underlying process, namely the PDFs, GPDs, and TMDs.

Distribution functions are classified according to their twist, indicating the order of a $1/Q$-expansion they appear in, with $Q$ the hard scale of the physical process. 
While extensive studies have been conducted on twist-2 contributions (leading twist), our understanding of twist-3 contributions remains very limited. 
However, twist-3 effects cannot be disregarded given the energy scales explored in current experiments; in fact, they are presumed to be significant, an aspect we address in this study for the case of GPDs. 

Generally, extracting GPDs from experimental data poses a very challenging inverse problem~\cite{Bertone:2021yyz, Moffat:2023svr}. 
This applies particularly to twist-3 GPDs.
But these objects, in addition to being important for reliable extraction of twist-2 GPDs, are interesting in their own right for a number of reasons.
First, there exists a non-trivial relation between a particular twist-3 GPD and the orbital angular momentum of quarks~\cite{Penttinen:2000dg}.
Twist-3 GPDs have also been shown to provide information about the (average) transverse force acting on a quark in a polarized nucleon~\cite{Burkardt:2008ps, Aslan:2019jis}.
Furthermore, certain spin-orbit correlations of the nucleon can be expressed through twist-3 GPDs~\cite{Lorce:2014mxa, Bhoonah:2017olu}. 
Finally, twist-3 GPDs can be related to Wigner functions~\cite{Meissner:2008ay, Meissner:2009ww, Lorce:2013pza, Rajan:2016tlg, Rajan:2017cpx}, the most general objects quantifying the parton structure of hadrons~\cite{Belitsky:2003nz}.
All these considerations serve as significant motivations for undertaking lattice-QCD calculations of twist-3 GPDs, a task that is highly demanding but holds tremendous promise for providing valuable insights into these quantities.

Direct calculations of GPDs in a Euclidean lattice are well known to be prohibited due to their light-cone definition. Consequently, for several decades, limited information on GPDs has been and continues to be, extracted from their Mellin moments. However, the development of alternative approaches for accessing GPDs in momentum ($x$) space has sparked a highly promising research program in lattice QCD. In this study, we employ the quasi-distributions method~\cite{Ji:2013dva}, which relies on calculating matrix elements with momentum-boosted hadrons coupled to non-local operators. To ensure gauge invariance for quark GPDs, the quark and anti-quark operators in position space include a straight Wilson line connecting the spatially separated fields. The resulting (momentum-space) quasi-GPDs are then obtained as the Fourier transform of these matrix elements. Finally, to establish a connection with the light-cone GPDs, these entities are related through the framework of Large-Momentum Effective Theory (LaMET)~\cite{Ji:2014gla}. Extensive reviews of the quasi-distribution approach, along with other methods for obtaining $x$-dependent distribution functions, can be found in Refs.~\cite{Cichy:2018mum,Ji:2020ect,Constantinou:2020pek,Cichy:2021lih,Cichy:2021ewm}. 
The dominating trend in references reported therein is calculations of PDFs, the simplest one-dimensional distributions, and hence natural starting points for novel lattice methods.
On the other hand, the applications concerning GPDs, three-dimensional distributions embracing several new aspects with respect to PDFs, have so far been relatively limited, see e.g.~Refs.\ \cite{Ji:2015qla, Xiong:2015nua, Bhattacharya:2018zxi, Liu:2019urm, Bhattacharya:2019cme, Chen:2019lcm, Radyushkin:2019owq, Ma:2019agv, Luo:2020yqj, Alexandrou:2020zbe,Alexandrou:2021bbo,CSSMQCDSFUKQCD:2021lkf,Dodson:2021rdq, Ma:2022ggj, Shastry:2022obb, Ma:2022gty,Ji:2022thb}.
In a recent publication~\cite{Bhattacharya:2022aob} (see also Refs.~\cite{Bhattacharya:2023tik,Constantinou:2022fqt,Cichy:2023dgk,Bhattacharya:2023ays} for summaries of this work and some additional results), we made significant progress in enhancing the efficiency and speed of lattice-QCD calculations pertaining to GPDs. This advancement was achieved by leveraging asymmetric frames (in contrast to the traditionally-used symmetric frames) and introducing flexibility in the distribution of transferred momentum. In that work, we presented the initial lattice results of the twist-2 unpolarized GPDs of quarks, $H$ and $E$, at zero skewness.

In the current work, the thread of lattice calculations of GPDs is combined with one of the determinations of twist-3 PDFs from the lattice.
For the latter, we have derived (simplified) matching relations in LaMET~\cite{Bhattacharya:2020xlt,Bhattacharya:2020jfj} (see Refs.~\cite{Braun:2021aon,Braun:2021gvv} for a full treatment including 3-parton correlators) and obtained lattice results for the $g_T$ and $h_L$ functions \cite{Bhattacharya:2020cen,Bhattacharya:2021moj} (see also Refs.~\cite{Bhattacharya:2022fai,Bhattacharya:2021rua,Bhattacharya:2022gfu,Constantinou:2021nbn}).
The present focus is on the chiral-even axial twist-3 GPDs of the proton, which is both computationally and theoretically challenging: \textbf{(a)} matrix elements with momentum-boosted proton states at off-forward kinematics have decreased signal compared to the forward limit; \textbf{(b)} non-local operators contribute to the signal decay and the enhancement of systematic uncertainties, such as finite-volume and discretization effects; \textbf{(c)} the definition of GPDs in the symmetric frame means that each value of momentum transfer requires a separate calculation; \textbf{(d)} the parametrization of each matrix element contains four twist-3 GPDs~\cite{Kiptily:2002nx}, which means that one needs four independent matrix elements to disentangle all GPDs. Such a cost at a fixed value of the momentum transfer is four times more computationally expensive than the PDF case and two times more than the (twist-2) helicity GPDs; \textbf{(e)}  at nonzero skewness ($\xi$), the perturbative formalism breaks down around $x = |\xi|$ due to divergent higher-twist contributions~\cite{Bhattacharya:2021grn}.

Twist-3 GPDs may exhibit discontinuities at $x = |\xi|$; see~\cite{Aslan:2018zzk, Aslan:2018tff} and references therein. 
Nonetheless, Ref.~\cite{Aslan:2018zzk} demonstrates that these discontinuities cancel out in the linear combinations of twist-3 GPDs involved in deeply-virtual Compton scattering amplitudes, ensuring the consistency of twist-3 amplitudes with QCD factorization. Although this intriguing feature warrants future investigation in lattice-QCD studies, our current focus will primarily be on the specific case of $\xi=0$.

We organize the manuscript as follows: In Section~\ref{sec:def}, we present the definition of light-cone and quasi twist-3 GPDs. Moving forward to Section~\ref{sec:method}, we provide the Euclidean decompositions of lattice-computable matrix elements in terms of the quasi-GPDs in position space. Additionally, we discuss the matching procedure required for extracting the light-cone GPDs from the quasi-GPDs. In Section~\ref{sec:lat_calc}, we discuss our lattice setup and outline the essential components of the renormalization process. Section~\ref{sec:results} is dedicated to our numerical results, accompanied by a detailed analysis of several consistency checks conducted to validate our findings. Finally, in Section~\ref{sec:summary}, we draw conclusions and discuss future prospects in light of our study.

\section{Definition of twist-3 GPDs}
\label{sec:def}

To begin, let us review the definition of light-cone GPDs of quarks for a spin-1/2 hadron. The relevant correlator is determined by the Fourier transform of off-forward matrix elements involving non-local quark fields (see, for instance Ref.~\cite{Diehl:2003ny}),
\begin{equation}
F^{[\Gamma]}(x, \Delta)=\frac{1}{2} \int \frac{dz^{-}}{2\pi} e^{i k \cdot z} \langle p_f;\lambda'| \bar{\psi} (-\tfrac{z}{2})\, \Gamma \, {\cal W}(-\tfrac{z}{2},\tfrac{z}{2}) \psi (\tfrac{z}{2})|p_i;\lambda \rangle \Big |_{z^{+}=0,\vec{z}_{\perp}=\vec{0}_{\perp}} \, ,
\label{e:corr_standard_GPD}
\end{equation} 
where $\Gamma$ represents a gamma matrix.
Color gauge invariance of this correlator is enforced by the Wilson line
\begin{equation}
{\cal W}(-\tfrac{z}{2},\tfrac{z}{2})\Big|_{z^{+}=0,\vec{z}_{\perp}=\vec{0}_{\perp}} = {\cal P} \, {\rm exp} \bigg ( -ig \int^{\tfrac{z^{-}}{2}}_{-\tfrac{z^{-}}{2}} \, dy^{-} A^{+}(0^{+},y^{-},\vec{0}_{\perp}) \bigg ) \,. 
\label{e: wilson_line_standard_GPD}
\end{equation}
In Eq.~(\ref{e: wilson_line_standard_GPD}), $g$ is the strong coupling constant, and $A^{+}$ is the light-cone plus component of the gluon field. The incoming (outgoing) hadronic state in Eq.~(\ref{e:corr_standard_GPD}) is characterized by its 4-momentum $p_i \, (p_f)$ and helicity $\lambda \, (\lambda')$. 
We define the kinematic variables.
\begin{equation}
P= \frac{1}{2}(p_i+p_f), \qquad
\Delta =p_f-p_i, \qquad 
\xi =\frac{p^{+}_i-p^{+}_f}{p^{+}_i+p^{+}_f}, \qquad
t = \Delta^{2} \,,
\end{equation}
where $P$ denotes the average 4-momentum of the hadrons, while $\Delta$ represents the 4-momentum transfer to the hadron. The skewness variable $\xi$, defined here for hadrons with a large light-cone plus-momentum, signifies the longitudinal momentum transfer to the hadron. 
In the ``symmetric frame" of reference, defined by $\vec{P}_{\perp} = 0$, we can obtain the following expression for $t$ for a nonzero transverse momentum transfer $\vec{\Delta}_{\perp}$,
\begin{eqnarray} 
t = -\frac{1}{1-\xi^{2}} (4\xi^{2}m^{2}+\vec{\Delta}^{2}_{\perp})\, ,
\label{e: t_s}
\end{eqnarray}
with $m$ denoting the nucleon mass. 
Within the scope of twist-3, the correlator involving $\Gamma = \gamma^\mu \gamma_5$ in Eq.~(\ref{e:corr_standard_GPD}) can be parameterized in terms of six distinct axial-vector GPDs. In momentum space, one finds~\cite{Kiptily:2002nx}
\begin{align}
F^{[\gamma^\mu \gamma_5]} (x, \Delta)&  = \frac{1}{2 P^+} \bar{u}(p_f, \lambda ') \bigg [ P^{\mu} \frac{\gamma^+ \gamma_5}{P^+} \widetilde{H} (x, \xi, t) +  P^{\mu} \frac{\Delta^+ \gamma_5}{2m P^+} \widetilde{E} (x, \xi, t) \nonumber \\
& \hspace{2.5cm} + \Delta^\mu_\perp \dfrac{\gamma_5}{2m} \big (\widetilde{E} (x, \xi, t)+\widetilde{G}_1 (x, \xi, t) \big ) + \gamma^\mu_\perp \gamma_5 \big (\widetilde{H}(x, \xi, t)+\widetilde{G}_2(x, \xi, t) \big ) \nonumber \\[0.1cm]
& \hspace{2.5cm} + \Delta^\mu_\perp \dfrac{\gamma^+ \gamma_5}{P^+} \widetilde{G}_3(x, \xi, t) + i \varepsilon^{\mu \nu}_\perp \Delta_{\nu} \dfrac{\gamma^+}{P^+} \widetilde{G}_4 (x, \xi, t)\bigg ] u(p_i, \lambda) \, .
\label{e:GPD_def}
\end{align}
The GPDs $\widetilde{H}$ and $\widetilde{E}$ characterize the twist-2 contribution of the correlator with $\mu=+$, while the GPDs $\widetilde{G}_i$ ($i=1, 2, 3, 4$) come into play at twist-3 accuracy if $\mu$ represents a transverse index. Additionally, $\varepsilon^{\mu \nu}_{\perp} \equiv \varepsilon^{- + \mu \nu} = \varepsilon^{03\mu \nu}$, where we adopt the convention $\varepsilon^{0123}=1$. 
We point out that Eq.~(\ref{e:GPD_def}) remains applicable in the position space, with the understanding that the arguments of the GPDs are transformed from $(x, \xi, t)$ to $(z, \xi, t)$. 
Note that different conventions for twist-3 light-cone GPDs are available in the literature~\cite{Belitsky:2001ns, Meissner:2009ww}; see also Ref.~\cite{Aslan:2018zzk} where the relations between several definitions can be found.

Now, let us shift our focus to quasi-GPDs, which are defined through the equal-time correlator~\cite{Ji:2013dva}
\begin{equation}
F^{[\Gamma]}(x, \Delta; P^{3}) = \frac{1}{2} \int \frac{dz^{3}}{2\pi}  e^{ik \cdot z} \langle p_f,\lambda '| \bar{\psi}(-\tfrac{z}{2}) \, \Gamma \, {\cal W}(-\tfrac{z}{2}, \tfrac{z}{2}) \psi (\tfrac{z}{2})|p_i, \lambda \rangle \Big |_{z^{0}=0, \vec{z}_{\perp}=\vec{0}_{\perp}} \, ,
\label{e: corr_quasi_GPD}
\end{equation}
where the Wilson line is given by
\begin{equation}
{\cal W}(-\tfrac{z}{2},\tfrac{z}{2})\Big|_{z^{0}=0,\vec{z}_{\perp}=\vec{0}_{\perp}} = {\cal P} \, {\rm exp} \bigg ( -ig \int^{\tfrac{z^{3}}{2}}_{-\tfrac{z^{3}}{2}} \, dy^{3} A^{3}(0, \vec{0}_{\perp}, y^{3}) \bigg )\, .
\end{equation}
The counterpart of Eq.~(\ref{e:GPD_def}) is,
\begin{align}
F^{[\gamma^\mu \gamma_5]} (x, \Delta; P^3)&  = \frac{1}{2 P^3} \bar{u}(p_f, \lambda ') \bigg [ P^\mu \dfrac{\gamma^3 \gamma_5}{P^0} F_{\widetilde{H}} (x, \xi, t; P^3) +  P^\mu \frac{\Delta^3 \gamma_5}{2m P^0} F_{\widetilde{E}} (x, \xi, t;P^3) \nonumber \\
& \hspace{2.5cm} + \Delta^\mu_\perp \dfrac{\gamma_5}{2m}  F_{\widetilde{E} +\widetilde{G}_1} (x, \xi, t;P^3) + \gamma^\mu_\perp \gamma_5 F_{\widetilde{H}+\widetilde{G}_2}(x, \xi, t;P^3)  \nonumber \\[0.1cm]
& \hspace{2.5cm} + \Delta^\mu_\perp \dfrac{\gamma^{3} \gamma_5}{P^{3}} F_{\widetilde{G}_3} (x, \xi, t;P^3) + i \varepsilon^{\mu \nu}_\perp \Delta_{\nu} \dfrac{\gamma^{3}}{P^{3}} F_{\widetilde{G}_4} (x, \xi, t;P^3)\bigg ] u(p_i, \lambda) \, ,
\label{e:tw3_axial}
\end{align}
where $F_{X}$ is the notation for quasi-GPDs. In Ref.~\cite{Constantinou:2017sej}, it was proposed to define the quasi-counterpart of $\widetilde{H}$ and $\widetilde{E}$ using $\Gamma = \gamma^3 \gamma_5$. The underlying argument was that $\gamma^{3}\gamma_{5}$ does not mix with other operators during renormalization, where this mixing is regarded as a lattice artifact. Ref.~\cite{Bhattacharya:2019cme} argued that with this definition, it is necessary to replace $\gamma^+ \gamma_5/P^+ \rightarrow \gamma^3\gamma_5/P^0$ in the prefactor of $\widetilde{H}$ to ensure consistency with the forward limit. This adjustment serves to justify the corresponding prefactors outlined in Eq.~(\ref{e:tw3_axial}). In the context of twist-3 GPDs, an important question arises concerning the selection of the prefactor for $\widetilde{G}_{3/4}$. Specifically, the debate centers around whether one should utilize $\gamma^+ \gamma_5/P^+$ ($\gamma^+/P^+$) or $\gamma^{3} \gamma_5/P^3$ ($\gamma^{3}/P^3$) as the appropriate choice for $\widetilde{G}_{3}$ ($\widetilde{G}_4)$. 
Following the reasoning presented in Ref.~\cite{Bhattacharya:2019cme}, we can conclude that due to the absence of forward counterparts for $\widetilde{G}_{3/4}$, there is no compelling reason to favor either of the two definitions. In our lattice results for $\widetilde{G}_4$, we will explore the $\gamma^{3}/P^3$ definition. It is worth noting that the two definitions for $\widetilde{G}_4$ are interconnected by a simple relation, namely
\begin{align}
F^{(3)}_{\widetilde{G}_4} & = \sqrt{1+ \dfrac{m^{2}-t/4}{(P^{3})^2}} \, F^{(+)}_{\widetilde{G}_4} \, ,
\label{e:simple_rel}
\end{align}
with $F^{(3)}_{\widetilde{G}_4}$ ($F^{(+)}_{\widetilde{G}_4}$) denoting the definition that employs $\gamma^{3}/P^3$ ($\gamma^{+}/P^+$) as the prefactor. In the next paragraph, we will delve into a discussion that shows the irrelevance of the definition for $\widetilde{G}_3$. Theoretically, it is expected to be zero for the specific kinematics that we are focusing on in this work.

In the following discussion, we briefly explore the symmetry properties of GPDs in position space, as this consideration plays a crucial role in leveraging symmetries to enhance statistics for lattice calculations. 
The hermiticity constraint provides insights into the symmetries of GPDs under the transformation $P^3 \rightarrow -P^3$ for a fixed value of $z^3$. 
Furthermore, for a fixed $P^3$ and with $\Delta \rightarrow -\Delta$, the hermiticity constraint reveals the symmetries of GPDs under the transformation $z^3 \rightarrow -z^3$. 
Notably, we find that the real part satisfies $\widetilde{G}_1(-P^3) = +\widetilde{G}_1(P^3)$, $\widetilde{G}_2(-P^3) = +\widetilde{G}_2(P^3)$, $\widetilde{G}_3(-P^3) = -\widetilde{G}_3(P^3)$, and $\widetilde{G}_4(-P^3) = +\widetilde{G}_4(P^3)$ (where we used that $\widetilde{H}(-P^3) = +\widetilde{H}(P^3)$ and likewise for $\widetilde{E}$). 
Additionally, we find $\widetilde{G}_1(-z^3) = +\widetilde{G}_1(z^3)$, $\widetilde{G}_2(-z^3) = +\widetilde{G}_2(z^3)$, $\widetilde{G}_3(-z^3) = -\widetilde{G}_3(z^3)$, and $\widetilde{G}_4(-z^3) = +\widetilde{G}_4(z^3)$ (where we used that $\widetilde{H}(-z^3) = +\widetilde{H}(z^3)$ and likewise for $\widetilde{E}$). 
By combining hermiticity with the time-reversal constraint, it follows that $\widetilde{G}_1$, $\widetilde{G}_2$, and $\widetilde{G}_4$ exhibit even behavior under $\xi \rightarrow -\xi$ (with the use of $\widetilde{H}$ and $\widetilde{E}$ are even, as demonstrated in Ref.~\cite{Bhattacharya:2019cme}), while $\widetilde{G}_3$ exhibits odd behavior. 
Furthermore, considering the requirement that matrix elements possess a well-defined forward limit, it can be inferred that $\widetilde{G}_3$ should exhibit at least linear scaling with respect to $\xi$ and should not exhibit a pole at $\xi=0$. Consequently, an expected outcome is that $\widetilde{G}_3$ assumes a value of zero when $\xi = 0$. This, of course, should be validated by our lattice results. 

The lattice data presented here offers an opportunity for additional consistency checks. In this regard, we highlight specific sum rules that convey the following information:
\begin{align}
G_P (t) & = \int^{1}_{-1} dx \, \big (\widetilde{E} (x, \xi, t)+\widetilde{G}_1 (x, \xi, t) \big ) \rightarrow \int^{1}_{-1} dx \, \widetilde{G}_1 (x, \xi, t) = 0\,,  \label{e:norm1} \\
G_A (t) & = \int^{1}_{-1} dx \, \big (\widetilde{H} (x, \xi, t)+\widetilde{G}_2 (x, \xi, t) \big ) \rightarrow \int^{1}_{-1} dx \, \widetilde{G}_2 (x, \xi, t) = 0\,, 
\end{align}
as well as 
\begin{align}
\int^{1}_{-1} dx \, \widetilde{G}_{3} (x, \xi, t) & = 0 \, , \\
 \int^{1}_{-1} dx \, \widetilde{G}_{4} (x, \xi, t) & = 0  \, .
\label{e:norm2}
\end{align}
These relations stem from the well-established fact, as documented in Ref.~\cite{Kiptily:2002nx,Diehl:2003ny,Lorce:2014mxa}, that integrating non-local operators over $x$ results in local operators. The above relations are commonly recognized as the non-forward extensions of the Burkhardt-Cottingham (BC) sum rules~\cite{Kiptily:2002nx}. Similar relations can be easily confirmed for the quasi-GPDs as well, albeit with the integration range over $x$ extended to $[-\infty, \infty]$, see Refs.~\cite{Bhattacharya:2019cme,Bhattacharya:2021boh}.

\section{Methodology}
\label{sec:method}
In this work, we implement the large momentum effective theory (LaMET) approach using matrix elements of non-local fermion operators where the fermion fields are spatially separated. 
Without any loss of generality, we choose the spatial separation to be along the $\hat{z}$ axis. 
The momentum boost must be in the same direction as the Wilson line, that is, $\vec{P}=(0,0,P_3)$~\footnote{Starting from this section, we will use lower indices for $P$ and $\Delta$, indicating that we are working in Euclidean space.}. 
For the case of GPDs, we calculate off-forward matrix elements with the direction of the momentum transfer perpendicular to $P_3$, $\vec{\Delta}=(\Delta_x,\Delta_y,0)$. 
The class of these values of $\vec{\Delta}$ corresponds to zero skewness, $\xi$. 
The latter is an important variable of GPDs, which is related to the momentum transfer in the direction of the boost. 
For the twist-3 GPDs under study, we construct the matrix elements
\begin{equation}
\label{eq:ME}
 h_j ( \Gamma_\kappa, z, p_f, p_i, \mu) = Z_{\gamma_j\gamma_5}(z,\mu) \bra{N(p_f)} \overline{\psi}(z) \, \gamma_j \,\gamma_5 \,\mathcal{W}(z,0) \psi(0) \ket{N (p_i)}\,, \quad j=1,\,2\,.
\end{equation}
Note that, for the twist-3 contributions, the index $j=1,\,2$ is spatial and perpendicular to the direction of the boost. 
GPDs require $\vec{p_f} - \vec{p_i} = \vec{\Delta} \ne 0$, which in the symmetric frame corresponds to $\vec{p}_f = \vec{P} + \frac{\vec{\Delta}}{2}$ and $\vec{p}_i= \vec{P} - \frac{\vec{\Delta}}{2}$. 
Here, we focus on zero skewness.
Unlike the matrix elements, $h_j$, GPDs depend on the 4-vector momentum transfer squared, $-t\equiv \vec{\Delta}^2 - (E_f-E_i)^2$, where $E_i$ and $E_f$ are the energies of the initial and final state ($E_{i/f}=\sqrt{m^2+\vec{p}_{i/f}^2}$). 
The multiplicative factor $Z_{\gamma_j\gamma_5}$ appearing in Eq.~\eqref{eq:ME} is the renormalization function for the operators under study and is calculated non-perturbatively. 
The renormalization function introduces a scheme and scale dependence in the matrix element, $h_j$, which is chosen to be the modified $\overline{\rm MS}$ (M$\overline{\rm MS}$) scheme at a scale of 2 GeV. 
Note that the final GPDs will be converted to the standard $\overline{\rm MS}$ scheme through the matching formalism. 
More discussion can be found towards the end of this section. 
The proton polarization of $h_j$ is chosen by the parity projector, $\Gamma_\kappa$. We distinguish between the unpolarized and polarized cases defined as
\begin{eqnarray}
\label{eq:P0}
\Gamma_0 &=& \frac{1}{4} \left( \mathbb{1} +\gamma^0\right) \\
\Gamma_{k} &=& \frac{1}{4} \left( \mathbb{1} +\gamma^0\right)\,i\gamma^5\gamma^k\,, \quad k=1,\,2,\,3 \,.
\label{eq:Pk}
\end{eqnarray}

The lattice matrix elements are parametrized according to the decomposition of Refs.~\cite{Kiptily:2002nx,Aslan:2018zzk}
\begin{equation}
 h_j = C\, {\rm Tr}\Big[\Gamma_\kappa \, \left(\frac{-ip_f \hspace*{-0.33cm}\slash \,\,+m}{2m} \right) \, {\cal F}^{[\gamma_j \gamma_5]} \left(\frac{-ip_i \hspace*{-0.33cm}\slash \,\,+m}{2m} \right) \Big]\,,
 \label{eq:tr} 
\end{equation}
where Eq.~\eqref{e:tw3_axial} in Euclidean space takes the form
\begin{eqnarray}
 \label{eq:a_decomp}
{\cal F}^{[\gamma_j \gamma_5]} = 
-i  \frac{\Delta_j\,\gamma_5}{2m} \,  F_{\widetilde{E}+ \widetilde{G}_1} + \gamma_{j}\,\gamma_5\  \, F_{\widetilde{H}+ \widetilde{G}_2}
+ \frac{\Delta_{j}\,  \gamma_3  \gamma_5}{P_3} \, \widetilde{G}_3
-   \frac{{\rm sign}[P_3]\,\varepsilon_\perp^{j\,\rho}\Delta_\rho\,\gamma_3}{P_3} \, \widetilde{G}_4\,,
\end{eqnarray}
For simplicity, we omit the arguments of the matrix elements, $ h_j \equiv h_j ( \Gamma_\kappa, z, p_f, p_i, \mu)$, $ F_X \equiv F_X(z,  \xi, t, P_3,\mu)$. 
The kinematic factor $C$ in the symmetric frame at zero skewness reads $C=\frac{2 m^2}{E (E+m)}$ ($E_i=E_f \equiv E$).
The appearance of ${\rm sign}[P_3]$ in the coefficient of $\widetilde{G}_4$ is to account for the antisymmetry of $\varepsilon_\perp^{j\,\rho}$ for negative values of $P_3$. 
Another important aspect of the parametrization in Euclidean space is the freedom to choose the temporal or the $z$ component of $\gamma_+/P_+$, which appears in the $\widetilde{G}_3$ and $\widetilde{G}_4$ terms. 
As discussed in Sec.~\ref{sec:def} one can relate the definition of $\widetilde{G}_3$ and $\widetilde{G}_4$ using $\gamma_3/P_3$, to the one obtained from $\gamma_+/P_+$, as discussed in Eq.~(\ref{e:simple_rel}). 
We also provide numerical results in Sec.~\ref{sec:results_LC}. 
The decomposition of Eq.~\eqref{eq:a_decomp} shows four combinations of the quasi GPDs, $F_X$: $F_{\widetilde{E}}+ F_{\widetilde{G}_1}$, $F_{\widetilde{H}}+\widetilde{G}_2$, $\widetilde{G}_3$ and $\widetilde{G}_4$. 
$F_{\widetilde{H}}$ and $F_{\widetilde{E}}$ are twist-2 contributions, while $F_{\widetilde{G}_i}$ enter at the twist-3 contributions, that is, they drop out of the decomposition of the operators $\gamma_3 \gamma_5$ and $\gamma_0 \gamma_5$. 
Based on Eq.~\eqref{eq:a_decomp}, the forward limit of the matrix elements $h_1$ and $h_2$ give the combination $F_{\widetilde{H}+ \widetilde{G}_2}$ called $g_T$. The latter is the 2-parton twist-3 PDF that we calculated in Ref.~\cite{Bhattacharya:2020cen}. 

To extract the twist-3 GPDs, one needs to calculate and disentangle four independent matrix elements. 
Thus, we use the unpolarized (Eq.~\eqref{eq:P0}), and three polarized (Eq.~\eqref{eq:Pk}) parity projectors that successfully disentangle the quasi-GPDs. At zero skewness, there are, in general, eight contributing matrix elements, that is
\begin{align}    
\label{eq:Pi1Gamma0}
{\Pi^1(\Gamma_0)}&=
 C \Bigg( -F_{\widetilde{H}+ \widetilde{G}_2} \frac{   P_3   \Delta_y }{4 m^2} 
 -F_{\widetilde{G}_4}   \frac{{\rm sign}[P_3]\, \Delta_y ( E +m)}{2
    m^2}   \Bigg) \,,\,\,\,\,\, \\[4ex]  
\label{eq:Pi1Gamma1}
{\Pi^1(\Gamma_1)}&=
i\, C \Bigg( F_{\widetilde{H}+ \widetilde{G}_2}  \frac{\left(4 m
    ( E +m)+ \Delta_y ^2\right)}{8 m^2}
- F_{\widetilde{E}+ \widetilde{G}_1}  \frac{  \Delta_x ^2
    ( E +m)}{8 m^3} + F_{\widetilde{G}_4}  \frac{{\rm sign}[P_3]\,
     \Delta_y ^2 ( E +m)}{4 m^2  P_3 }
         \Bigg) \,,\,\,\,\,\, \qquad  \\[4ex]  
\label{eq:Pi1Gamma2}
{\Pi^1(\Gamma_2)}&=
i\, C \Bigg( -F_{\widetilde{H}+ \widetilde{G}_2}\frac{ \Delta_x  \Delta_y }{8
    m^2}
-F_{\widetilde{E}+ \widetilde{G}_1} \frac{ \Delta_x  \Delta_y ( E +m)}{8
    m^3}-F_{\widetilde{G}_4} \frac{{\rm sign}[P_3]\,\Delta_x  \Delta_y ( E +m)}{4 m^2
     P_3 }
        \Bigg) \,,\,\,\,\,\, \phantom{.......} \\[4ex]  
\label{eq:Pi1Gamma3}
{\Pi^1(\Gamma_3)}&=
C \Bigg( -
 F_{\widetilde{G}_3} \frac{ E     \Delta_x ( E +m)}{2 m^2      P_3 }
        \Bigg) \,,\,\,\,\,\, \\[4ex]  
\label{eq:Pi2Gamma0}
{\Pi^2(\Gamma_0)}&=
C \Bigg(F_{\widetilde{H}+ \widetilde{G}_2} \frac{   P_3   \Delta_x }{4
    m^2} +
F_{\widetilde{G}_4} \frac{ {\rm sign}[P_3]\,  \Delta_x 
    ( E +m)}{2 m^2}
        \Bigg) \,,\,\,\,\,\,  
\end{align}
\begin{align}
\label{eq:Pi2Gamma1}
{\Pi^2(\Gamma_1)}&=
i\, C \Bigg( 
- F_{\widetilde{H}+ \widetilde{G}_2}\frac{
     \Delta_x  \Delta_y }{8 m^2}
-F_{\widetilde{E}+ \widetilde{G}_1} \frac{  \Delta_x 
     \Delta_y ( E +m)}{8 m^3}-F_{\widetilde{G}_4} \frac{ {\rm sign}[P_3]\,  \Delta_x 
     \Delta_y ( E +m)}{4 m^2  P_3 }
        \Bigg) \,,\,\,\,\,\, \qquad \\[4ex]  
\label{eq:Pi2Gamma2}
{\Pi^2(\Gamma_2)}&=
i\, C \Bigg( 
F_{\widetilde{H}+ \widetilde{G}_2}\frac{ \left(
     4 m (E  + m) + \Delta_x^2 \right)}{8
    m^2}
- F_{\widetilde{E}+ \widetilde{G}_1} \frac{
     \Delta_y ^2 ( E +m)}{8 m^3}+F_{\widetilde{G}_4} \frac{  {\rm sign}[P_3]\,  \Delta_x ^2 ( E +m)}{4 m^2
     P_3 }
        \Bigg) \,,\,\,\,\,\, \\[4ex]  
\label{eq:Pi2Gamma3}
{\Pi^2(\Gamma_3)}&=
C \Bigg(-
F_{\widetilde{G}_3} \frac{ E    \Delta_y 
    ( E +m)}{2 m^2  P_3}
        \Bigg)  \,,\,\,\,\,\, 
\end{align}
where $\Pi^j(\Gamma_\kappa)$ is the ground state contribution to the matrix element $h_j( \Gamma_\kappa, z, p_f, p_i, \mu)$. 
The extraction of the ground state is discussed in Sec.~\ref{sec:comp} (Eq.~\eqref{eq:ratio}). 
As a demonstration of how we isolate GPDs, let us consider momentum transfer of the class $\vec{\Delta}=(\Delta_x,0,0)$. 
In this case, the matrix elements of Eqs.~\eqref{eq:Pi1Gamma1},~\eqref{eq:Pi1Gamma3},~\eqref{eq:Pi2Gamma0},~\eqref{eq:Pi2Gamma2} are non-vanishing and independent. 
Thus, $F_{\widetilde{G}_3}$ is given by $\Pi^1(\Gamma_3)$, while $F_{\widetilde{E}+ \widetilde{G}_1}$, $F_{\widetilde{H}+ \widetilde{G}_2}$ and $F_{\widetilde{G}_4}$ are extracted from combining $\Pi^1(\Gamma_1)$, $\Pi^2(\Gamma_0)$ and $\Pi^2(\Gamma_2)$. 
The inversion of Eqs.~\eqref{eq:Pi1Gamma0} - \eqref{eq:Pi2Gamma3} can be done analytically for a general form of $\vec{\Delta}$ but leads to lengthy expressions, so we do not present it here.
In practice, we implement a numerical inversion for each value of $P_3$ and $\vec{\Delta}$ separately, which leads to identical coefficients as the analytical case.

\medskip
The renormalization of the axial operator is extracted non-perturbatively in the RI$'$ scheme~\cite{Martinelli:1994ty}, using the momentum source method~\cite{Gockeler:1998ye,Alexandrou:2015sea} that offers high statistical accuracy. The prescription for the nonlocal operator, $Z_{\gamma_j\gamma_5}$, is
\bea
\label{renorm}
Z^{-1}_{\gamma_j\gamma_5}(z,\mu_0) =\left( {\rm Tr} \left[S^{-1}\, S^{\rm tree} \right] \right)^{-1}  {\rm Tr} \left[ {\cal V}^j_A(p,z) \left({\cal V}_A^{j, \rm tree}(p,z)\right)^{-1}\right]_{p^2{=}\bar\mu_0^2} \,,
\eea
where the first trace determines the fermion field renormalization, ${\cal V}(p,z)$ ($S\equiv S(p)$) is the amputated vertex function of the operator (fermion propagator), and ${\cal V}^{{\rm tree}}$ ($S^{{\rm tree}}(p)$) is its tree-level value. 
$Z_{\gamma_j\gamma_5}$ is calculated on various pion mass ensembles, and a chiral extrapolation is required to extract the appropriate mass-independent estimate ${Z}^{\rm RI}_{\gamma_j\gamma_5,0}(z,\mu_0)$. Here we use the fit
\begin{equation}
\label{eq:Zchiral_fit}
Z^{\rm RI}_{\gamma_j\gamma_5}(z,\bar\mu_0,m_\pi) = Z^{\rm RI}_{\gamma_j\gamma_5,0}(z,\mu_0) + m_\pi^2 \,Z^{\rm RI}_{\gamma_j\gamma_5,1}(z,\mu_0) \,,
\end{equation}
which successfully eliminates the pion mass dependence in the renormalization functions~\cite{Alexandrou:2019lfo}.
The chirally extrapolated values are converted to the $\overline{\rm MS}$ scheme and evolved to $\mu{=}2$ GeV using the results of Ref.~\cite{Constantinou:2017sej}. 
We obtain ${Z}^{\overline{\rm MS}}_{\gamma_j\gamma_5,0}(z,2 {\rm GeV})$ by extrapolating $(a\,p)^2 \to 0$ using a linear fit and data in the region $(a\,\bar\mu_0)^2 \in [1,2.6]$. 
In this work, we also average the estimates for $Z_{\gamma_1\gamma_5}$ and $Z_{\gamma_2\gamma_5}$. 
We give our final estimates in the so-called modified $\MSb$ ($\MMSb$) scheme~\cite{Alexandrou:2019lfo}, which is also used in the matching kernel discussed below.

\medskip
The quasi-GPDs in coordinate space, $F_X(z)$, must be reconstructed in the momentum space, $F_X(x)$, which is defined through the Fourier transform
 \begin{equation}
 F_X(x,\xi,t,P_3,\mu) = \int\frac{dz}{4\pi} e^{-ixP_3z}F_X(z,  \xi, t, P_3,\mu)\,. \label{e:convo}
   \end{equation} 
The integration over $z$ has, in principle, limits $\in (-\infty,+\infty)$. 
However, the lattice discretization and periodicity give access to a limited number of $z$ values, that is, $z \in [0,L/2]$, where $L$ is the spatial extent of the lattice. 
Thus, the integral of Eq.~\eqref{e:convo} becomes a summation and is truncated to some $z_{\rm max}$, causing an ill-posed inverse problem in the reconstruction of the $x$-dependence that does not have a unique solution (see Ref.~\cite{Karpie:2019eiq} for a detailed discussion of this aspect). 
The simplest assumption one can make is that $F_X$ is zero beyond some $z_{\rm max}$, which introduces a bias and systematic uncertainties in the final GPDs. 
To avoid imposing such an assumption, we use the model-independent Backus-Gilbert reconstruction method~\cite{BackusGilbert}. 
More details on the implementation can be found in Ref.~\cite{Alexandrou:2021bbo}. 

While the quasi-GPDs are defined in momentum space, they have a dependence on the finite momentum $P_3$ inherited from the matrix elements. 
As $P_3$ increases, the matching formalism, which is based on large momentum effective theory~\cite{Ji:2014gla}, relates the quasi-GPDs to the light-cone GPDs, $G_X$, with a better convergence. 
The matching kernel, $C_X$, is calculated order by order in perturbation theory and, at one-loop for zero skewness, reads
  \begin{equation}
  \label{eq:matching}
     F_X^{\MMSb}(x,t,P_3,\mu) = \int_{-1}^1 \frac{dy}{|y|} \,C_{\gamma_j\gamma_5}^{\MMSb, \MSb}\left(\frac{x}{y},\frac{\mu}{yP_3}\right) \,G_X^\MSb(y,t,\mu) \,\,+\, \mathcal{O}\left(\frac{m^2}{P_3^2}, \frac{t}{P_3^2}, \frac{\Lambda_{\rm QCD}^2}{x^2P_{3_{\phantom{L}}}^2}\right)\,.
      \end{equation}
(In this equation, we have simplified the arguments of $F_X$, $C$, and $G$ since we focus on $\xi =0$.) Note that since the calculation is performed at $\xi = 0$, we expect that the matching equations coincide with those of the GPD's forward limit, as discussed in Ref.~\cite{Liu:2019urm} in the context of twist-2 GPDs. We use the same matching kernel for $\widetilde{H}+\widetilde{G}_2$, $\widetilde{E}+\widetilde{G}_1$, $\widetilde{G}_3$, and $\widetilde{G}_4$, as they are extracted by the same operator $\gamma_j \gamma_5$. 
Based on this observation, we use a single $C_{\gamma_j\gamma_5}$ in Eq.~\eqref{eq:matching}, which is the (simplified) matching kernel calculated for the twist-3 PDF $g_T$.
The kernel reads~\cite{Bhattacharya:2020xlt}
\begin{eqnarray*}
C^{(1)}_{\rm{M}\overline{\rm{MS}}}\left(\xi,\frac{\mu^2}{p_3^2}\right)  
= \dfrac{\alpha_s C_F}{2\pi}
\begin{dcases}
0 \\[1cm]
\delta(\xi) \\[1cm]
0 \\
\end{dcases} 
& \, \, + \, \, & \dfrac{\alpha_s C_F}{2\pi}
\begin{dcases}
\left[\dfrac{-\xi^2 + 2\xi +1}{1-\xi} \, \ln \dfrac{\xi}{\xi - 1}  + \dfrac{\xi}{1-\xi} + \dfrac{3}{2\xi}\right]_+ 
& \, \xi > 1  \\[0.2cm]
\left[  \dfrac{-\xi^2 + 2\xi +1}{1-\xi} \, \ln \dfrac{4\xi(1 - \xi) (x P_{3})^2}{\mu^2} +
\dfrac{\xi^2-\xi-1}{1-\xi}  \right]_+
& \, 0 < \xi < 1 \\[0.2cm]
\left[\dfrac{-\xi^2 + 2\xi +1}{1-\xi} \, \ln \dfrac{\xi - 1}{\xi} - \dfrac{\xi}{1-\xi} + \dfrac{3}{2(1-\xi)} \right]_+ 
& \, \xi < 0 \, ,
\end{dcases}
\end{eqnarray*}
and connects the quasi-GPDs evaluated in the M$\overline{\rm MS}$ scheme, to the light-cone GPDs in the standard $\overline{\rm MS}$ scheme. 
(For a complete matching result for $g_T$, including three-parton correlators, we refer to~\cite{Braun:2021aon}.)
Note that in the aforementioned equation, the first term arises as a result of perturbative ``zero-mode contributions". Upon performing the convolution integral in (the inversion of) Eq.~(\ref{eq:matching}), it becomes evident that this zero-mode term necessitates the evaluation of the quasi-GPD at $x=\infty$, a point that remains inaccessible in lattice QCD. Hence, we will focus on implementing the second term of the matching coefficient.
The renormalization scale is chosen to be 2 GeV for both quantities. 

\section{Lattice calculation} 
\label{sec:lat_calc}

\subsection{Computational setup}
\label{sec:comp}

In this calculation, we use one ensemble of two dynamical degenerate light quarks, a strange and charm quark ($N_f=2+1+1$). 
The strange and charm quarks have a physical mass, while the mass of the light quarks has been tuned to correspond to a pion mass of about 260 MeV. 
These gauge configurations~\cite{Alexandrou:2021gqw} employ the Iwasaki improved gauge action~\cite{Iwasaki:1985we} and twisted-mass fermions at maximal twist with a clover improvement~\cite{Sheikholeslami:1985ij}. 
The lattice spacing is $a\simeq 0.093$ fm, and the lattice volume is $32^3\times 64$ ($L\approx 3$ fm). 
The parameters of the ensemble are given in Table~\ref{tab:ensemble}. 

\begin{table}[h!]
\begin{center}
\renewcommand{\arraystretch}{1.4}
\renewcommand{\tabcolsep}{6pt}
\begin{tabular}{c|c c c c c c }
\hline
Name & $\beta$ & $N_f$ & $L^3\times T$ & $a$ [fm] & $M_\pi$ & $m_\pi L$ \\
\hline 
cA211.32 & $1.726$ & $u,d,s,c$ & $32^3\times 64$  & 0.093 & 260 MeV & 4 \\
\hline
\end{tabular}
\begin{minipage}{15cm}
\caption{Parameters of the ensemble used in this work: $\beta$ is the bare coupling and $L,T$ are the size of the lattice along the spatial and temporal directions.}
\label{tab:ensemble}
\end{minipage}
\end{center}
\end{table}

\begin{figure}[h!]
\includegraphics[scale=0.85]{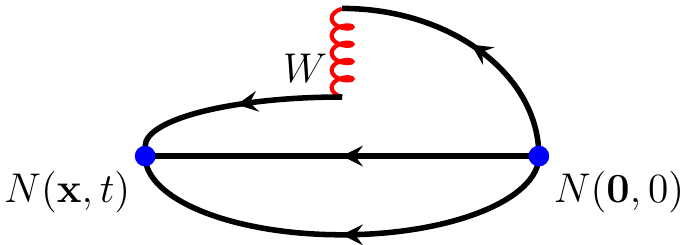}
\caption{Pictorial representation of the connected contributions to the three-point functions. The initial and final states with the quantum numbers of the nucleon are indicated by $N(\vec{0},0)$ and $N(\vec{x},t)$, respectively. The red curly line indicates the Wilson line, $W$, of the non-local operator.}
\label{fig:diagram}
\end{figure}

The construction of the matrix element $h_j$ combines the proton two-point and three-point correlation functions,
\begin{equation}
 C^{\rm 2pt}(\textbf{P}, t_s, 0) = (\Gamma_0)_{\alpha\beta} \sum_\textbf{x} e^{-i \textbf{P} \cdot \textbf{x}} \bra 0 N_\alpha(\textbf{x}, t_s) N_\beta(\textbf{0},0)\ket 0\,, 
\end{equation}
\begin{equation}
 C^{\rm 3pt}_j (\vec{p_f}, \vec{p_i}, t_s, \tau, 0)  = (\Gamma_k)_{\alpha \beta} \sum_{\vec{x},\vec{y}} e^{-i (\vec{p_f} - \vec{p_i}) \cdot \vec{y}} e^{-i \vec{p_f} \cdot \vec{x}} \bra 0 N_\alpha(\vec{x}, t_s)\mathcal{O}_{\gamma^j \gamma^5}(\vec{y},\tau;z) N_\beta(\vec{0},0)\ket 0\,,
\end{equation}
where $N_\alpha,N_\beta$ are the proton interpolating fields, $\tau$ is the current insertion time, and $t_s$ is the time separation between the source and the sink (the source is taken at $t=0$). 
We use the sequential method that requires fixing the source-sink time separation, $t_s$, but allows one to couple any operator to the proton states at a very small computational cost compared to the cost for a single operator. 
Here, we choose $t_s=10a$, as the signal-to-noise ratio decays faster in the twist-3 matrix elements of $\gamma_1\gamma_5$ and $\gamma_2 \gamma_5$, as compared to  $\gamma_3 \gamma_5$. 
We calculate the $u-d$ isovector flavor combination, which is extracted from the connected contributions to the three-point functions shown in Fig.~\ref{fig:diagram}; the disconnected component of $u-d$ is zero. 

An appropriate ratio must be constructed to cancel the time dependence of the exponentials introduced by the non-zero momentum transfer between the initial and final states. The overlap from the interpolating fields must also be eliminated. The ratio reads
\begin{eqnarray}
\label{eq:ratio}
R_j (\Gamma_\kappa, p_f, p_i; t_s, \tau) = \frac{C^{\rm 3pt}_j (\Gamma_\kappa, p_f, p_i; t_s, \tau)}{C^{\rm 2pt}( p_f;t_s)} \sqrt{\frac{C^{\rm 2pt}( p_i, t_s{-}\tau)C^{\rm 2pt}( p_f, \tau)C^{\rm 2pt}( p_f, t_s)}{C^{\rm 2pt}( p_f, t_s{-}\tau)C^{\rm 2pt}( P_i, \tau)C^{\rm 2pt}( p_i, t_s)}}\, 
\stackrel{t_s<\tau<t'} \to \Pi^j(\Gamma_\kappa)  \equiv h_j\,.\qquad
 \end{eqnarray}
Note that the two-point function is projected with $\Gamma_0$.
We extract the ground-state matrix elements from $R^j$ by taking a plateau fit with respect to $\tau$ in a region of convergence, indicated by $\Pi^j(\Gamma_\kappa)$. 
We identify plateaus for $\tau \in [3a-7a]$.
For simplicity, the dependence on $z$, $p_f,\, p_i$, and the renormalization scale $\mu$ is not shown explicitly in $\Pi^j$.
 
We calculate $h_1$ and $h_2$, for a class of momenta of the form $\vec{\Delta} = (\pm q,0,0)$, $\vec{\Delta} = (0,\pm q,0)$, and $\vec{\Delta} = (\pm q,\pm q,0)$; these correspond to zero skewness, $\Delta_3=0$. 
The average nucleon boost is nonzero along the $z$-axis, and we consider both plus and minus directions, $\vec{P}=(0,0,\pm P_3)$ for each value of $\vec{\Delta}$. 
This leads to an overall factor of eight more statistics for all cases but also a factor of eight more computational costs due to the constraints of the symmetric frame.
An additional factor of 2 in statistics for the for $\vec{\Delta} = (\pm q,\pm q,0)$ case is obtained from the averaging of $h_1$ and $h_2$, which are equivalent up to a minus sign for this kinematic setup. 
We note that a novel method to extract twist-2 GPDs from any kinematic frame via a Lorentz-covariant decomposition has been proposed~\cite{Bhattacharya:2022aob}. 
While the latter can be extended to twist-3 GPDs, it is beyond the scope of this work. 
The parameters of the calculation and the number of measurements are given in Table~\ref{tab:params_GPD} for both the PDF and the GPDs. 
  
\begin{table}[h!]
\centering
\renewcommand{\arraystretch}{1.5}
\begin{tabular}{ccc|cccc}
\hline
$P_3\, [\,{\rm GeV}\,]$ & $\vec{q}\, [\,\frac{2\pi}{L}\,]$  & $-t\, [\,{\rm GeV}^2\,]$ & \,\,$N_{\rm ME}$\,\,  & \,\,$N_{\rm confs}$\,\, & \,\,$N_{\rm src}$\,\, & \,\,$N_{\rm total}$\,\, \\ \hline
$\pm 0.83$ & $(0,0,0)$      & 0   &2   &194   &8    &3104 \\ 
$\pm 1.25$ & $(0,0,0)$      & 0    &2   &731   &16   &23392 \\ 
$\pm 1.67$  & $(0,0,0)$      & 0   &2   &1644  &64   &210432 \\ \hline
$\pm 0.83$ & $(\pm 2,0,0)$      & 0.69  &8    &67   &8  & 4288 \\ 
$\pm 1.25$ & $(\pm 2,0,0)$      & 0.69  &8    &249   &8  & 15936 \\ 
$\pm 1.67$ & $(\pm 2,0,0)$      & 0.69  &8    &294  &32 &  75264\\ 
$\pm 1.25$ & $(\pm 2,\pm 2,0)$  & 1.38  &16    &224  &8  & 28672 \\ 
$\pm 1.25$ & $(\pm 4,0,0)$      & 2.76  &8    &329     &32   &84224  \\ \hline
\end{tabular}
\caption{Statistics used in this calculation. $N_{\rm ME}$, $N_{\rm confs}$, $N_{\rm src}$ and $N_{\rm total}$ is the number of matrix elements, configurations, source positions per configuration and total statistics, respectively.}
    \label{tab:params_GPD}
\end{table}

\subsection{Renormalization}

We calculate $Z_{\gamma_j\gamma_5}$ following the methodology described in the previous section implemented on five $N_f=4$ ensembles, which are given in Tab.~\ref{Table:Z_ensembles}. 
The ensembles correspond to the same $\beta$ value as for the $N_f=2+1+1$ ensemble used for the extraction of the matrix elements. 
The light and heavy quarks are mass-degenerate so that the chiral limit can be taken. 
\begin{table}[h]
\begin{center}
\renewcommand{\arraystretch}{1.5}
\renewcommand{\tabcolsep}{5.5pt}
\begin{tabular}{ccc}
\hline\hline 
$\beta=1.726$, & $c_{\rm SW} = 1.74$, & $a=0.093$~fm \\
\hline\hline\\[-3ex]
{$24^3\times 48$}  & {$\,\,a\mu = 0.0060$}  & $\,\,m_\pi = 357.84$~MeV     \\
\hline
{$24^3\times 48$}  & $\,\,a\mu = 0.0080$     & $\,\,m_\pi = 408.11$~MeV     \\
\hline
{$24^3\times 48$}  & $\,\,a\mu = 0.0100$    & $\,\,m_\pi = 453.48$~MeV    \\
\hline
{$24^3\times 48$}  & $\,\,a\mu = 0.0115$    & $\,\,m_\pi = 488.41$~MeV    \\
\hline
{$24^3\times 48$}  & $\,\,a\mu = 0.0130$    & $\,\,m_\pi = 518.02$~MeV    \\
\hline\hline
\end{tabular}
\vspace*{-0.25cm}
\begin{center}
\caption{\small{Parameters of the $N_f=4$ ensembles used for the calculation and chiral extrapolation of the renormalization functions}.}
\label{Table:Z_ensembles}
\end{center}
\end{center}
\vspace*{-0.2cm}
\end{table} 

The values of the vertex momentum $p$ entering Eq.~\eqref{renorm} are carefully chosen to reduce discretization effects in the $Z_{\gamma_j\gamma_5}$, as $p$ defines the RI$'$ renormalization scale, $\bar\mu_0$. 
More details can be found in Ref.~\cite{Alexandrou:2015sea}. 
To this end, we choose spatially isotropic momenta, that is $p=(p_0,p_1,p_1,p_1)$, so that the ratio $\frac{p^4}{(p^2)^2}$ is less than 0.35, as suggested in Ref.~\cite{Constantinou:2010gr}. 
We use 17 different values of $\bar\mu_0$, between the range $(a\,\bar\mu_0)^2 \in [0.7,2.6]$. 
Finally, ${Z}^{\MMSb}_{\gamma_j\gamma_5,0}(z,2 {\rm GeV})$ is obtained by extrapolating $(a\,p)^2 {\to} 0$ using a linear fit and data in the region $(a\,\bar\mu_0)^2 \in [1,2.6]$. 
An extensive investigation of systematic uncertainties related to the renormalization functions can be found in Ref.~\cite{Alexandrou:2019lfo}.

\section{Results}
\label{sec:results}

\subsection{Bare matrix elements}
\begin{figure}[h!]
\includegraphics[scale=0.375]{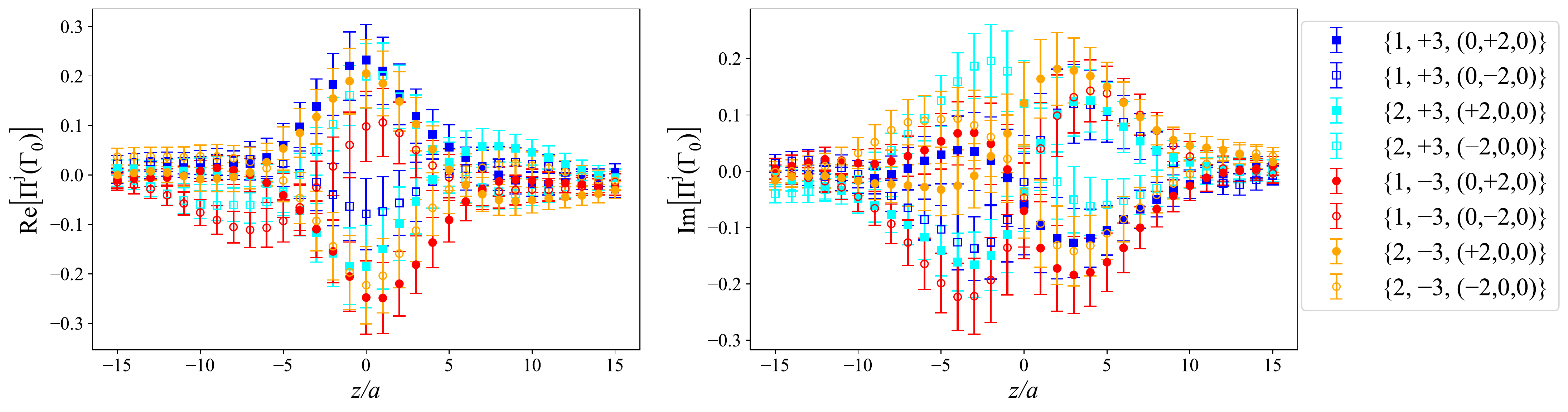}
\vskip -0.3cm
\caption{Real (left) and imaginary (right) parts of the matrix elements $\Pi^j(\Gamma_0)$ ($j=1,2$) for all kinematic cases corresponding to $-t=0.69$ GeV$^2$ and $|P_3|=1.25$ GeV. The data are indicated by $\{j,P_3,\vec{\Delta}\}$, where $P_3=\pm3$, and $\vec{\Delta}=(\pm2,0,0),\,(0,\pm2,0)$, given in units of $\frac{2\pi}{L}$. The errors correspond to the statistical uncertainties.}
\label{fig:ME_j_0}     
\end{figure}

\begin{figure}[h!]
\includegraphics[scale=0.39]{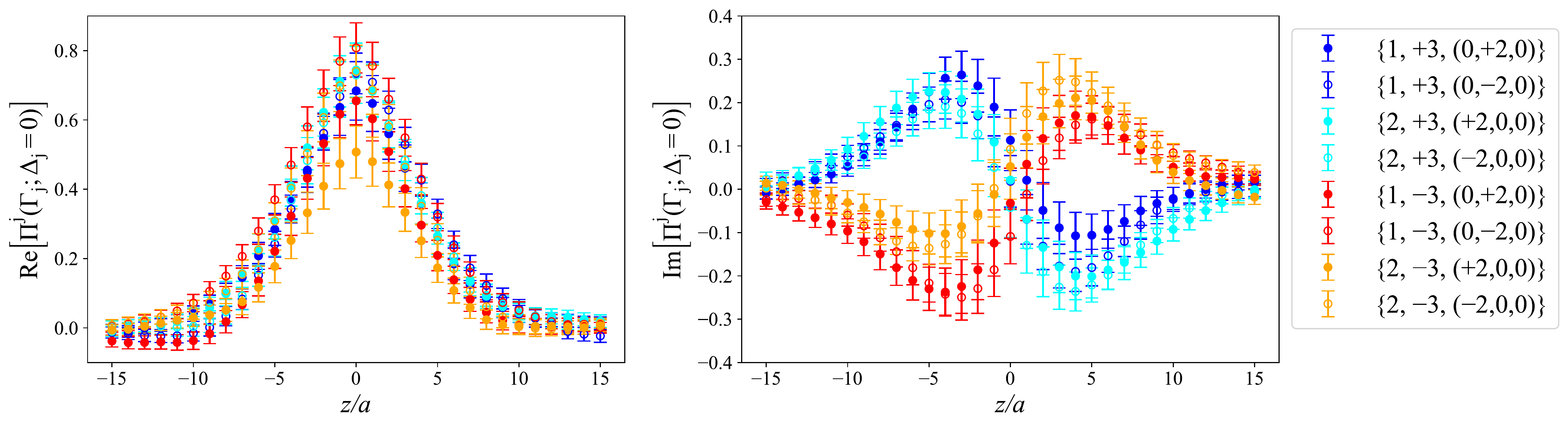}\\
\includegraphics[scale=0.39]{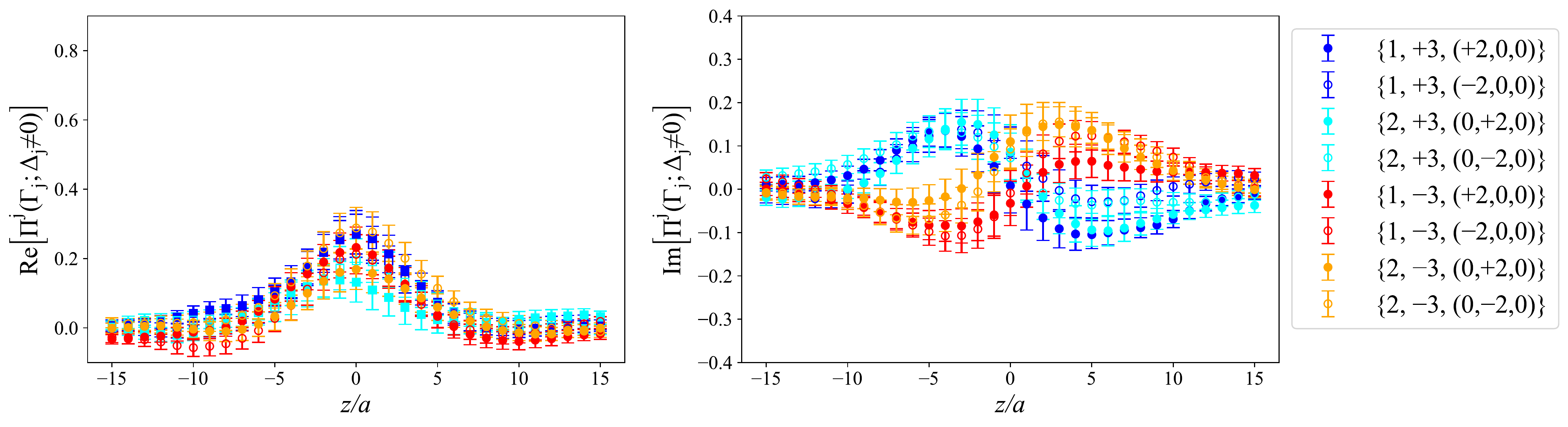}
\vskip -0.3cm
\caption{Top: Real (left) and imaginary (right) parts of the matrix elements $\Pi^j(\Gamma_j)$ ($j=1,2$) for all kinematic cases corresponding to $\Delta_j=0$ at $-t=0.69$ GeV$^2$ and $|P_3|=1.25$ GeV. The data are indicated by $\{j,P_3,\vec{\Delta}\}$, where $P_3=\pm3$, and $\vec{\Delta}=(\pm2,0,0),\,(0,\pm2,0)$. The momenta are given in units of $\frac{2\pi}{L}$. Bottom: Same as the top panel, but for $\Delta_j\neq0$. The errors correspond to the statistical uncertainties.}
\label{fig:ME_j_j}     
\end{figure}

\begin{figure}[h!]
\includegraphics[scale=0.39]{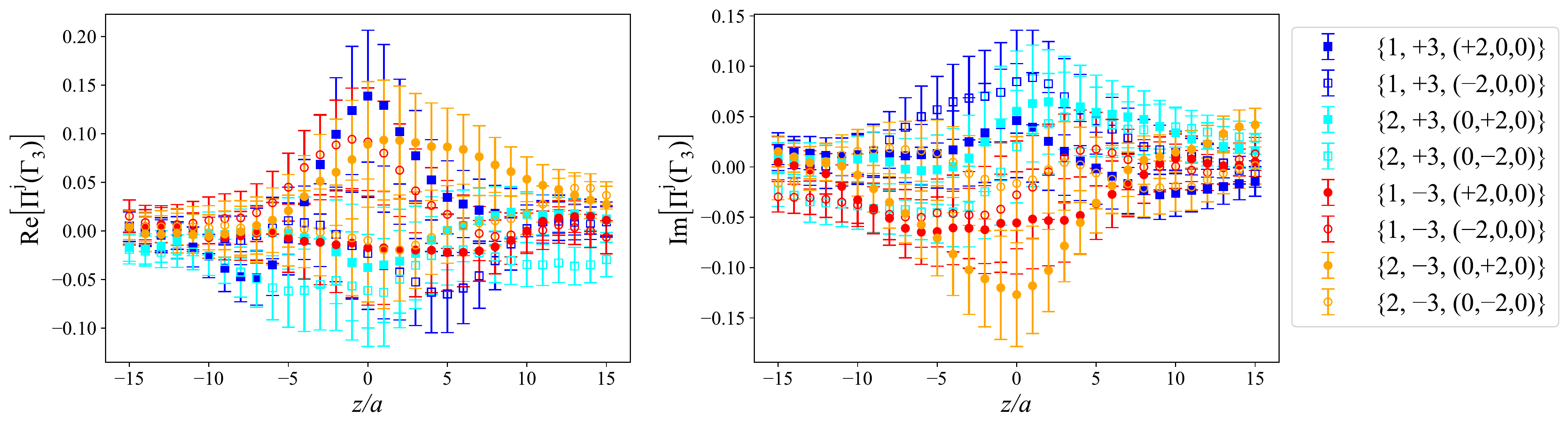}
\vskip -0.3cm
\caption{Real (left) and imaginary (right) parts of the matrix elements $\Pi^j(\Gamma_3)$ ($j=1,2$) for all kinematic cases corresponding to $-t=0.69$ GeV$^2$ and $|P_3|=1.25$ GeV. The data are indicated by $\{j,P_3,\vec{\Delta}\}$, where $P_3=\pm3$, and $\vec{\Delta}=(\pm2,0,0),\,(0,\pm2,0)$. The momenta are given in units of $\frac{2\pi}{L}$. The errors correspond to the statistical uncertainties.}
\label{fig:ME_j_3}     
\end{figure}

Since we have performed the calculation for $\pm P_3$ and $\pm \vec{\Delta}$, it is interesting to observe the symmetries of the matrix elements. 
For example, it can be seen from Eqs.~\eqref{eq:Pi1Gamma0} - \eqref{eq:Pi2Gamma3} that
\begin{eqnarray}
\Pi^1(\Gamma_0)\Big{|}_{\vec{\Delta}=(0,\pm \Delta,0)} = - \,\Pi^2(\Gamma_0)\Big{|}_{ \vec{\Delta}=(\pm \Delta,0,0)},\,\qquad
\Pi^1(\Gamma_3)\Big{|}_{\vec{\Delta}=(\pm \Delta,0,0)} = + \Pi^2(\Gamma_3)\Big{|}_{\vec{\Delta}=(0,\pm \Delta,0)}\,.
\end{eqnarray}
Similarly, 
\begin{eqnarray}
\Pi^1(\Gamma_0)\Big{|}_{\vec{\Delta}=(0,+ \Delta,0)} = - \,\Pi^1(\Gamma_0)\Big{|}_{\vec{\Delta}=(0,-\Delta,0)}  ,\,\qquad 
\Pi^2(\Gamma_0)\Big{|}_{\vec{\Delta}=(+\Delta,0,0)} = - \,\Pi^2(\Gamma_0)\Big{|}_{\vec{\Delta}=(-\Delta,0,0)} \,.
\end{eqnarray}
The above symmetries hold because (a) the kinematic coefficients of $F_X$ are either symmetric or antisymmetric with $\pm z$, $\pm P_3$ and $\pm \vec{\Delta}$; (b) the amplitudes $F_X$ have definite symmetries. 
For example, $\widetilde{G}_1$, $\widetilde{G}_2$, and $\widetilde{G}_4$ are symmetric in $P_3$ and $z$, while $\widetilde{G}_3$ is anti-symmetric, as discussed in Sec.~\ref{sec:def}.
Therefore, we utilize all eight combinations of $\pm P$ and $\pm \vec{\Delta}$ to decrease statistical uncertainties by combining the kinematically equivalent matrix elements. 
In Figs.~\ref{fig:ME_j_0} - \ref{fig:ME_j_3}, we present the non-vanishing matrix elements for $-t=0.69$ GeV$^2$ and $P_3=1.25$ GeV. 
Fig.~\ref{fig:ME_j_0} shows the unpolarized projector, $\Pi^j(\Gamma_0)$, which contains the twist-2 contribution $F_{\widetilde{H}}$. 
Fig.~\ref{fig:ME_j_j} gives the matrix element $\Pi^j(\Gamma_j)$ for two cases, that is $\Delta_j=0$ and $\Delta_j\neq0$.
These lead to kinematically independent matrix elements, as can be seen in Eq.~\eqref{eq:Pi1Gamma1} and Eq.~\eqref{eq:Pi2Gamma2}.
We find that the $\Delta_j=0$ case is about twice as large as the case of $\Delta_j\neq0$.
$\Pi^j(\Gamma_3)$ only has twist-3 contribution, with $F_{\widetilde{G}_3}$ found to be very small and noisy, as shown in Fig.~\ref{fig:ME_j_3} for $-t=0.69$ GeV$^2$ and $|P_3|=1.25$ GeV.
For the kinematic setup of $-t=0.69$ GeV$^2$, both $\Pi^1(\Gamma_2)$ and $\Pi^1(\Gamma_2)$ are zero due to the presence of $\Delta_x \Delta_y$ that vanishes, and are not shown here. 
However, they contribute to the GPDs for $-t=1.38$ GeV$^2$.

It is interesting to compare the magnitude of the matrix elements from Figs.~\ref{fig:ME_j_0} - \ref{fig:ME_j_3}. 
We find that $\Pi^j(\Gamma_j;\Delta_j=0)$ dominates in magnitude, followed by $\Pi^j(\Gamma_j;\Delta_j\neq0)$ and $\Pi^j(\Gamma_0)$ that are similar. Finally, $\Pi^j(\Gamma_3)$ is suppressed 
and compatible with zero, a conclusion that holds for all values of $-t$ we explore here. 
This impacts directly the extraction of $\widetilde{G}_3$, as $\Pi^j(\Gamma_3)$ is proportional to $F_{\widetilde{G}_3}$. 
The vanishing signal of $F_{\widetilde{G}_3}$ is connected to zero skewness, where we expect that $\widetilde{G}_3$ is zero (see, e.g., Eq.~\eqref{eq:int_xG3}).

\begin{figure}[h!]
\includegraphics[scale=0.39]{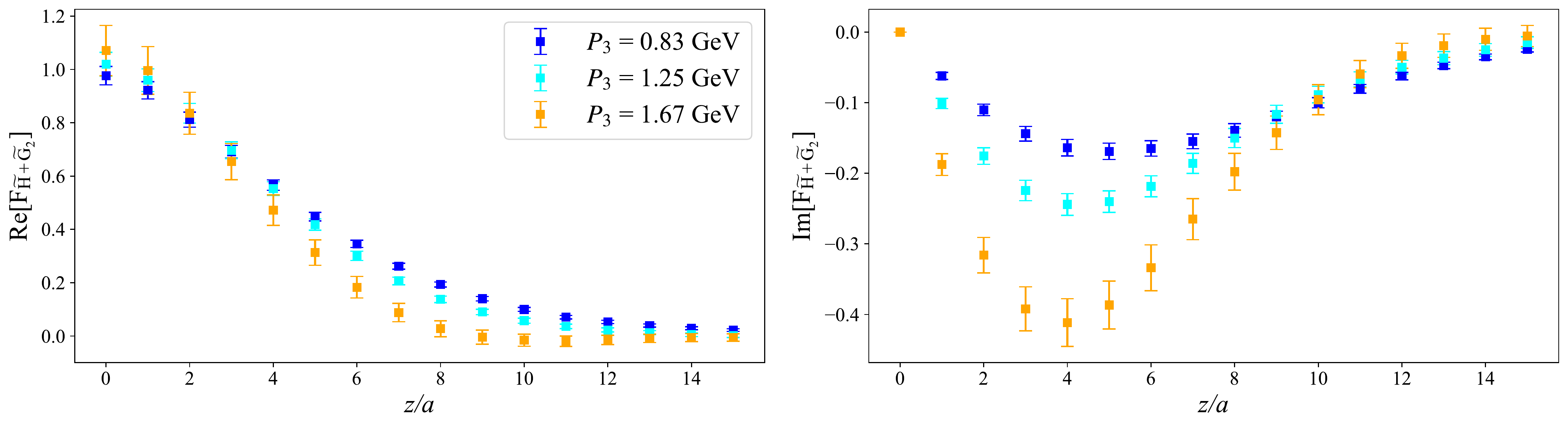}\\
\includegraphics[scale=0.39]{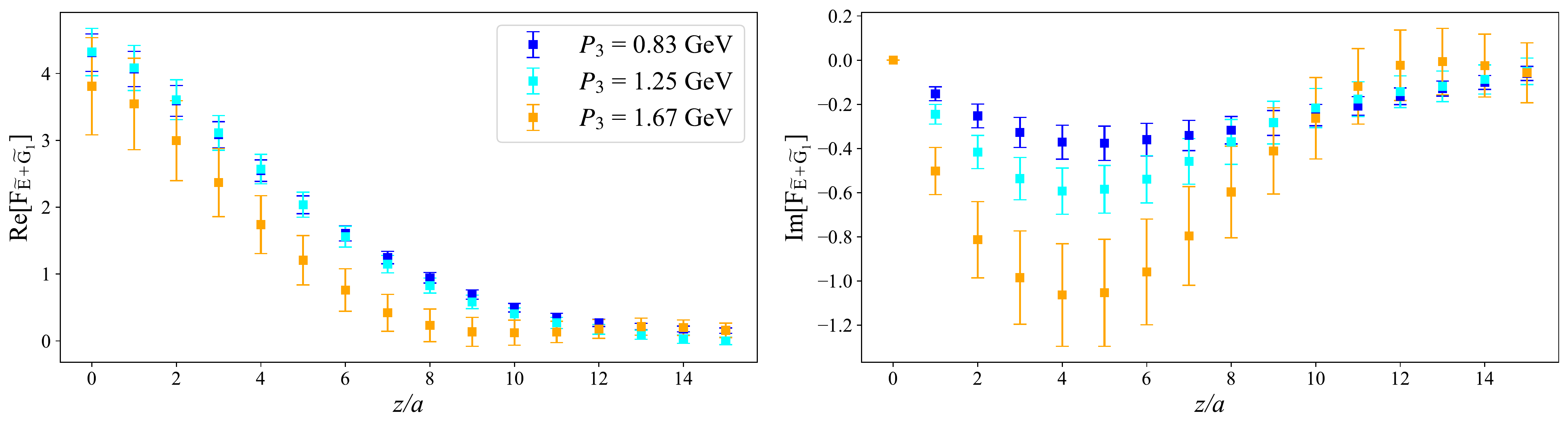}
\vskip -0.3cm
\caption{Real (left) and imaginary (right) parts of $F_{\widetilde{H}+\widetilde{G}_2}$ (top) and $F_{\widetilde{E}+\widetilde{G}_1}$ (bottom) at $-t=0.69$ GeV$^2$ and $P_3=0.83,\,1.25,\,1.67$ GeV. All kinematically equivalent cases have been averaged. The errors correspond to the statistical uncertainties.}
\label{fig:ME_P3_G1G2}     
\end{figure}

Using Eqs.~\eqref{eq:Pi1Gamma0} - \eqref{eq:Pi2Gamma3}, we decomposed the amplitudes $F_X$ where $X=\widetilde{H}+\widetilde{G}_2,\,\widetilde{E}+\widetilde{G}_1,\,\widetilde{G}_3,\,\widetilde{G}_4$.
We first focus on the numerically dominant GPDs containing the twist-2 counterparts, that is $F_{\widetilde{H}+\widetilde{G}_2}$ and $F_{\widetilde{E}+\widetilde{G}_1}$.
We examine their dependence on the momentum boost in Fig.~\ref{fig:ME_P3_G1G2} at $-t=0.69$ GeV$^2$, for which have $P_3=0.83,\,1.25,\,1.67$ GeV. 

\begin{figure}[h!]
\includegraphics[scale=0.39]{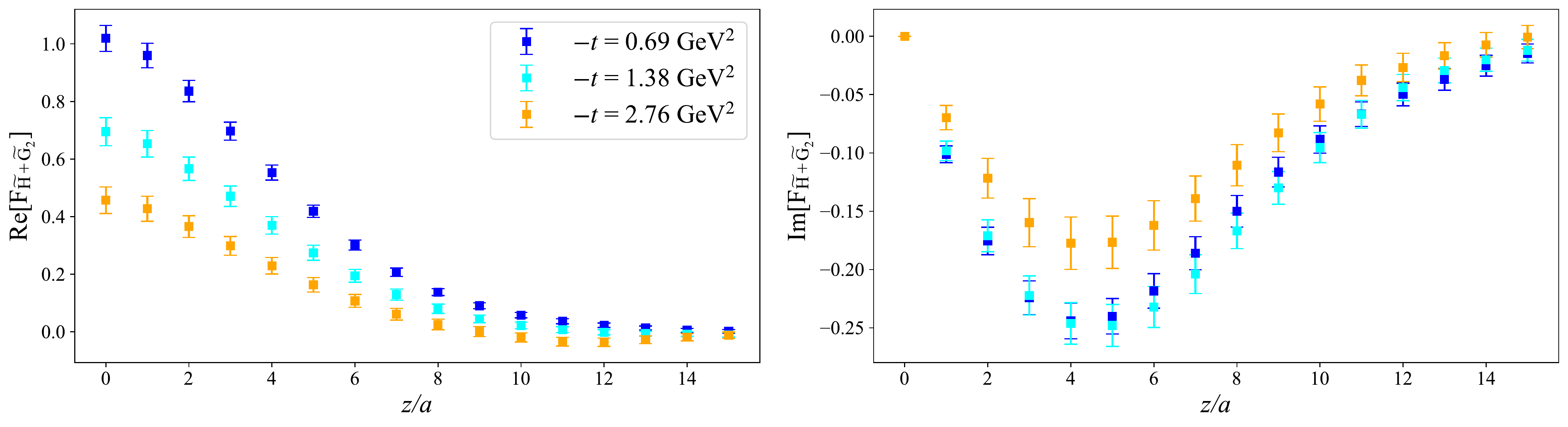}\\
\includegraphics[scale=0.39]{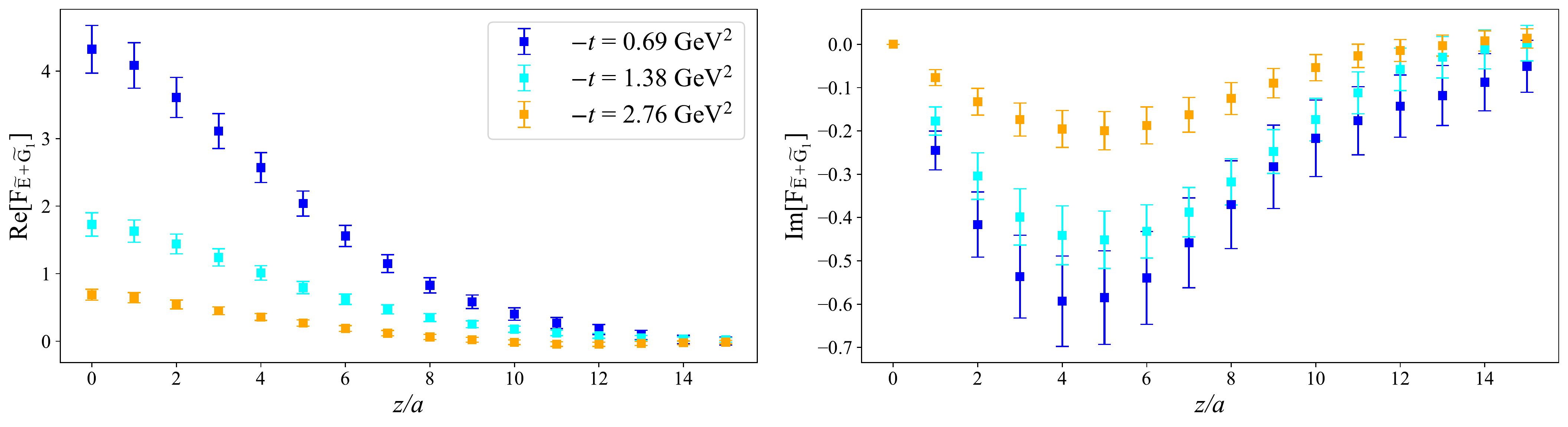}
\vskip -0.3cm
\caption{Real (left) and imaginary (right) parts of $F_{\widetilde{H}+\widetilde{G}_2}$ (top) and $F_{\widetilde{E}+\widetilde{G}_1}$ (bottom) at $P_3=1.25$ GeV and $-t=0.69,\,1.38,\,2.76$ GeV$^2$. All kinematically equivalent cases have been averaged. The errors correspond to the statistical uncertainties.}
\label{fig:ME_t_G1G2}     
\end{figure}

The dependence of $F_X$ on the momentum transfer is shown in Fig.~\ref{fig:ME_t_G1G2}, where the momentum boost is fixed to $|P_3|=1.25$ GeV. 
For $F_{\widetilde{H}+\widetilde{G}_2}$, we find a very mild $P_3$ dependence, with some difference in the intermediate $z$ region. 
However, a more noticeable dependence on the momentum boost is observed in the case of $F_{\widetilde{E}+\widetilde{G}_1}$.
These conclusions hold for both the real and imaginary parts of these quantities.
Another observation is that $F_{\widetilde{E}+\widetilde{G}_1}$ is noisier than $F_{\widetilde{H}+\widetilde{G}_2}$.
We note that the quasi-GPDs are defined at a finite momentum; a non-trivial momentum dependence is expected. 
Therefore, one may not draw conclusions on the $P_3$ dependence of the quasi-GPDs. 
On the contrary, one should examine convergence in the final light-cone GPDs. 
Overall, $F_{\tilde{E}} + F_{\tilde{G_1}}$ has the largest magnitude followed by $F_{\widetilde{H}+\widetilde{G}_2}$, in particular at $z=0$. This is expected from the behavior of the axial, $G_A$, and induced pseudoscalar, $G_P$, form factors~\cite{Alexandrou:2020okk}, which are related to the GPDs through the generalization of Burkhardt-Cottingham sum rules~\cite{Ji:1996ek,Kiptily:2002nx} (see also Sec.~\ref{sec:def})
\begin{equation}
\int_{-1}^1 dx \, \widetilde{H}(x,\xi,t)= G_A(t)\,, \quad\int_{-1}^1 dx \, \widetilde{E}(x,\xi,t)=G_P(t)\,,  
\label{eq:sum_rule}
\end{equation}
coupled to
\begin{equation}
\int_{-1}^1 dx \, \widetilde{G}_{i}(x,\xi,t)= 0\,,\quad i=1,2,3,4\,,
\label{eq:Gi_sum_rule}
\end{equation}
which also holds for quasi-GPDs.
The large magnitude of $F_{\widetilde{E}+\widetilde{G}_1}$ is in accordance with the findings of the twist-2 case~\cite{Alexandrou:2020zbe}, $F_{\tilde{E}}$. 

Regarding the $-t$ dependence in the quasi-GPDs, our complete set of values for the momentum transfer is available only for $P_3=1.25$ GeV. 
Increasing further the momentum boost leads to very noisy data, as can be seen from Fig.~\ref{fig:ME_P3_G1G2}. 
The noise increase becomes more prominent for $-t=1.38$ GeV$^2$ and $-t=2.76$ GeV$^2$. 
As observed in Fig.~\ref{fig:ME_t_G1G2}, there is a good signal for all cases and a hierarchy among the various momentum transfers, where the quasi-GPDs are decaying towards zero as $-t$ increases. 
The difference between $-t=0.69$ GeV$^2$ and $-t=1.38$ GeV$^2$ is very small for the imaginary part of both $F_{\widetilde{H}+\widetilde{G}_2}$  and $F_{\widetilde{E}+\widetilde{G}_1}$.

\begin{figure}[h!]
\includegraphics[scale=0.39]{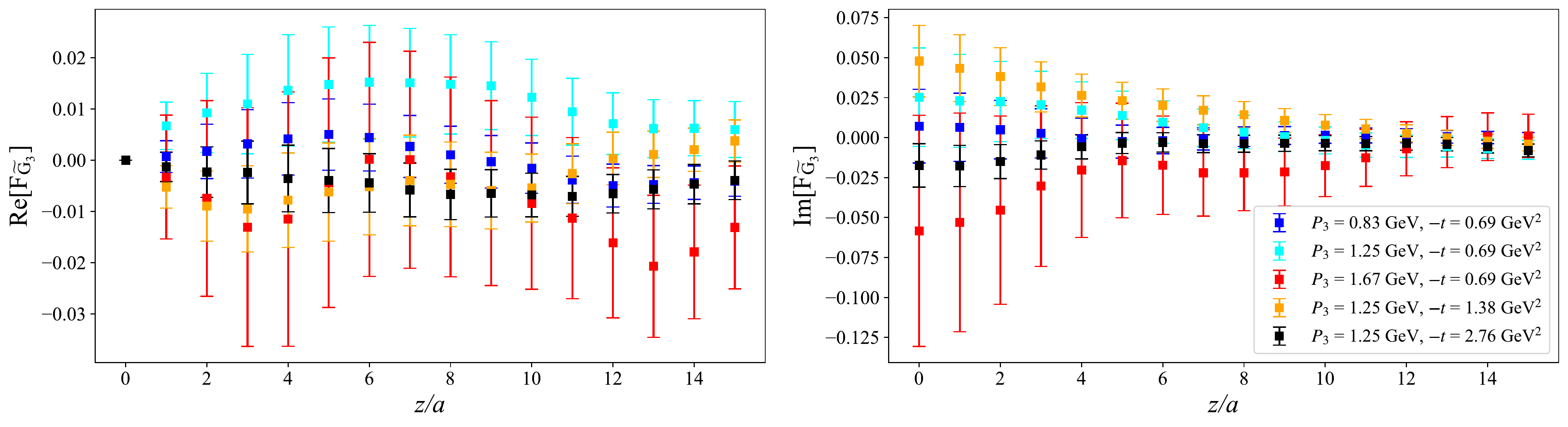}
\vskip -0.3cm
\caption{Real (left) and imaginary (right) parts of $F_{\widetilde{G}_3}$ for all available values of $P_3$ and $-t$. The errors correspond to the statistical uncertainties.}
\label{fig:ME_G3}     
\end{figure}

\begin{figure}[h!]
\includegraphics[scale=0.39]{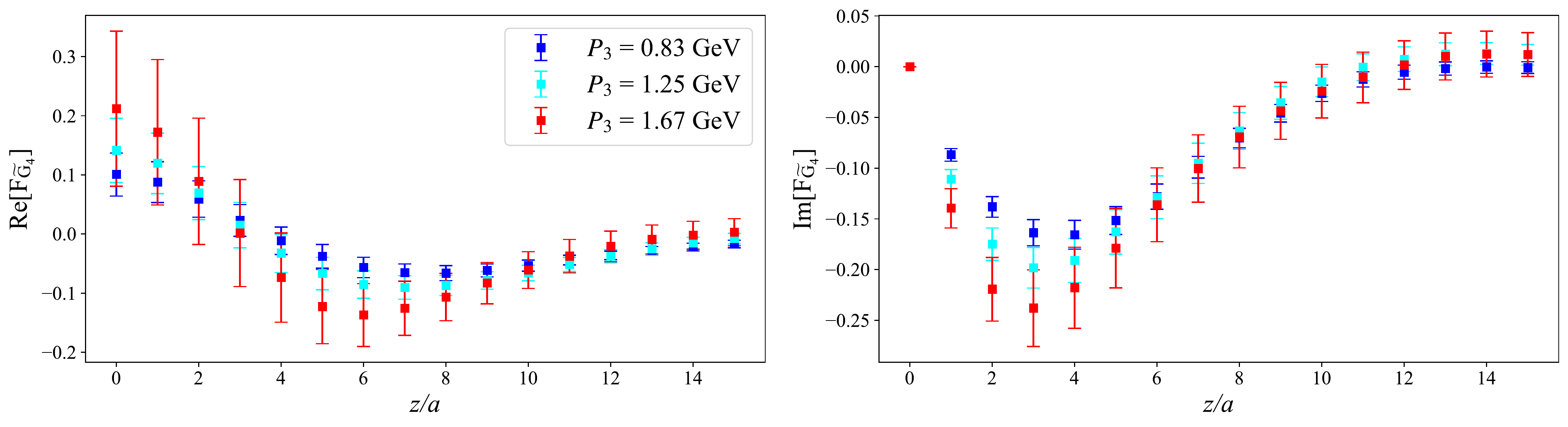}
\vskip -0.3cm
\caption{Real (left) and imaginary (right) parts of $F_{\widetilde{G}_4}$ for $P_3=0.83,\,1.25,\,1.67$ GeV at $-t=0.69$ GeV$^2$. The errors correspond to the statistical uncertainties.}
\label{fig:ME_G4_P3}     
\end{figure}

\begin{figure}[h!]
\includegraphics[scale=0.39]{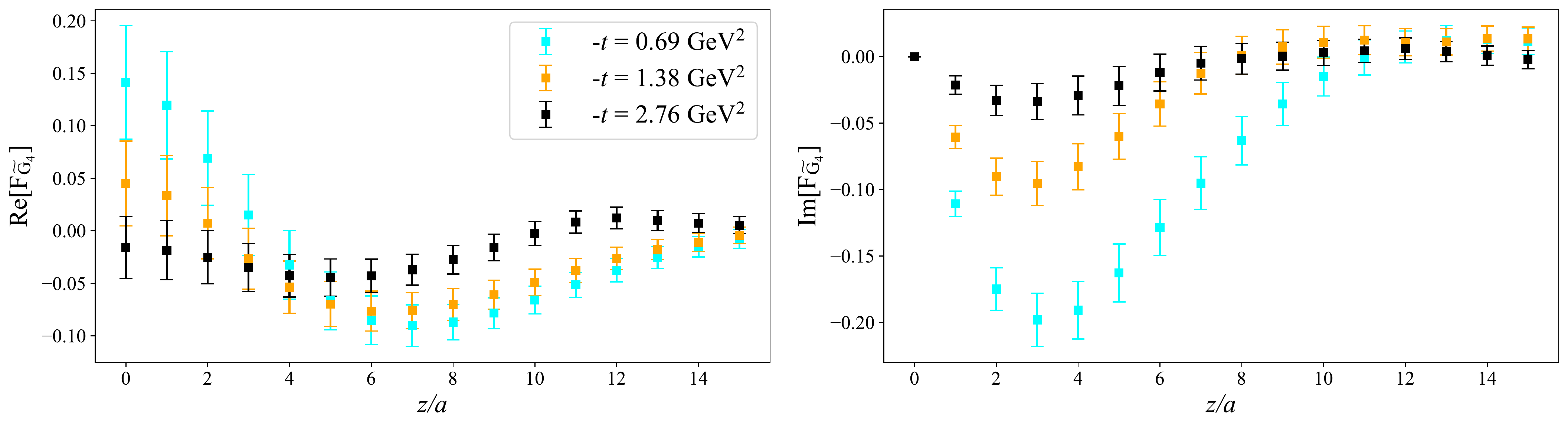}
\vskip -0.3cm
\caption{Real (left) and imaginary (right) parts of $F_{\widetilde{G}_4}$ for $P_3=1.25$ GeV and $-t=0.69,\,1.38,\,2.76$ GeV$^2$. The errors correspond to the statistical uncertainties.}
\label{fig:ME_G4_t}     
\end{figure}

We now turn our attention to $F_{\widetilde{G}_3}$ and $F_{\widetilde{G}_4}$, which are plotted in Figs.~\ref{fig:ME_G3} - \ref{fig:ME_G4_t} for all available data of Table.~\ref{tab:params_GPD}.
As explained in Section~\ref{sec:def}, the calculation of $F_{\widetilde{G}_3}$ yields a value of zero, which aligns with the expected theoretical outcome based on the hermiticity and time-reversal constraints imposed on GPDs. Also, the generalization of the Efremov-Leader-Teryaev sum rules~\cite{Efremov:1996hd} indicates that 
\begin{equation}
\label{eq:int_xG3}
\int dx\, x\, \widetilde{G}_3=\frac{\xi}{4} G_E(t)\,,    
\end{equation}
which is zero in our zero skewness calculation ($\xi=0$).  

\noindent
On the other hand, $F_{\widetilde{G}_4}$ is small but not negligible. 
This finding is consistent with the sum rule connecting it to the electric Sachs form factor $G_E$,
\begin{equation}
\int_{-1}^1 dx \, x\, \widetilde{G}_{4}(x,\xi,t)= \frac{1}{4} G_E\,.  
\end{equation}
The dependence on the momentum boost is hidden within the statistical uncertainties. 
Nevertheless, there is a tendency for the real part of $F_{\widetilde{G}_4}$ to increase for higher momentum, while the imaginary part becomes more negative, as can be seen from Fig.~\ref{fig:ME_G4_P3}.
Despite the increase of the noise, $F_{\widetilde{G}_4}$ remains nonzero for all values of $P_3$.
The momentum transfer dependence of $F_{\widetilde{G}_4}$ is shown in Fig.~\ref{fig:ME_G4_t}.
The dominant contribution corresponds to $-t=0.69$ GeV$^2$ with the increase of $-t$ suppressing its magnitude.

\subsection{Light-cone GPDs}
\label{sec:results_LC}

\begin{figure}[h!]
\includegraphics[scale=0.44]{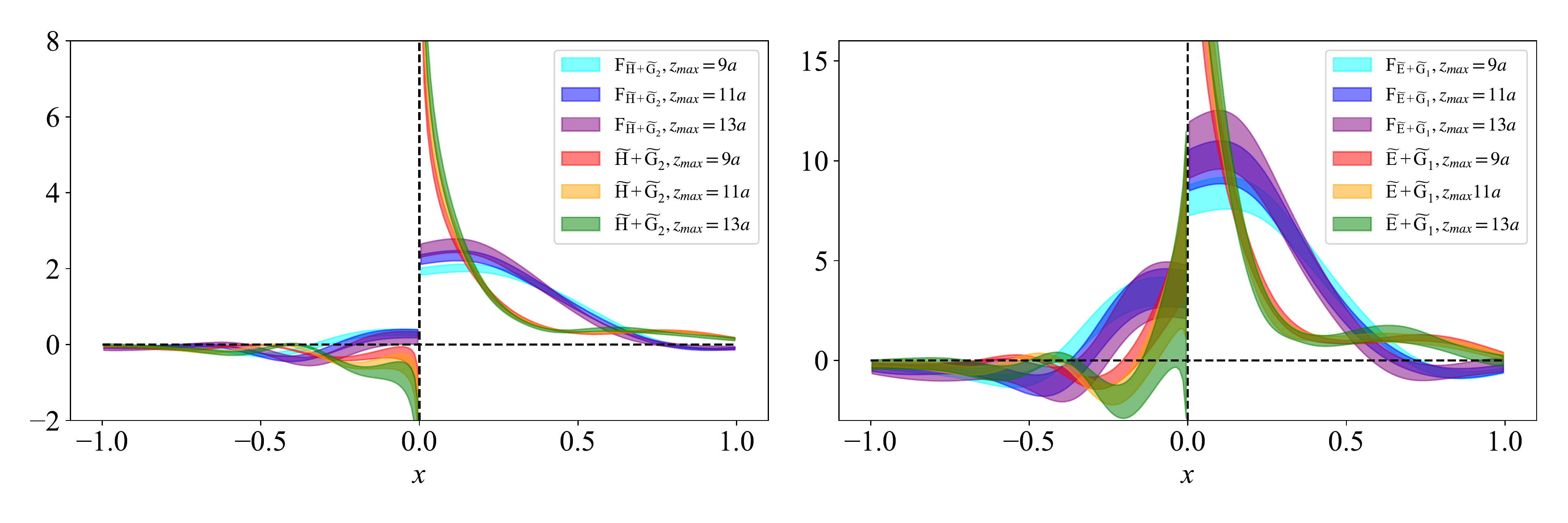}
\vskip -0.3cm
\caption{$\zmax$ dependence of $F_{{\widetilde{H}} + {\widetilde{G}_2}}$ and $\widetilde{H}+\widetilde{G}_2$ (left), as well as $F_{{\widetilde{E}} + {\widetilde{G}_1}}$ and $\widetilde{E}+\widetilde{G}_1$ (right) at $-t=0.69$ GeV$^2$ and $P_3=1.25$ GeV. Results are given in the $\MSb$ scheme at a scale of 2 GeV. The errors correspond to the statistical uncertainties.}
\label{fig:G1G2_zmax}     
\end{figure}

\begin{figure}[h!]
\includegraphics[scale=0.44]{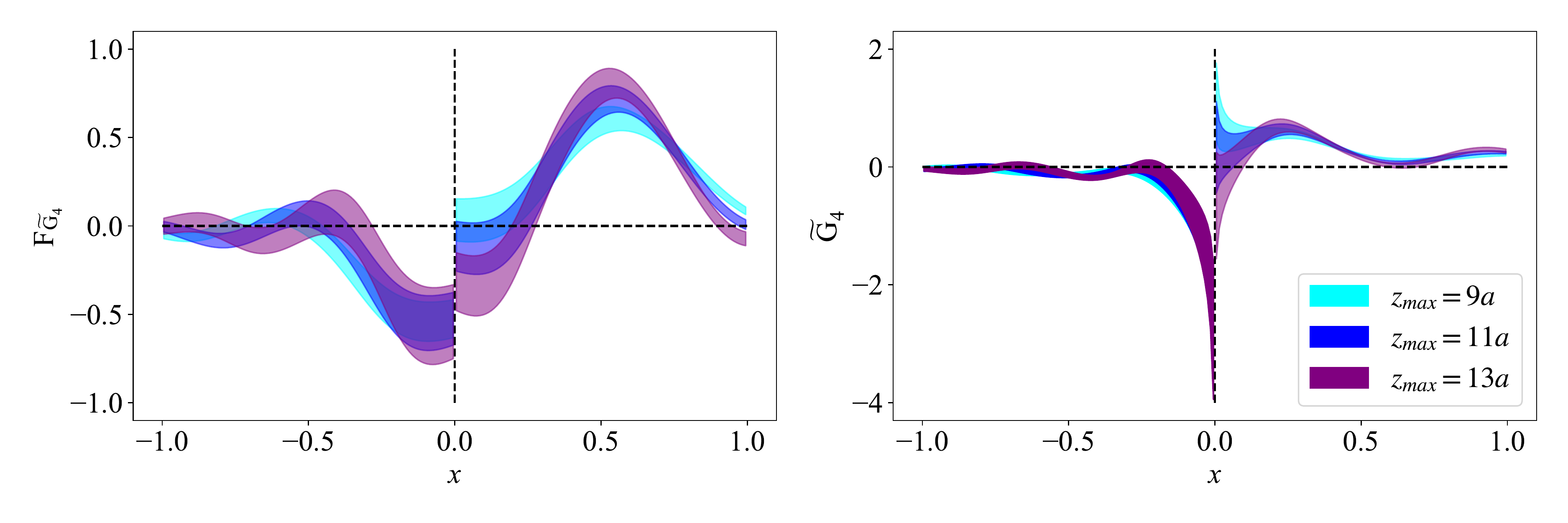} 
\vskip -0.3cm
\caption{$\zmax$ dependence of $F_{\widetilde{G}_4}$ and $\widetilde{G}_4$ at $-t=0.69$ GeV$^2$ and $P_3=1.25$ GeV. Results are given in $\MSb$ scheme at a scale of 2 GeV. The errors correspond to the statistical uncertainties.}
\label{fig:G4_zmax}     
\end{figure}

In this paragraph, we present the reconstruction of the $x$-dependence of the light-cone GPDs, for which we use the Backus-Gilbert method on each $F_X$ and then apply the matching kernel to obtain the light-cone GPDs as a function of $x$. 
Since we perform the calculation at zero skewness, it is anticipated that the matching formalism of GPDs is the same as for PDFs~\cite{Liu:2019urm}. 
Thus, we use the results of Ref.~\cite{Bhattacharya:2020xlt}, which correspond to the forward limit of the twist-3 axial GPDs, $g_T$. 
A dedicated analytic calculation for the twist-3 case is required to prove this argument, which we leave as future work.
In this presentation, we neglect $\widetilde{G}_3$, which was found to be zero. 
We first examine the dependence of the quasi-GPDs on the maximum value of $z$ that enters the Backus-Gilbert reconstruction. 
In particular, we test $\zmax = 9a,\,11a,\,13a$, and present the quasi-GPDs as well as the light-cone GPDs in Figs.~\ref{fig:G1G2_zmax} - \ref{fig:G4_zmax} for $P_3=1.25$ GeV and $-t=0.69$ GeV$^2$.
The function $F_{{\widetilde{H}} + {\widetilde{G}_2}}$ obtained in momentum space for $\zmax=11a$ and $\zmax=13a$ is compatible for all values of $x$; similarly for the light-cone ${\widetilde{H}} + {\widetilde{G}_2}$. 
Some tension is observed between $\zmax=9a$ and $\zmax=11a$ for $F_{{\widetilde{H}} + {\widetilde{G}_2}}$ at small $|x|$ values. 
In the case of the $F_{{\widetilde{E}} + {\widetilde{G}_1}}$ and ${\widetilde{E}} + {\widetilde{G}_1}$ we find compatibility between all three values of $\zmax$. 
We note that the statistical uncertainties are enhanced compared to ${\widetilde{H}} + {\widetilde{G}_2}$. 
For $F_{\widetilde{G}_4}$, we find a very small $\zmax$ dependence that is within the statistical uncertainties and can be neglected.
The same conclusions hold for other values of $P_3$ and $-t$. 
Thus, we choose $\zmax=11a$ for the final results, and we add a systematic error equal to one-half of the difference of the results between $\zmax=9a$ and $\zmax=13a$ to reflect the uncertainty in the choice of the optimal truncation $\zmax$.

\begin{figure}[h!]
\includegraphics[scale=0.45]{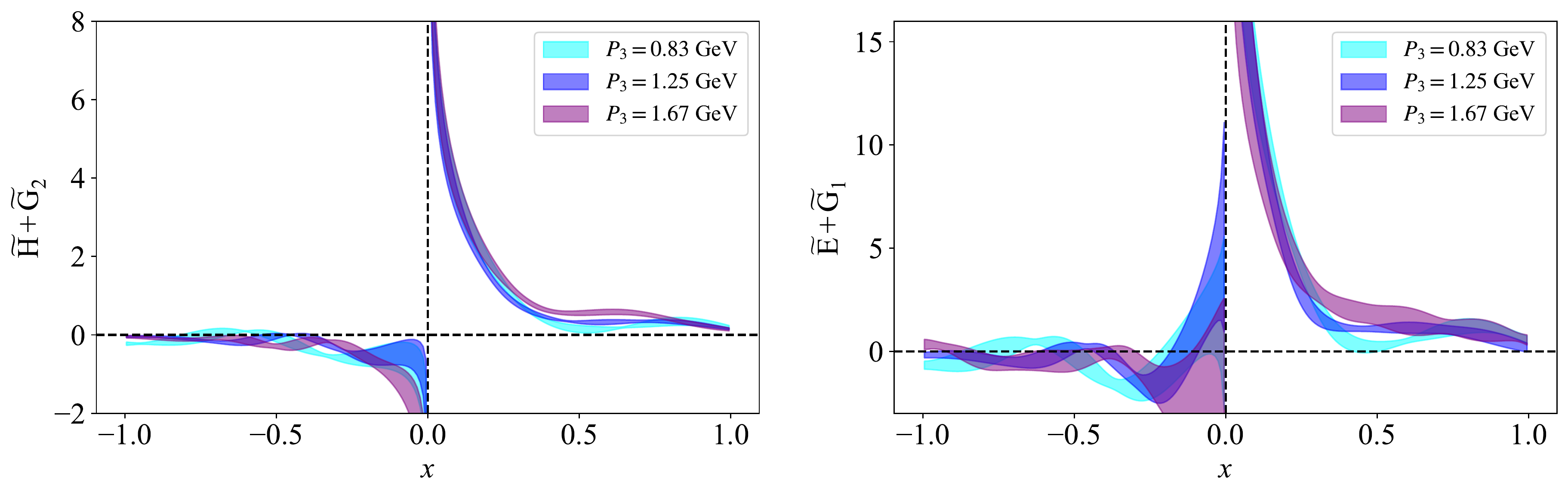}
\vskip -0.3cm
\caption{$P_3$ dependence of $\widetilde{H}+\widetilde{G}_2$ (left) and $\widetilde{E}+\widetilde{G}_1$ (right) at $-t=0.69$ GeV$^2$ and $P_3=0.83,\,1.25,\,1.67$ GeV. The results correspond to $z_{\rm max}=11a$. Results are given in $\MSb$ scheme at a scale of 2 GeV. The bands correspond to the statistical errors and the systematic uncertainty due to the $x$-dependent reconstruction.}
\label{fig:G1G2_P3}     
\end{figure}

\begin{figure}[h!]
\includegraphics[scale=0.45]{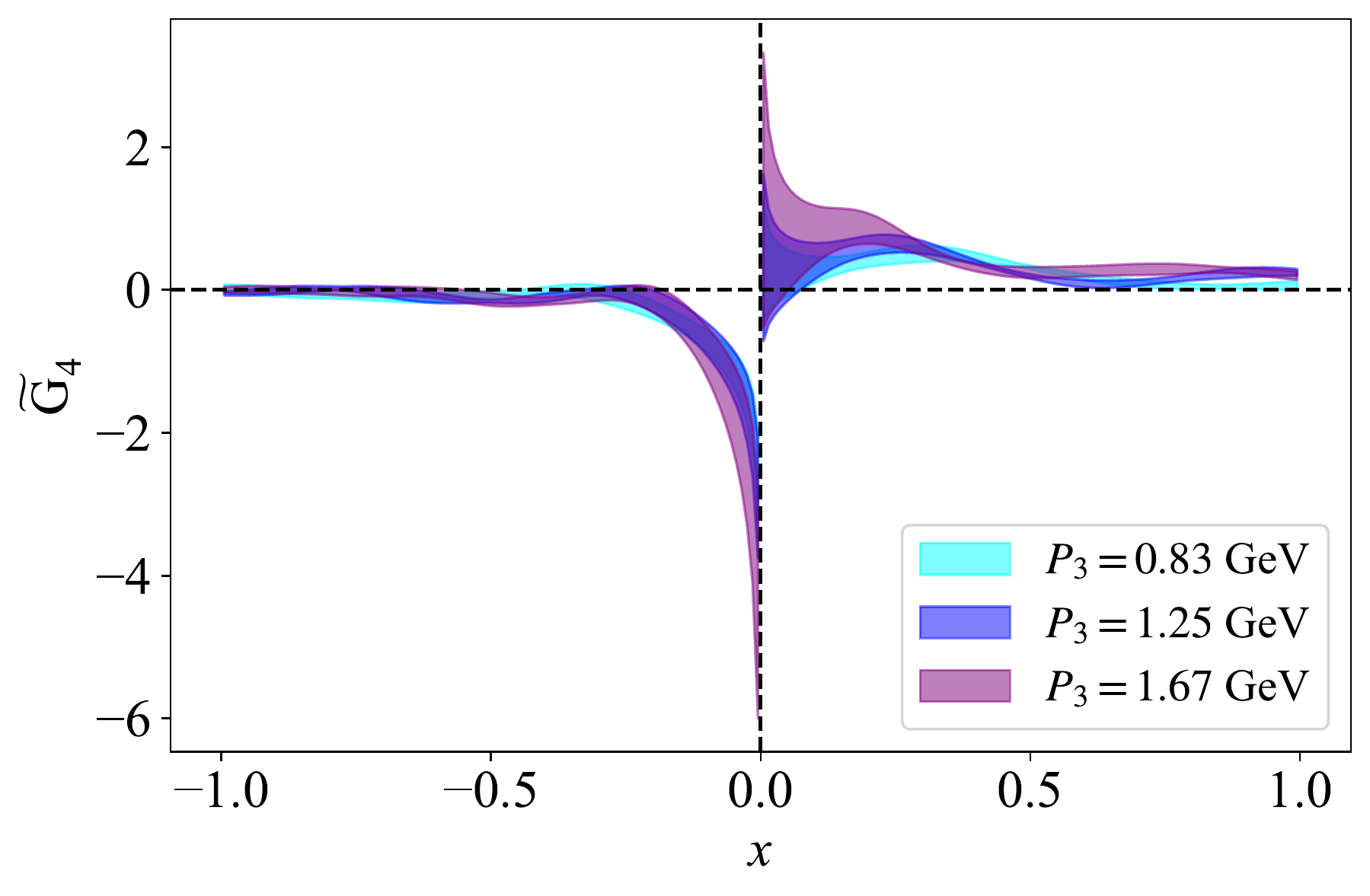}
\vskip -0.3cm
\caption{$P_3$ dependence of $\widetilde{G}_4$ at $-t=0.69$ GeV$^2$ and $P_3=0.83,\,1.25,\,1.67$ GeV. The results correspond to $z_{\rm max}=11a$. Results are given in the $\MSb$ scheme at a scale of 2 GeV. The bands correspond to the statistical errors and the systematic uncertainty due to the $x$-dependent reconstruction.}
\label{fig:G4_P3}     
\end{figure}

The $P_3$-dependence of the light-cone GPDs is shown in Figs.~\ref{fig:G1G2_P3} - \ref{fig:G4_P3} for $-t=0.69$ GeV$^2$. Both
${\widetilde{H}} + {\widetilde{G}_2}$ and ${\widetilde{E}} + {\widetilde{G}_1}$ are found to have similar results between $P_3=0.83,\,1.25,\,1.67$ GeV in the region $x \in [0,0.4]$. 
There are some differences in the region $x \in [0.5,0.7]$, while all $P_3$ converge to the same behavior in the large $x$ region.
The $P_3$ behavior of $\widetilde{G}_4$ is within uncertainties in the negative $x$ region, as well as the positive region up to about $x=0.4$. 
Beyond that point, the approach to $x=1$ exhibits some differences.
We note that the convergence in $P_3$ is satisfactory within the reported statistical and reconstruction uncertainties.
However, this finding applies to the present calculation, and an investigation of various systematic effects is required before reaching general conclusions on such a convergence.
Furthermore, currently one cannot quantitatively estimate the uncertainty related to the momentum-boost convergence. How much the results for our highest value of $P_3$ differ from the results in the limit $P_3 \to \infty$ ultimately depends on the QCD dynamics. We note, though, that model calculations suggest sufficient convergence of the quasi-GPDs at $P_3 \sim 2 \, \textrm{GeV}$, for a wide range of the GPD variables~\cite{Bhattacharya:2018zxi, Bhattacharya:2019cme}.

\begin{figure}[h!]
\includegraphics[scale=0.45]{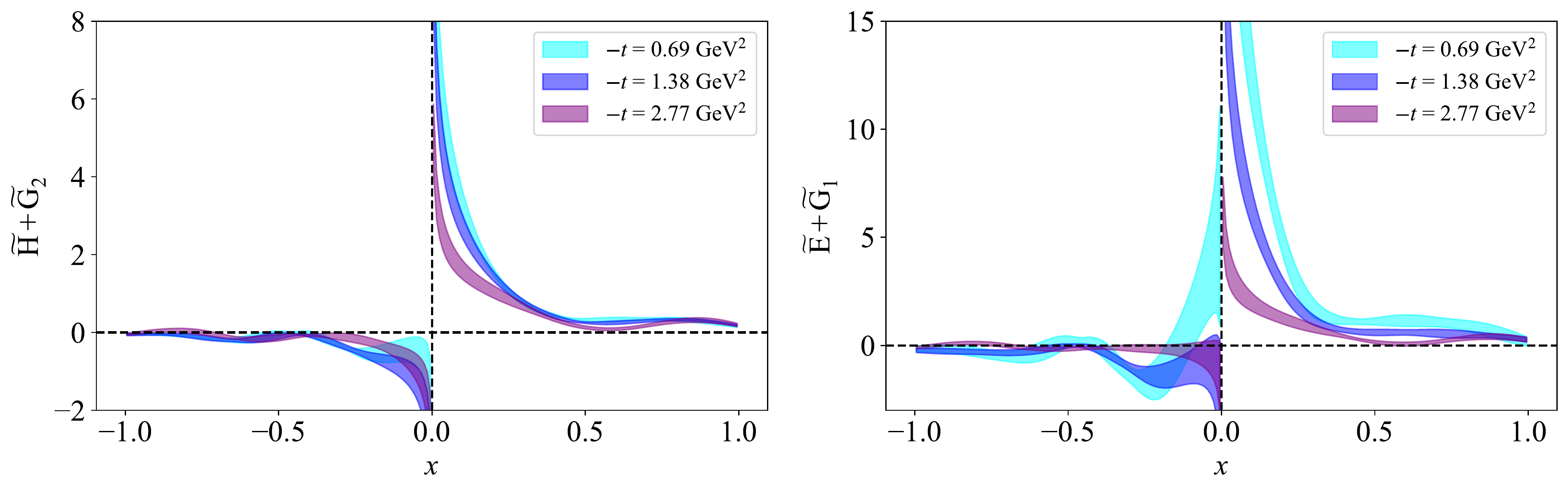}
\vskip -0.3cm
\caption{$\widetilde{H}+\widetilde{G}_2$ (left) and $\widetilde{E}+\widetilde{G}_1$ (right) and $P_3=1.25$ GeV for various values of $-t$. Results are given in the $\MSb$ scheme at a scale of 2 GeV. The bands correspond to the statistical errors and the systematic uncertainty due to the $x$-dependent reconstruction.}
\label{fig:G1G2_final}     
\end{figure}
\begin{figure}[h!]
\includegraphics[scale=0.45]{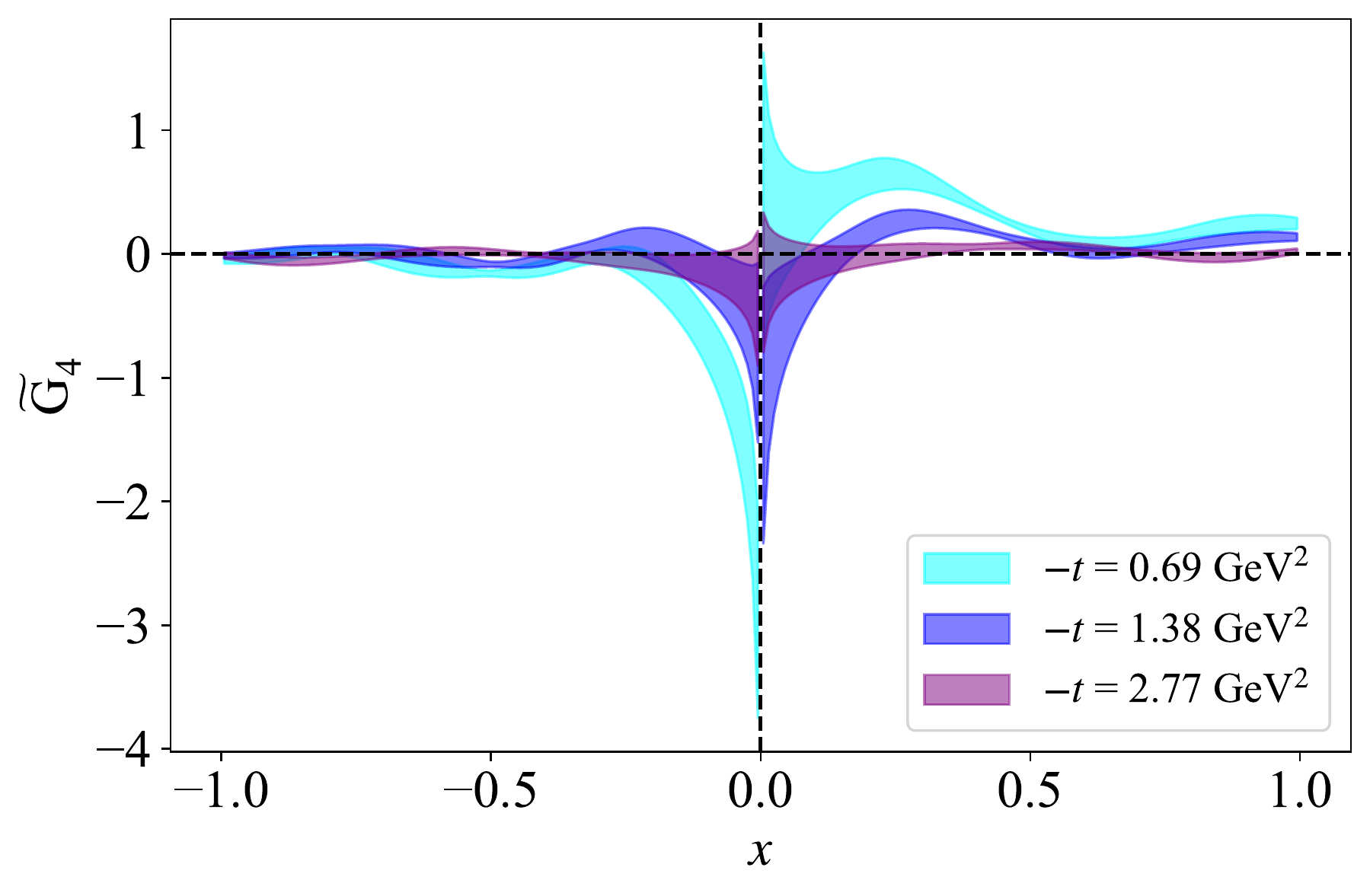}
\vskip -0.3cm
\caption{$\widetilde{G}_4$ at $P_3=1.25$ GeV for various values of $-t$. Results are given in the $\MSb$ scheme at a scale of 2 GeV. The bands correspond to the statistical errors and the systematic uncertainty due to the $x$-dependent reconstruction.}
\label{fig:G4_final}     
\end{figure}

Fig.~\ref{fig:G1G2_final} compares $\widetilde{H}+\widetilde{G}_2$ and $\widetilde{E}+\widetilde{G}_1$ at $-t=0.69,\,1.38,\,2.76$ GeV$^2$. 
It is found that as $-t$ increases, the numerical value of GPDs decrease in the $x\in[0,0.8]$ region, while the large-$x$ region is not very sensitive to $-t$. 
It is also notable that the difference between $-t=0.69$ GeV$^2$ and $-t=1.38$ GeV$^2$ is smaller than the difference between $-t=1.38$ GeV$^2$ and $-t=2.76$ GeV$^2$. 
We observe that the $-t$ dependence of $\widetilde{E}+\widetilde{G}_1$ is stronger than for $\widetilde{H}+\widetilde{G}_2$.
This may be an indication of the pion pole expected in $\widetilde{E}$~\cite{Penttinen:1999th}, which cannot be isolated from the lattice data at zero skewness. 
Further investigation is required to address the pion pole indication.
The $-t$ dependence of $\widetilde{G}_4$ is shown in Fig.~\ref{fig:G4_final}, which follows the same hierarchy as $\widetilde{H}+\widetilde{G}_2$ and $\widetilde{E}+\widetilde{G}_1$. 
We remind the reader that, presently, lattice QCD calculations are not reliable for extracting the small-$x$ region ($x\lesssim 0.15$), nor the antiquark region, regardless of the twist level. 
Therefore, conclusions about these regions are not reliable.

\begin{figure}[h!]
\includegraphics[scale=0.45]{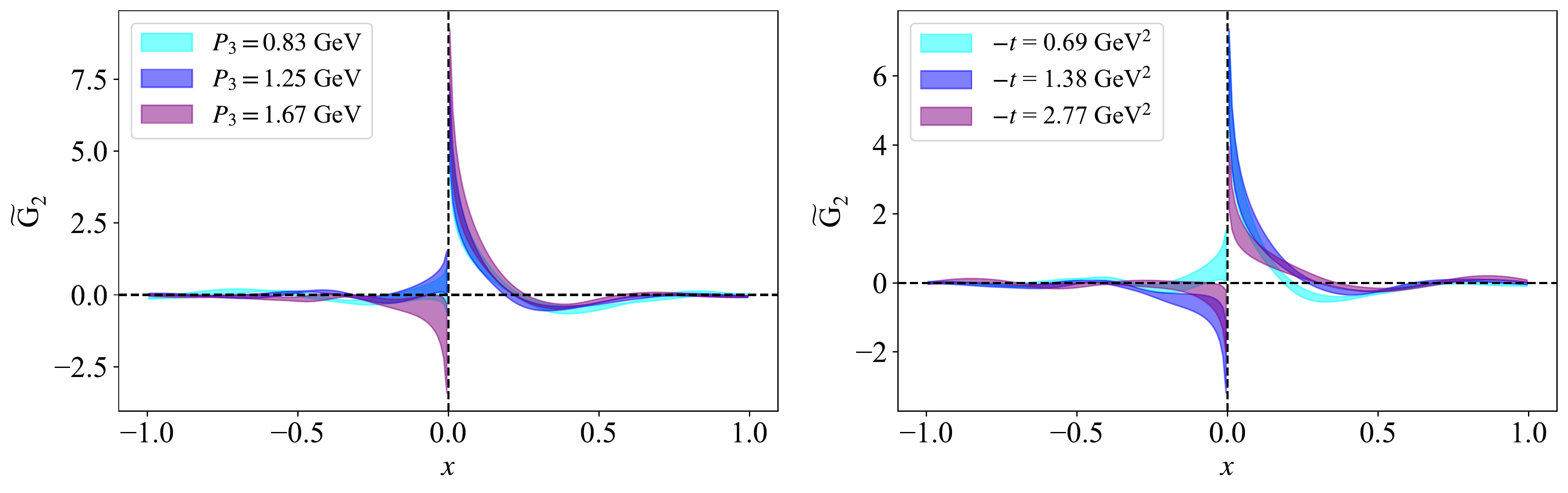}
\vskip -0.3cm
\caption{Left: $\widetilde{G}_2$ at $-t=0.69$ GeV$^2$ and various $P_3$. Right:  $\widetilde{G}_2$ at $P_3=1.25$ GeV and various $-t$.  Results are given in the $\MSb$ scheme at a scale of 2 GeV. The bands correspond to the statistical errors and the systematic uncertainty due to the $x$-dependent reconstruction.}
\label{fig:G2}     
\end{figure}

One can extract $\widetilde{G}_2$, which cannot be obtained directly from lattice data but may be isolated from the data on $\widetilde{H}+\widetilde{G}_2$ and  its twist-2 counterpart, $\widetilde{H}$, which we also calculate in this work following the same setup. 
The $P_3$ and $-t$ dependence of $\widetilde{G}_2$ is shown in Fig.~\ref{fig:G2}. 
In summary, there is no $P_3$ dependence, while the $-t$ dependence is non-monotonic and depends on the range of $x$.
Furthermore, it is observed that $\widetilde{G}_2$ becomes negative in the positive $x$ region. 
Such observation is not an artifact of the matching, as the same feature holds for $F_{\widetilde{G}_2}$ in the $x$ space.
In fact, this is not an unphysical behavior because the norm of both the quasi and light-cone $\widetilde{G}_2$ should vanish, indicating that negative regions must exist. 
We remind the reader that extracting $\widetilde{G}_1$ is not possible for zero skewness because $\widetilde{E}$ is not accessible from the $\gamma_3\gamma_5$ operator due to a vanishing kinematic factor. 
Nevertheless, this calculation gives a glimpse into $\widetilde{E}$ through its twist-3 counterpart, $\widetilde{E}+\widetilde{G}_1$. 
One can also extract the Mellin moments of $\widetilde{E}$ directly from our twist-3 data using the sum rule of Eq.~\eqref{eq:sum_rule} and using the fact that the integral of $G_1$ is zero, as given in Eq.~\eqref{eq:Gi_sum_rule}.
\begin{figure}[h!]
\includegraphics[scale=0.5]{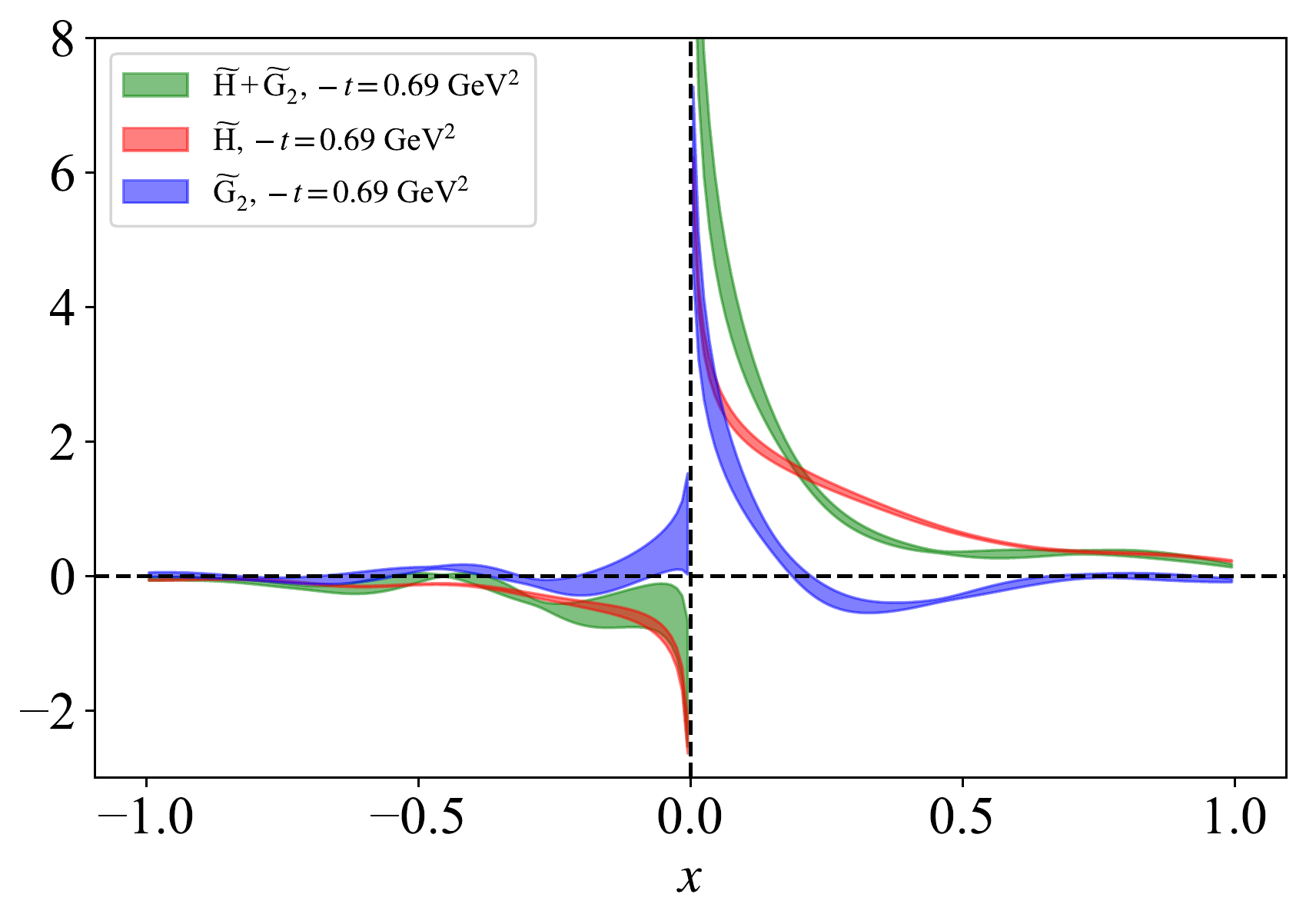} 
\caption{Comparison of $\widetilde{H}$,  $\widetilde{H}+\widetilde{G}_2$, and $\widetilde{G}_2$ at $-t=0.69$ GeV$^2$. Results are given in the $\MSb$ scheme at a scale of 2 GeV. The bands correspond to the statistical errors and the systematic uncertainty due to the $x$-dependent reconstruction.}
\label{fig:GPDs_twist2_twist3}
\end{figure}     

It is useful to compare $\widetilde{H}+\widetilde{G}_2$, $\widetilde{H}$, as well as the extracted $\widetilde{G}_2$. These quantities are shown in Fig.~\ref{fig:GPDs_twist2_twist3} for $-t=0.69$ GeV$^2$.
The difference of $\widetilde{H}+\widetilde{G}_2$ and $\widetilde{H}$ is sizeable, which gives rise to a non-negligible $\widetilde{G}_2$. 
Another observation is that $\widetilde{H}+\widetilde{G}_2$ and $\widetilde{G}_2$ approach zero much faster than $\widetilde{H}$.

\subsection{Consistency Checks}
\label{s:sum_rules}

There are a number of consistency checks one can perform using the lattice data presented above. 
Here, we examine the following aspects expected by theory.
\begin{itemize}
\item[\textbf{1.}] The local limit of $\widetilde{H}+\widetilde{G}_2$ and $\widetilde{H}$ should coincide. 
We find numerically that this holds for all our data, as can be seen in Table~\ref{tab:data}.
A similar property should hold for $\widetilde{E}+\widetilde{G}_1$ and $\widetilde{E}$ even though $\widetilde{E}$  is not directly accessible from lattice data at zero skewness.
\begin{table}[h!]
\begin{center}
\renewcommand{\arraystretch}{1.4}
\renewcommand{\tabcolsep}{6pt}
\begin{tabular}{|l|c| c| c| c| c|  }
\hline
GPD & $P_3=0.83$ [GeV] & $P_3=1.25$ [GeV]  & $P_3=1.67$ [GeV]  & $P_3=1.25$ [GeV] & $P_3=1.25$ [GeV]   \\
    & $-t=0.69$ [GeV$^2$] &  $-t=0.69$ [GeV$^2$] &  $-t=0.69$ [GeV$^2$] &  $-t=1.38$ [GeV$^2$] & $-t=2.76$ [GeV$^2$]  \\
\hline 
$\widetilde{H}$ &0.741(21)   &0.712(27)   &0.802(48)    &0.499(21)  &0.281(18)   \\
$\widetilde{H}+\widetilde{G}_2$ &0.719(25)   &0.750(33)   &0.788(70)   &0.511(36)  &0.336(34)   \\
\hline
\end{tabular}
\begin{minipage}{15cm}
\caption{The local limit of $\widetilde{H}+\widetilde{G}_2$ and $\widetilde{H}$ for the various $P_3$ and $-t$ used in this work. The number in the parenthesis indicates the statistical uncertainty.}
\label{tab:data}
\end{minipage}
\end{center}
\end{table}
\item[\textbf{2.}] The norm of $\widetilde{G}_i$ is expected to be zero. (See Eqs.~(\ref{e:norm1}) - (\ref{e:norm2}).)
Here, we find that the norm of $\widetilde{G}_2$ and $\widetilde{G}_4$ is about a factor of 10 smaller than the norm of $\widetilde{H}+\widetilde{G}_2$ and $\widetilde{E}+\widetilde{G}_1$. 
The norm of $\widetilde{G}_3$ is trivially zero, as the GPD vanishes at zero skewness. 
\item[\textbf{3.}] The norms and Mellin moments of both the quasi and light-cone GPDs should be momentum-boost independent. (See~\cite{Bhattacharya:2019cme,Bhattacharya:2021boh}.)
This holds for all results presented in this work.
\item[\textbf{4.}]
As the momentum transfer squared increases, the norm of the GPDs should decrease, as observed in the form factors. 
Our data satisfy this expectation.
\item[\textbf{5.}]
As the number of Mellin moments of $\widetilde{H}+\widetilde{G}_2$ and $\widetilde{E}+\widetilde{G}_1$ increases, its magnitude should decrease compared to the lower moments; that is
\begin{eqnarray}
\int_{-1}^1 dx \, x^{n^\prime}\,X(x,\xi,t) &<& \int_{-1}^1 dx \, x^n\,X(x,\xi,t)\,, \quad n^{\prime}> n\,\,\quad X=\widetilde{H}+\widetilde{G}_2,\,\widetilde{E}+\widetilde{G}_1.
\end{eqnarray}
This is because higher moments have their support at higher values of $x$.
We have tested this argument for $n\le3$ and found that, indeed, our data maintain this feature. Such an expectation does not hold for $\widetilde{G}_i$, which have zero norm. 
In fact, the Mellin moments of $\widetilde{G}_2$ alternate in sign, while for $\widetilde{G}_4$ they are approximately constant. 
\end{itemize}

\section{Summary}
\label{sec:summary}

In this work, we presented results on the axial twist-3 GPDs, $\widetilde{H}+\widetilde{G}_2,\,\widetilde{E}+\widetilde{G}_1,\,\widetilde{G}_3,\,\widetilde{G}_4$, using the quasi-GPD method that gives access to the $x$-dependence of GPDs. 
The method requires the evaluation of matrix elements of nonlocal operators and momentum-boosted hadrons. 
We use the axial-vector operator with spatial indices that are perpendicular to the direction of the boost, $P_3$, that is, $\gamma_1\gamma_5$ and $\gamma_2\gamma_5$; on the light cone, this corresponds to the twist-3 counterpart of the helicity GPD. 
Our kinematic setup uses the symmetric frame, which is computationally expensive and requires separate calculations for every value of the momentum transfer. 
In this work, we are able to obtain three values of the momentum transfer, $-t=0.69,\,1.38,\,2.76$ GeV$^2$ at $P_3=1.25$ GeV. 
We also check for convergence in $P_3$ at $-t=0.69$ GeV$^2$, with three values, that is, $P_3=0.83,\,1.25,\,1.67$ GeV. 
All lattice data correspond to zero skewness.
We obtain four independent matrix elements by using unpolarized and three polarized parity projectors that can successfully disentangle the four GPDs. 
We find a very good signal for $\widetilde{H}+\widetilde{G}_2$ and $\widetilde{E}+\widetilde{G}_1$ for all values of $P_3$ and $-t$. $\widetilde{G}_4$ is smaller in magnitude and has a higher relative error than the aforementioned GPDs. 
$\widetilde{G}_3$ is found to be compatible with zero and very noisy; this is a consequence of the calculation being performed at zero skewness.
We note that, in general, the noise-to-signal ratio is increased for the matrix elements $\gamma_1\gamma_5$ and $\gamma_2\gamma_5$, compared to the twist-2 case of $\gamma^5\gamma^+$, computed from $\gamma_0\gamma_5$ and $\gamma_3\gamma_5$.
With the current uncertainties, a momentum boost of $P_3=1.25$ GeV is sufficient to match the lattice data to the light-cone GPDs. 
We have examined the effect of the reconstruction of the $x$-dependence by applying the Backus-Gilbert method using the data up to different values of $\zmax$.
We found that the optimal choice is $z=11a\sim 1$~fm, which is compatible with a larger value of $\zmax$.
The final results for the light-cone GPDs $\widetilde{H}+\widetilde{G}_2$, $\widetilde{E}+\widetilde{G}_1$ and $\widetilde{G}_4$ are shown in Figs.~\ref{fig:G1G2_final} -~\ref{fig:G4_final} for different values of the momentum transfer. 
We combine $\widetilde{H}+\widetilde{G}_2$ with the twist-2 $\widetilde{H}$ GPD to extract $\widetilde{G}_2$, which is shown in Fig.~\ref{fig:G2}. 
This function is not negligible in magnitude and has interesting features, such as negative values at positive $x$.
This is in line with the theoretical expectation that the norm of $\widetilde{G}_2$ is zero.
Lastly, we perform a number of consistency checks: the local limit of twist-3 GPDs, their norms, and the $P_3$ independence of the norms.
These tests show encouraging results, but further investigation is needed to provide, e.g., the Mellin moments of the quantities under study.

We emphasize that this is the first calculation of twist-3 GPDs from lattice QCD, and as such, there are a number of systematic uncertainties that should be addressed in the future in order to gradually move twist-3 calculations toward precision results. 
One of the aspects to consider is excited-states contamination. Here, we use a source-sink time separation of $10a$, as we are calculating matrix elements that have higher statistical noise compared to the twist-2 counterparts (see, e.g., Fig.~\ref{fig:ME_j_0}). 
The above investigation can be tackled using a single ensemble, unlike other effects such as discretization and volume effects.
As in the twist-2 case, another source of systematic uncertainty is related to the momentum boost that enters the matching formalism. 
Addressing this systematic effect is quite challenging due to constraints in reaching a high momentum boost in lattice QCD calculations (see, e.g., ~\cite{Alexandrou:2019lfo}).  Nevertheless, we believe that reliable future calculations in the range $P_3 \sim 2.0 - 2.5 \, \textrm{GeV}$, which may be within reach, could already provide significant new insights in that regard.
Another systematic uncertainty for twist-3 GPDs, which presently can hardly be quantified, is due to the mixing between the two-parton and the three-parton correlators.
This should be addressed at the level of the matching as discussed in Ref.~\cite{Braun:2021aon} for the axial current evaluated in the forward limit. 
To fully and rigorously pursue this program requires massive new developments, though. Specifically, the matching for three-parton quasi-correlators needs to be derived, and these objects must be computed in lattice QCD through pioneering studies of four-point functions. 
Two-loop matching results are also desirable to further scrutinize the convergence of the LaMET approach.  However, because of the complexity of this matter for twist-3, such results most likely will not be available in the foreseeable future.

Additional improvements and investigations will follow this work.
We will examine the intricate nuances associated with the Wandzura-Wilczek approximation, which poses substantial challenges in the realm of GPDs, encompassing both zero skewness and nonzero skewness scenarios. 
In our future calculations, we aim to broaden our scope by exploring a wider range of momentum transfer and nonzero skewness. 
Another direction is the study of systematic uncertainties, such as excited states, discretization effects, volume effects, and pion mass dependence.
In addition to the study of the axial twist-3 GPDs, we will extend our calculations to other cases, that is, the scalar, vector, and tensor twist-3 GPDs.

\vspace*{1cm}
\centerline{\textbf{Acknowledgements}}
The authors are grateful to Joshua Miller for helping in the presentation of results in the manuscript.
S.~B. has been supported by the U.S. Department of Energy under Contract No. DE-SC0012704, and also by  Laboratory Directed Research and Development (LDRD) funds from Brookhaven Science Associates. 
The work of S.B.~and A.M.~has been supported by the National Science Foundation under grant number PHY-2110472. A.M.~has also been supported by the U.S. Department of Energy, Office of Science, Office of Nuclear Physics, within the framework of the TMD Topical Collaboration. K.C.\ is supported by the National Science Centre (Poland) grants SONATA BIS no.\ 2016/22/E/ST2/00013 and OPUS no.\ 2021/43/B/ST2/00497. M.C., J.D. and A.S. acknowledge financial support by the U.S. Department of Energy, Office of Nuclear Physics, Early Career Award under Grant No.\ DE-SC0020405. F.S.\ was funded by the NSFC and the Deutsche Forschungsgemeinschaft (DFG, German Research
Foundation) through the funds provided to the Sino-German Collaborative Research Center TRR110 “Symmetries and the Emergence of Structure in QCD” (NSFC Grant No. 12070131001, DFG Project-ID 196253076 - TRR 110). 
We acknowledge partial support by the U.S. Department of Energy, Office of Science, Office of Nuclear Physics under the Quark-Gluon Tomography (QGT) Topical Collaboration with Award DE-SC0023646. 
Computations for this work were carried out in part on facilities of the USQCD Collaboration, which are funded by the Office of Science of the U.S. Department of Energy. 
This research was supported in part by PLGrid Infrastructure (Prometheus supercomputer at AGH Cyfronet in Cracow).
Computations were also partially performed at the Poznan Supercomputing and Networking Center (Eagle supercomputer), the Interdisciplinary Centre for Mathematical and Computational Modelling of the Warsaw University (Okeanos supercomputer), and the Academic Computer Centre in Gda\'nsk (Tryton supercomputer). The gauge configurations have been generated by the Extended Twisted Mass Collaboration on the KNL (A2) Partition of Marconi at CINECA, through the Prace project Pra13\_3304 ``SIMPHYS".
Inversions were performed using the DD-$\alpha$AMG solver~\cite{Frommer:2013fsa} with twisted mass
  support~\cite{Alexandrou:2016izb}.

\bibliography{references.bib}

\begin{thebibliography}{79}%
\makeatletter
\providecommand \@ifxundefined [1]{%
 \@ifx{#1\undefined}
}%
\providecommand \@ifnum [1]{%
 \ifnum #1\expandafter \@firstoftwo
 \else \expandafter \@secondoftwo
 \fi
}%
\providecommand \@ifx [1]{%
 \ifx #1\expandafter \@firstoftwo
 \else \expandafter \@secondoftwo
 \fi
}%
\providecommand \natexlab [1]{#1}%
\providecommand \enquote  [1]{``#1''}%
\providecommand \bibnamefont  [1]{#1}%
\providecommand \bibfnamefont [1]{#1}%
\providecommand \citenamefont [1]{#1}%
\providecommand \href@noop [0]{\@secondoftwo}%
\providecommand \href [0]{\begingroup \@sanitize@url \@href}%
\providecommand \@href[1]{\@@startlink{#1}\@@href}%
\providecommand \@@href[1]{\endgroup#1\@@endlink}%
\providecommand \@sanitize@url [0]{\catcode `\\12\catcode `\$12\catcode
  `\&12\catcode `\#12\catcode `\^12\catcode `\_12\catcode `\%12\relax}%
\providecommand \@@startlink[1]{}%
\providecommand \@@endlink[0]{}%
\providecommand \url  [0]{\begingroup\@sanitize@url \@url }%
\providecommand \@url [1]{\endgroup\@href {#1}{\urlprefix }}%
\providecommand \urlprefix  [0]{URL }%
\providecommand \Eprint [0]{\href }%
\providecommand \doibase [0]{http://dx.doi.org/}%
\providecommand \selectlanguage [0]{\@gobble}%
\providecommand \bibinfo  [0]{\@secondoftwo}%
\providecommand \bibfield  [0]{\@secondoftwo}%
\providecommand \translation [1]{[#1]}%
\providecommand \BibitemOpen [0]{}%
\providecommand \bibitemStop [0]{}%
\providecommand \bibitemNoStop [0]{.\EOS\space}%
\providecommand \EOS [0]{\spacefactor3000\relax}%
\providecommand \BibitemShut  [1]{\csname bibitem#1\endcsname}%
\let\auto@bib@innerbib\@empty
\bibitem [{\citenamefont {Mueller}\ \emph {et~al.}(1994)\citenamefont
  {Mueller}, \citenamefont {Robaschik}, \citenamefont {Geyer}, \citenamefont
  {Dittes},\ and\ \citenamefont {Horejsi}}]{Mueller:1998fv}%
  \BibitemOpen
  \bibfield  {author} {\bibinfo {author} {\bibfnamefont {D.}~\bibnamefont
  {Mueller}}, \bibinfo {author} {\bibfnamefont {D.}~\bibnamefont {Robaschik}},
  \bibinfo {author} {\bibfnamefont {B.}~\bibnamefont {Geyer}}, \bibinfo
  {author} {\bibfnamefont {F.~M.}\ \bibnamefont {Dittes}}, \ and\ \bibinfo
  {author} {\bibfnamefont {J.}~\bibnamefont {Horejsi}},\ }\href {\doibase
  10.1002/prop.2190420202} {\bibfield  {journal} {\bibinfo  {journal} {Fortsch.
  Phys.}\ }\textbf {\bibinfo {volume} {42}},\ \bibinfo {pages} {101} (\bibinfo
  {year} {1994})},\ \Eprint {http://arxiv.org/abs/hep-ph/9812448}
  {arXiv:hep-ph/9812448 [hep-ph]} \BibitemShut {NoStop}%
\bibitem [{\citenamefont {Ji}(1997)}]{Ji:1996ek}%
  \BibitemOpen
  \bibfield  {author} {\bibinfo {author} {\bibfnamefont {X.-D.}\ \bibnamefont
  {Ji}},\ }\href {\doibase 10.1103/PhysRevLett.78.610} {\bibfield  {journal}
  {\bibinfo  {journal} {Phys. Rev. Lett.}\ }\textbf {\bibinfo {volume} {78}},\
  \bibinfo {pages} {610} (\bibinfo {year} {1997})},\ \Eprint
  {http://arxiv.org/abs/hep-ph/9603249} {arXiv:hep-ph/9603249 [hep-ph]}
  \BibitemShut {NoStop}%
\bibitem [{\citenamefont {Radyushkin}(1996)}]{Radyushkin:1996nd}%
  \BibitemOpen
  \bibfield  {author} {\bibinfo {author} {\bibfnamefont {A.~V.}\ \bibnamefont
  {Radyushkin}},\ }\href {\doibase 10.1016/0370-2693(96)00528-X} {\bibfield
  {journal} {\bibinfo  {journal} {Phys. Lett.}\ }\textbf {\bibinfo {volume}
  {B380}},\ \bibinfo {pages} {417} (\bibinfo {year} {1996})},\ \Eprint
  {http://arxiv.org/abs/hep-ph/9604317} {arXiv:hep-ph/9604317 [hep-ph]}
  \BibitemShut {NoStop}%
\bibitem [{\citenamefont {Collins}\ \emph {et~al.}(1989)\citenamefont
  {Collins}, \citenamefont {Soper},\ and\ \citenamefont
  {Sterman}}]{Collins:1989gx}%
  \BibitemOpen
  \bibfield  {author} {\bibinfo {author} {\bibfnamefont {J.~C.}\ \bibnamefont
  {Collins}}, \bibinfo {author} {\bibfnamefont {D.~E.}\ \bibnamefont {Soper}},
  \ and\ \bibinfo {author} {\bibfnamefont {G.~F.}\ \bibnamefont {Sterman}},\
  }\href {\doibase 10.1142/9789814503266_0001} {\bibfield  {journal} {\bibinfo
  {journal} {Adv. Ser. Direct. High Energy Phys.}\ }\textbf {\bibinfo {volume}
  {5}},\ \bibinfo {pages} {1} (\bibinfo {year} {1989})},\ \Eprint
  {http://arxiv.org/abs/hep-ph/0409313} {arXiv:hep-ph/0409313} \BibitemShut
  {NoStop}%
\bibitem [{\citenamefont {Bertone}\ \emph {et~al.}(2021)\citenamefont
  {Bertone}, \citenamefont {Dutrieux}, \citenamefont {Mezrag}, \citenamefont
  {Moutarde},\ and\ \citenamefont {Sznajder}}]{Bertone:2021yyz}%
  \BibitemOpen
  \bibfield  {author} {\bibinfo {author} {\bibfnamefont {V.}~\bibnamefont
  {Bertone}}, \bibinfo {author} {\bibfnamefont {H.}~\bibnamefont {Dutrieux}},
  \bibinfo {author} {\bibfnamefont {C.}~\bibnamefont {Mezrag}}, \bibinfo
  {author} {\bibfnamefont {H.}~\bibnamefont {Moutarde}}, \ and\ \bibinfo
  {author} {\bibfnamefont {P.}~\bibnamefont {Sznajder}},\ }\href {\doibase
  10.1103/PhysRevD.103.114019} {\bibfield  {journal} {\bibinfo  {journal}
  {Phys. Rev. D}\ }\textbf {\bibinfo {volume} {103}},\ \bibinfo {pages}
  {114019} (\bibinfo {year} {2021})},\ \Eprint
  {http://arxiv.org/abs/2104.03836} {arXiv:2104.03836 [hep-ph]} \BibitemShut
  {NoStop}%
\bibitem [{\citenamefont {Moffat}\ \emph {et~al.}(2023)\citenamefont {Moffat},
  \citenamefont {Freese}, \citenamefont {Clo\"et}, \citenamefont {Donohoe},
  \citenamefont {Gamberg}, \citenamefont {Melnitchouk}, \citenamefont {Metz},
  \citenamefont {Prokudin},\ and\ \citenamefont {Sato}}]{Moffat:2023svr}%
  \BibitemOpen
  \bibfield  {author} {\bibinfo {author} {\bibfnamefont {E.}~\bibnamefont
  {Moffat}}, \bibinfo {author} {\bibfnamefont {A.}~\bibnamefont {Freese}},
  \bibinfo {author} {\bibfnamefont {I.}~\bibnamefont {Clo\"et}}, \bibinfo
  {author} {\bibfnamefont {T.}~\bibnamefont {Donohoe}}, \bibinfo {author}
  {\bibfnamefont {L.}~\bibnamefont {Gamberg}}, \bibinfo {author} {\bibfnamefont
  {W.}~\bibnamefont {Melnitchouk}}, \bibinfo {author} {\bibfnamefont
  {A.}~\bibnamefont {Metz}}, \bibinfo {author} {\bibfnamefont {A.}~\bibnamefont
  {Prokudin}}, \ and\ \bibinfo {author} {\bibfnamefont {N.}~\bibnamefont
  {Sato}},\ }\href@noop {} {\  (\bibinfo {year} {2023})},\ \Eprint
  {http://arxiv.org/abs/2303.12006} {arXiv:2303.12006 [hep-ph]} \BibitemShut
  {NoStop}%
\bibitem [{\citenamefont {Penttinen}\ \emph
  {et~al.}(2000{\natexlab{a}})\citenamefont {Penttinen}, \citenamefont
  {Polyakov}, \citenamefont {Shuvaev},\ and\ \citenamefont
  {Strikman}}]{Penttinen:2000dg}%
  \BibitemOpen
  \bibfield  {author} {\bibinfo {author} {\bibfnamefont {M.}~\bibnamefont
  {Penttinen}}, \bibinfo {author} {\bibfnamefont {M.~V.}\ \bibnamefont
  {Polyakov}}, \bibinfo {author} {\bibfnamefont {A.~G.}\ \bibnamefont
  {Shuvaev}}, \ and\ \bibinfo {author} {\bibfnamefont {M.}~\bibnamefont
  {Strikman}},\ }\href {\doibase 10.1016/S0370-2693(00)01035-2} {\bibfield
  {journal} {\bibinfo  {journal} {Phys. Lett. B}\ }\textbf {\bibinfo {volume}
  {491}},\ \bibinfo {pages} {96} (\bibinfo {year} {2000}{\natexlab{a}})},\
  \Eprint {http://arxiv.org/abs/hep-ph/0006321} {arXiv:hep-ph/0006321}
  \BibitemShut {NoStop}%
\bibitem [{\citenamefont {Burkardt}(2013)}]{Burkardt:2008ps}%
  \BibitemOpen
  \bibfield  {author} {\bibinfo {author} {\bibfnamefont {M.}~\bibnamefont
  {Burkardt}},\ }\href {\doibase 10.1103/PhysRevD.88.114502} {\bibfield
  {journal} {\bibinfo  {journal} {Phys. Rev.}\ }\textbf {\bibinfo {volume}
  {D88}},\ \bibinfo {pages} {114502} (\bibinfo {year} {2013})},\ \Eprint
  {http://arxiv.org/abs/0810.3589} {arXiv:0810.3589 [hep-ph]} \BibitemShut
  {NoStop}%
\bibitem [{\citenamefont {Aslan}\ \emph {et~al.}(2019)\citenamefont {Aslan},
  \citenamefont {Burkardt},\ and\ \citenamefont {Schlegel}}]{Aslan:2019jis}%
  \BibitemOpen
  \bibfield  {author} {\bibinfo {author} {\bibfnamefont {F.~P.}\ \bibnamefont
  {Aslan}}, \bibinfo {author} {\bibfnamefont {M.}~\bibnamefont {Burkardt}}, \
  and\ \bibinfo {author} {\bibfnamefont {M.}~\bibnamefont {Schlegel}},\ }\href
  {\doibase 10.1103/PhysRevD.100.096021} {\bibfield  {journal} {\bibinfo
  {journal} {Phys. Rev. D}\ }\textbf {\bibinfo {volume} {100}},\ \bibinfo
  {pages} {096021} (\bibinfo {year} {2019})},\ \Eprint
  {http://arxiv.org/abs/1904.03494} {arXiv:1904.03494 [hep-ph]} \BibitemShut
  {NoStop}%
\bibitem [{\citenamefont {Lorc\'e}(2014)}]{Lorce:2014mxa}%
  \BibitemOpen
  \bibfield  {author} {\bibinfo {author} {\bibfnamefont {C.}~\bibnamefont
  {Lorc\'e}},\ }\href {\doibase 10.1016/j.physletb.2014.06.068} {\bibfield
  {journal} {\bibinfo  {journal} {Phys. Lett. B}\ }\textbf {\bibinfo {volume}
  {735}},\ \bibinfo {pages} {344} (\bibinfo {year} {2014})},\ \Eprint
  {http://arxiv.org/abs/1401.7784} {arXiv:1401.7784 [hep-ph]} \BibitemShut
  {NoStop}%
\bibitem [{\citenamefont {Bhoonah}\ and\ \citenamefont
  {Lorc\'e}(2017)}]{Bhoonah:2017olu}%
  \BibitemOpen
  \bibfield  {author} {\bibinfo {author} {\bibfnamefont {A.}~\bibnamefont
  {Bhoonah}}\ and\ \bibinfo {author} {\bibfnamefont {C.}~\bibnamefont
  {Lorc\'e}},\ }\href {\doibase 10.1016/j.physletb.2017.10.003} {\bibfield
  {journal} {\bibinfo  {journal} {Phys. Lett. B}\ }\textbf {\bibinfo {volume}
  {774}},\ \bibinfo {pages} {435} (\bibinfo {year} {2017})},\ \Eprint
  {http://arxiv.org/abs/1703.08322} {arXiv:1703.08322 [hep-ph]} \BibitemShut
  {NoStop}%
\bibitem [{\citenamefont {Meissner}\ \emph {et~al.}(2008)\citenamefont
  {Meissner}, \citenamefont {Metz}, \citenamefont {Schlegel},\ and\
  \citenamefont {Goeke}}]{Meissner:2008ay}%
  \BibitemOpen
  \bibfield  {author} {\bibinfo {author} {\bibfnamefont {S.}~\bibnamefont
  {Meissner}}, \bibinfo {author} {\bibfnamefont {A.}~\bibnamefont {Metz}},
  \bibinfo {author} {\bibfnamefont {M.}~\bibnamefont {Schlegel}}, \ and\
  \bibinfo {author} {\bibfnamefont {K.}~\bibnamefont {Goeke}},\ }\href
  {\doibase 10.1088/1126-6708/2008/08/038} {\bibfield  {journal} {\bibinfo
  {journal} {JHEP}\ }\textbf {\bibinfo {volume} {08}},\ \bibinfo {pages} {038}
  (\bibinfo {year} {2008})},\ \Eprint {http://arxiv.org/abs/0805.3165}
  {arXiv:0805.3165 [hep-ph]} \BibitemShut {NoStop}%
\bibitem [{\citenamefont {Meissner}\ \emph {et~al.}(2009)\citenamefont
  {Meissner}, \citenamefont {Metz},\ and\ \citenamefont
  {Schlegel}}]{Meissner:2009ww}%
  \BibitemOpen
  \bibfield  {author} {\bibinfo {author} {\bibfnamefont {S.}~\bibnamefont
  {Meissner}}, \bibinfo {author} {\bibfnamefont {A.}~\bibnamefont {Metz}}, \
  and\ \bibinfo {author} {\bibfnamefont {M.}~\bibnamefont {Schlegel}},\ }\href
  {\doibase 10.1088/1126-6708/2009/08/056} {\bibfield  {journal} {\bibinfo
  {journal} {JHEP}\ }\textbf {\bibinfo {volume} {08}},\ \bibinfo {pages} {056}
  (\bibinfo {year} {2009})},\ \Eprint {http://arxiv.org/abs/0906.5323}
  {arXiv:0906.5323 [hep-ph]} \BibitemShut {NoStop}%
\bibitem [{\citenamefont {Lorc\'e}\ and\ \citenamefont
  {Pasquini}(2013)}]{Lorce:2013pza}%
  \BibitemOpen
  \bibfield  {author} {\bibinfo {author} {\bibfnamefont {C.}~\bibnamefont
  {Lorc\'e}}\ and\ \bibinfo {author} {\bibfnamefont {B.}~\bibnamefont
  {Pasquini}},\ }\href {\doibase 10.1007/JHEP09(2013)138} {\bibfield  {journal}
  {\bibinfo  {journal} {JHEP}\ }\textbf {\bibinfo {volume} {09}},\ \bibinfo
  {pages} {138} (\bibinfo {year} {2013})},\ \Eprint
  {http://arxiv.org/abs/1307.4497} {arXiv:1307.4497 [hep-ph]} \BibitemShut
  {NoStop}%
\bibitem [{\citenamefont {Rajan}\ \emph {et~al.}(2016)\citenamefont {Rajan},
  \citenamefont {Courtoy}, \citenamefont {Engelhardt},\ and\ \citenamefont
  {Liuti}}]{Rajan:2016tlg}%
  \BibitemOpen
  \bibfield  {author} {\bibinfo {author} {\bibfnamefont {A.}~\bibnamefont
  {Rajan}}, \bibinfo {author} {\bibfnamefont {A.}~\bibnamefont {Courtoy}},
  \bibinfo {author} {\bibfnamefont {M.}~\bibnamefont {Engelhardt}}, \ and\
  \bibinfo {author} {\bibfnamefont {S.}~\bibnamefont {Liuti}},\ }\href
  {\doibase 10.1103/PhysRevD.94.034041} {\bibfield  {journal} {\bibinfo
  {journal} {Phys. Rev. D}\ }\textbf {\bibinfo {volume} {94}},\ \bibinfo
  {pages} {034041} (\bibinfo {year} {2016})},\ \Eprint
  {http://arxiv.org/abs/1601.06117} {arXiv:1601.06117 [hep-ph]} \BibitemShut
  {NoStop}%
\bibitem [{\citenamefont {Rajan}\ \emph {et~al.}(2018)\citenamefont {Rajan},
  \citenamefont {Engelhardt},\ and\ \citenamefont {Liuti}}]{Rajan:2017cpx}%
  \BibitemOpen
  \bibfield  {author} {\bibinfo {author} {\bibfnamefont {A.}~\bibnamefont
  {Rajan}}, \bibinfo {author} {\bibfnamefont {M.}~\bibnamefont {Engelhardt}}, \
  and\ \bibinfo {author} {\bibfnamefont {S.}~\bibnamefont {Liuti}},\ }\href
  {\doibase 10.1103/PhysRevD.98.074022} {\bibfield  {journal} {\bibinfo
  {journal} {Phys. Rev. D}\ }\textbf {\bibinfo {volume} {98}},\ \bibinfo
  {pages} {074022} (\bibinfo {year} {2018})},\ \Eprint
  {http://arxiv.org/abs/1709.05770} {arXiv:1709.05770 [hep-ph]} \BibitemShut
  {NoStop}%
\bibitem [{\citenamefont {Belitsky}\ \emph {et~al.}(2004)\citenamefont
  {Belitsky}, \citenamefont {Ji},\ and\ \citenamefont
  {Yuan}}]{Belitsky:2003nz}%
  \BibitemOpen
  \bibfield  {author} {\bibinfo {author} {\bibfnamefont {A.~V.}\ \bibnamefont
  {Belitsky}}, \bibinfo {author} {\bibfnamefont {X.-d.}\ \bibnamefont {Ji}}, \
  and\ \bibinfo {author} {\bibfnamefont {F.}~\bibnamefont {Yuan}},\ }\href
  {\doibase 10.1103/PhysRevD.69.074014} {\bibfield  {journal} {\bibinfo
  {journal} {Phys. Rev. D}\ }\textbf {\bibinfo {volume} {69}},\ \bibinfo
  {pages} {074014} (\bibinfo {year} {2004})},\ \Eprint
  {http://arxiv.org/abs/hep-ph/0307383} {arXiv:hep-ph/0307383} \BibitemShut
  {NoStop}%
\bibitem [{\citenamefont {Ji}(2013)}]{Ji:2013dva}%
  \BibitemOpen
  \bibfield  {author} {\bibinfo {author} {\bibfnamefont {X.}~\bibnamefont
  {Ji}},\ }\href {\doibase 10.1103/PhysRevLett.110.262002} {\bibfield
  {journal} {\bibinfo  {journal} {Phys. Rev. Lett.}\ }\textbf {\bibinfo
  {volume} {110}},\ \bibinfo {pages} {262002} (\bibinfo {year} {2013})},\
  \Eprint {http://arxiv.org/abs/1305.1539} {arXiv:1305.1539 [hep-ph]}
  \BibitemShut {NoStop}%
\bibitem [{\citenamefont {Ji}(2014)}]{Ji:2014gla}%
  \BibitemOpen
  \bibfield  {author} {\bibinfo {author} {\bibfnamefont {X.}~\bibnamefont
  {Ji}},\ }\href {\doibase 10.1007/s11433-014-5492-3} {\bibfield  {journal}
  {\bibinfo  {journal} {Sci. China Phys. Mech. Astron.}\ }\textbf {\bibinfo
  {volume} {57}},\ \bibinfo {pages} {1407} (\bibinfo {year} {2014})},\ \Eprint
  {http://arxiv.org/abs/1404.6680} {arXiv:1404.6680 [hep-ph]} \BibitemShut
  {NoStop}%
\bibitem [{\citenamefont {Cichy}\ and\ \citenamefont
  {Constantinou}(2019)}]{Cichy:2018mum}%
  \BibitemOpen
  \bibfield  {author} {\bibinfo {author} {\bibfnamefont {K.}~\bibnamefont
  {Cichy}}\ and\ \bibinfo {author} {\bibfnamefont {M.}~\bibnamefont
  {Constantinou}},\ }\href {\doibase 10.1155/2019/3036904} {\bibfield
  {journal} {\bibinfo  {journal} {Adv. High Energy Phys.}\ }\textbf {\bibinfo
  {volume} {2019}},\ \bibinfo {pages} {3036904} (\bibinfo {year} {2019})},\
  \Eprint {http://arxiv.org/abs/1811.07248} {arXiv:1811.07248 [hep-lat]}
  \BibitemShut {NoStop}%
\bibitem [{\citenamefont {Ji}\ \emph {et~al.}(2021)\citenamefont {Ji},
  \citenamefont {Liu}, \citenamefont {Liu}, \citenamefont {Zhang},\ and\
  \citenamefont {Zhao}}]{Ji:2020ect}%
  \BibitemOpen
  \bibfield  {author} {\bibinfo {author} {\bibfnamefont {X.}~\bibnamefont
  {Ji}}, \bibinfo {author} {\bibfnamefont {Y.-S.}\ \bibnamefont {Liu}},
  \bibinfo {author} {\bibfnamefont {Y.}~\bibnamefont {Liu}}, \bibinfo {author}
  {\bibfnamefont {J.-H.}\ \bibnamefont {Zhang}}, \ and\ \bibinfo {author}
  {\bibfnamefont {Y.}~\bibnamefont {Zhao}},\ }\href {\doibase
  10.1103/RevModPhys.93.035005} {\bibfield  {journal} {\bibinfo  {journal}
  {Rev. Mod. Phys.}\ }\textbf {\bibinfo {volume} {93}},\ \bibinfo {pages}
  {035005} (\bibinfo {year} {2021})},\ \Eprint
  {http://arxiv.org/abs/2004.03543} {arXiv:2004.03543 [hep-ph]} \BibitemShut
  {NoStop}%
\bibitem [{\citenamefont {Constantinou}(2021)}]{Constantinou:2020pek}%
  \BibitemOpen
  \bibfield  {author} {\bibinfo {author} {\bibfnamefont {M.}~\bibnamefont
  {Constantinou}},\ }\bibfield  {booktitle} {\emph {\bibinfo {booktitle} {{38th
  International Symposium on Lattice Field Theory}}},\ }\href {\doibase
  10.1140/epja/s10050-021-00353-7} {\bibfield  {journal} {\bibinfo  {journal}
  {Eur. Phys. J. A}\ }\textbf {\bibinfo {volume} {57}},\ \bibinfo {pages} {77}
  (\bibinfo {year} {2021})},\ \Eprint {http://arxiv.org/abs/2010.02445}
  {arXiv:2010.02445 [hep-lat]} \BibitemShut {NoStop}%
\bibitem [{\citenamefont {Cichy}(2021)}]{Cichy:2021lih}%
  \BibitemOpen
  \bibfield  {author} {\bibinfo {author} {\bibfnamefont {K.}~\bibnamefont
  {Cichy}},\ }in\ \href@noop {} {\emph {\bibinfo {booktitle} {{38th
  International Symposium on Lattice Field Theory}}}}\ (\bibinfo {year}
  {2021})\ \Eprint {http://arxiv.org/abs/2110.07440} {arXiv:2110.07440
  [hep-lat]} \BibitemShut {NoStop}%
\bibitem [{\citenamefont {Cichy}(2022)}]{Cichy:2021ewm}%
  \BibitemOpen
  \bibfield  {author} {\bibinfo {author} {\bibfnamefont {K.}~\bibnamefont
  {Cichy}},\ }\href {\doibase 10.1051/epjconf/202225801005} {\bibfield
  {journal} {\bibinfo  {journal} {EPJ Web Conf.}\ }\textbf {\bibinfo {volume}
  {258}},\ \bibinfo {pages} {01005} (\bibinfo {year} {2022})},\ \Eprint
  {http://arxiv.org/abs/2111.04552} {arXiv:2111.04552 [hep-lat]} \BibitemShut
  {NoStop}%
\bibitem [{\citenamefont {Ji}\ \emph {et~al.}(2015)\citenamefont {Ji},
  \citenamefont {Sch{\"a}fer}, \citenamefont {Xiong},\ and\ \citenamefont
  {Zhang}}]{Ji:2015qla}%
  \BibitemOpen
  \bibfield  {author} {\bibinfo {author} {\bibfnamefont {X.}~\bibnamefont
  {Ji}}, \bibinfo {author} {\bibfnamefont {A.}~\bibnamefont {Sch{\"a}fer}},
  \bibinfo {author} {\bibfnamefont {X.}~\bibnamefont {Xiong}}, \ and\ \bibinfo
  {author} {\bibfnamefont {J.-H.}\ \bibnamefont {Zhang}},\ }\href {\doibase
  10.1103/PhysRevD.92.014039} {\bibfield  {journal} {\bibinfo  {journal} {Phys.
  Rev.}\ }\textbf {\bibinfo {volume} {D92}},\ \bibinfo {pages} {014039}
  (\bibinfo {year} {2015})},\ \Eprint {http://arxiv.org/abs/1506.00248}
  {arXiv:1506.00248 [hep-ph]} \BibitemShut {NoStop}%
\bibitem [{\citenamefont {Xiong}\ and\ \citenamefont
  {Zhang}(2015)}]{Xiong:2015nua}%
  \BibitemOpen
  \bibfield  {author} {\bibinfo {author} {\bibfnamefont {X.}~\bibnamefont
  {Xiong}}\ and\ \bibinfo {author} {\bibfnamefont {J.-H.}\ \bibnamefont
  {Zhang}},\ }\href {\doibase 10.1103/PhysRevD.92.054037} {\bibfield  {journal}
  {\bibinfo  {journal} {Phys. Rev.}\ }\textbf {\bibinfo {volume} {D92}},\
  \bibinfo {pages} {054037} (\bibinfo {year} {2015})},\ \Eprint
  {http://arxiv.org/abs/1509.08016} {arXiv:1509.08016 [hep-ph]} \BibitemShut
  {NoStop}%
\bibitem [{\citenamefont {Bhattacharya}\ \emph {et~al.}(2019)\citenamefont
  {Bhattacharya}, \citenamefont {Cocuzza},\ and\ \citenamefont
  {Metz}}]{Bhattacharya:2018zxi}%
  \BibitemOpen
  \bibfield  {author} {\bibinfo {author} {\bibfnamefont {S.}~\bibnamefont
  {Bhattacharya}}, \bibinfo {author} {\bibfnamefont {C.}~\bibnamefont
  {Cocuzza}}, \ and\ \bibinfo {author} {\bibfnamefont {A.}~\bibnamefont
  {Metz}},\ }\href {\doibase 10.1016/j.physletb.2018.09.061} {\bibfield
  {journal} {\bibinfo  {journal} {Phys. Lett. B}\ }\textbf {\bibinfo {volume}
  {788}},\ \bibinfo {pages} {453} (\bibinfo {year} {2019})},\ \Eprint
  {http://arxiv.org/abs/1808.01437} {arXiv:1808.01437 [hep-ph]} \BibitemShut
  {NoStop}%
\bibitem [{\citenamefont {Liu}\ \emph {et~al.}(2019)\citenamefont {Liu},
  \citenamefont {Wang}, \citenamefont {Xu}, \citenamefont {Zhang},
  \citenamefont {Zhang}, \citenamefont {Zhao},\ and\ \citenamefont
  {Zhao}}]{Liu:2019urm}%
  \BibitemOpen
  \bibfield  {author} {\bibinfo {author} {\bibfnamefont {Y.-S.}\ \bibnamefont
  {Liu}}, \bibinfo {author} {\bibfnamefont {W.}~\bibnamefont {Wang}}, \bibinfo
  {author} {\bibfnamefont {J.}~\bibnamefont {Xu}}, \bibinfo {author}
  {\bibfnamefont {Q.-A.}\ \bibnamefont {Zhang}}, \bibinfo {author}
  {\bibfnamefont {J.-H.}\ \bibnamefont {Zhang}}, \bibinfo {author}
  {\bibfnamefont {S.}~\bibnamefont {Zhao}}, \ and\ \bibinfo {author}
  {\bibfnamefont {Y.}~\bibnamefont {Zhao}},\ }\href {\doibase
  10.1103/PhysRevD.100.034006} {\bibfield  {journal} {\bibinfo  {journal}
  {Phys. Rev. D}\ }\textbf {\bibinfo {volume} {100}},\ \bibinfo {pages}
  {034006} (\bibinfo {year} {2019})},\ \Eprint
  {http://arxiv.org/abs/1902.00307} {arXiv:1902.00307 [hep-ph]} \BibitemShut
  {NoStop}%
\bibitem [{\citenamefont {Bhattacharya}\ \emph
  {et~al.}(2020{\natexlab{a}})\citenamefont {Bhattacharya}, \citenamefont
  {Cocuzza},\ and\ \citenamefont {Metz}}]{Bhattacharya:2019cme}%
  \BibitemOpen
  \bibfield  {author} {\bibinfo {author} {\bibfnamefont {S.}~\bibnamefont
  {Bhattacharya}}, \bibinfo {author} {\bibfnamefont {C.}~\bibnamefont
  {Cocuzza}}, \ and\ \bibinfo {author} {\bibfnamefont {A.}~\bibnamefont
  {Metz}},\ }\href {\doibase 10.1103/PhysRevD.102.054021} {\bibfield  {journal}
  {\bibinfo  {journal} {Phys. Rev. D}\ }\textbf {\bibinfo {volume} {102}},\
  \bibinfo {pages} {054021} (\bibinfo {year} {2020}{\natexlab{a}})},\ \Eprint
  {http://arxiv.org/abs/1903.05721} {arXiv:1903.05721 [hep-ph]} \BibitemShut
  {NoStop}%
\bibitem [{\citenamefont {Chen}\ \emph {et~al.}(2020)\citenamefont {Chen},
  \citenamefont {Lin},\ and\ \citenamefont {Zhang}}]{Chen:2019lcm}%
  \BibitemOpen
  \bibfield  {author} {\bibinfo {author} {\bibfnamefont {J.-W.}\ \bibnamefont
  {Chen}}, \bibinfo {author} {\bibfnamefont {H.-W.}\ \bibnamefont {Lin}}, \
  and\ \bibinfo {author} {\bibfnamefont {J.-H.}\ \bibnamefont {Zhang}},\ }\href
  {\doibase 10.1016/j.nuclphysb.2020.114940} {\bibfield  {journal} {\bibinfo
  {journal} {Nucl. Phys. B}\ }\textbf {\bibinfo {volume} {952}},\ \bibinfo
  {pages} {114940} (\bibinfo {year} {2020})},\ \Eprint
  {http://arxiv.org/abs/1904.12376} {arXiv:1904.12376 [hep-lat]} \BibitemShut
  {NoStop}%
\bibitem [{\citenamefont {Radyushkin}(2019)}]{Radyushkin:2019owq}%
  \BibitemOpen
  \bibfield  {author} {\bibinfo {author} {\bibfnamefont {A.~V.}\ \bibnamefont
  {Radyushkin}},\ }\href {\doibase 10.1103/PhysRevD.100.116011} {\bibfield
  {journal} {\bibinfo  {journal} {Phys. Rev. D}\ }\textbf {\bibinfo {volume}
  {100}},\ \bibinfo {pages} {116011} (\bibinfo {year} {2019})},\ \Eprint
  {http://arxiv.org/abs/1909.08474} {arXiv:1909.08474 [hep-ph]} \BibitemShut
  {NoStop}%
\bibitem [{\citenamefont {Ma}\ \emph {et~al.}(2020)\citenamefont {Ma},
  \citenamefont {Zhu},\ and\ \citenamefont {Lu}}]{Ma:2019agv}%
  \BibitemOpen
  \bibfield  {author} {\bibinfo {author} {\bibfnamefont {Z.-L.}\ \bibnamefont
  {Ma}}, \bibinfo {author} {\bibfnamefont {J.-Q.}\ \bibnamefont {Zhu}}, \ and\
  \bibinfo {author} {\bibfnamefont {Z.}~\bibnamefont {Lu}},\ }\href {\doibase
  10.1103/PhysRevD.101.114005} {\bibfield  {journal} {\bibinfo  {journal}
  {Phys. Rev. D}\ }\textbf {\bibinfo {volume} {101}},\ \bibinfo {pages}
  {114005} (\bibinfo {year} {2020})},\ \Eprint
  {http://arxiv.org/abs/1912.12816} {arXiv:1912.12816 [hep-ph]} \BibitemShut
  {NoStop}%
\bibitem [{\citenamefont {Luo}\ and\ \citenamefont {Sun}(2020)}]{Luo:2020yqj}%
  \BibitemOpen
  \bibfield  {author} {\bibinfo {author} {\bibfnamefont {X.}~\bibnamefont
  {Luo}}\ and\ \bibinfo {author} {\bibfnamefont {H.}~\bibnamefont {Sun}},\
  }\href {\doibase 10.1140/epjc/s10052-020-8402-z} {\bibfield  {journal}
  {\bibinfo  {journal} {Eur. Phys. J. C}\ }\textbf {\bibinfo {volume} {80}},\
  \bibinfo {pages} {828} (\bibinfo {year} {2020})},\ \Eprint
  {http://arxiv.org/abs/2005.09832} {arXiv:2005.09832 [hep-ph]} \BibitemShut
  {NoStop}%
\bibitem [{\citenamefont {Alexandrou}\ \emph {et~al.}(2020)\citenamefont
  {Alexandrou}, \citenamefont {Cichy}, \citenamefont {Constantinou},
  \citenamefont {Hadjiyiannakou}, \citenamefont {Jansen}, \citenamefont
  {Scapellato},\ and\ \citenamefont {Steffens}}]{Alexandrou:2020zbe}%
  \BibitemOpen
  \bibfield  {author} {\bibinfo {author} {\bibfnamefont {C.}~\bibnamefont
  {Alexandrou}}, \bibinfo {author} {\bibfnamefont {K.}~\bibnamefont {Cichy}},
  \bibinfo {author} {\bibfnamefont {M.}~\bibnamefont {Constantinou}}, \bibinfo
  {author} {\bibfnamefont {K.}~\bibnamefont {Hadjiyiannakou}}, \bibinfo
  {author} {\bibfnamefont {K.}~\bibnamefont {Jansen}}, \bibinfo {author}
  {\bibfnamefont {A.}~\bibnamefont {Scapellato}}, \ and\ \bibinfo {author}
  {\bibfnamefont {F.}~\bibnamefont {Steffens}},\ }\href {\doibase
  10.1103/PhysRevLett.125.262001} {\bibfield  {journal} {\bibinfo  {journal}
  {Phys. Rev. Lett.}\ }\textbf {\bibinfo {volume} {125}},\ \bibinfo {pages}
  {262001} (\bibinfo {year} {2020})},\ \Eprint
  {http://arxiv.org/abs/2008.10573} {arXiv:2008.10573 [hep-lat]} \BibitemShut
  {NoStop}%
\bibitem [{\citenamefont {Alexandrou}\ \emph {et~al.}(2022)\citenamefont
  {Alexandrou}, \citenamefont {Cichy}, \citenamefont {Constantinou},
  \citenamefont {Hadjiyiannakou}, \citenamefont {Jansen}, \citenamefont
  {Scapellato},\ and\ \citenamefont {Steffens}}]{Alexandrou:2021bbo}%
  \BibitemOpen
  \bibfield  {author} {\bibinfo {author} {\bibfnamefont {C.}~\bibnamefont
  {Alexandrou}}, \bibinfo {author} {\bibfnamefont {K.}~\bibnamefont {Cichy}},
  \bibinfo {author} {\bibfnamefont {M.}~\bibnamefont {Constantinou}}, \bibinfo
  {author} {\bibfnamefont {K.}~\bibnamefont {Hadjiyiannakou}}, \bibinfo
  {author} {\bibfnamefont {K.}~\bibnamefont {Jansen}}, \bibinfo {author}
  {\bibfnamefont {A.}~\bibnamefont {Scapellato}}, \ and\ \bibinfo {author}
  {\bibfnamefont {F.}~\bibnamefont {Steffens}},\ }\href {\doibase
  10.1103/PhysRevD.105.034501} {\bibfield  {journal} {\bibinfo  {journal}
  {Phys. Rev. D}\ }\textbf {\bibinfo {volume} {105}},\ \bibinfo {pages}
  {034501} (\bibinfo {year} {2022})},\ \Eprint
  {http://arxiv.org/abs/2108.10789} {arXiv:2108.10789 [hep-lat]} \BibitemShut
  {NoStop}%
\bibitem [{\citenamefont {Hannaford-Gunn}\ \emph {et~al.}(2022)\citenamefont
  {Hannaford-Gunn}, \citenamefont {Can}, \citenamefont {Horsley}, \citenamefont
  {Nakamura}, \citenamefont {Perlt}, \citenamefont {Rakow}, \citenamefont
  {St\"uben}, \citenamefont {Schierholz}, \citenamefont {Young},\ and\
  \citenamefont {Zanotti}}]{CSSMQCDSFUKQCD:2021lkf}%
  \BibitemOpen
  \bibfield  {author} {\bibinfo {author} {\bibfnamefont {A.}~\bibnamefont
  {Hannaford-Gunn}}, \bibinfo {author} {\bibfnamefont {K.~U.}\ \bibnamefont
  {Can}}, \bibinfo {author} {\bibfnamefont {R.}~\bibnamefont {Horsley}},
  \bibinfo {author} {\bibfnamefont {Y.}~\bibnamefont {Nakamura}}, \bibinfo
  {author} {\bibfnamefont {H.}~\bibnamefont {Perlt}}, \bibinfo {author}
  {\bibfnamefont {P.~E.~L.}\ \bibnamefont {Rakow}}, \bibinfo {author}
  {\bibfnamefont {H.}~\bibnamefont {St\"uben}}, \bibinfo {author}
  {\bibfnamefont {G.}~\bibnamefont {Schierholz}}, \bibinfo {author}
  {\bibfnamefont {R.~D.}\ \bibnamefont {Young}}, \ and\ \bibinfo {author}
  {\bibfnamefont {J.~M.}\ \bibnamefont {Zanotti}} (\bibinfo {collaboration}
  {CSSM/QCDSF/UKQCD}),\ }\href {\doibase 10.1103/PhysRevD.105.014502}
  {\bibfield  {journal} {\bibinfo  {journal} {Phys. Rev. D}\ }\textbf {\bibinfo
  {volume} {105}},\ \bibinfo {pages} {014502} (\bibinfo {year} {2022})},\
  \Eprint {http://arxiv.org/abs/2110.11532} {arXiv:2110.11532 [hep-lat]}
  \BibitemShut {NoStop}%
\bibitem [{\citenamefont {Dodson}\ \emph {et~al.}(2022)\citenamefont {Dodson},
  \citenamefont {Bhattacharya}, \citenamefont {Cichy}, \citenamefont
  {Constantinou}, \citenamefont {Metz}, \citenamefont {Scapellato},\ and\
  \citenamefont {Steffens}}]{Dodson:2021rdq}%
  \BibitemOpen
  \bibfield  {author} {\bibinfo {author} {\bibfnamefont {J.}~\bibnamefont
  {Dodson}}, \bibinfo {author} {\bibfnamefont {S.}~\bibnamefont
  {Bhattacharya}}, \bibinfo {author} {\bibfnamefont {K.}~\bibnamefont {Cichy}},
  \bibinfo {author} {\bibfnamefont {M.}~\bibnamefont {Constantinou}}, \bibinfo
  {author} {\bibfnamefont {A.}~\bibnamefont {Metz}}, \bibinfo {author}
  {\bibfnamefont {A.}~\bibnamefont {Scapellato}}, \ and\ \bibinfo {author}
  {\bibfnamefont {F.}~\bibnamefont {Steffens}},\ }\href {\doibase
  10.22323/1.396.0054} {\bibfield  {journal} {\bibinfo  {journal} {PoS}\
  }\textbf {\bibinfo {volume} {LATTICE2021}},\ \bibinfo {pages} {054} (\bibinfo
  {year} {2022})},\ \Eprint {http://arxiv.org/abs/2112.05538} {arXiv:2112.05538
  [hep-lat]} \BibitemShut {NoStop}%
\bibitem [{\citenamefont {Ma}\ \emph {et~al.}(2022)\citenamefont {Ma},
  \citenamefont {Pang},\ and\ \citenamefont {Zhang}}]{Ma:2022ggj}%
  \BibitemOpen
  \bibfield  {author} {\bibinfo {author} {\bibfnamefont {J.~P.}\ \bibnamefont
  {Ma}}, \bibinfo {author} {\bibfnamefont {Z.~Y.}\ \bibnamefont {Pang}}, \ and\
  \bibinfo {author} {\bibfnamefont {G.~P.}\ \bibnamefont {Zhang}},\ }\href
  {\doibase 10.1007/JHEP08(2022)130} {\bibfield  {journal} {\bibinfo  {journal}
  {JHEP}\ }\textbf {\bibinfo {volume} {08}},\ \bibinfo {pages} {130} (\bibinfo
  {year} {2022})},\ \Eprint {http://arxiv.org/abs/2202.07116} {arXiv:2202.07116
  [hep-ph]} \BibitemShut {NoStop}%
\bibitem [{\citenamefont {Shastry}\ \emph {et~al.}(2022)\citenamefont
  {Shastry}, \citenamefont {Broniowski},\ and\ \citenamefont
  {Ruiz~Arriola}}]{Shastry:2022obb}%
  \BibitemOpen
  \bibfield  {author} {\bibinfo {author} {\bibfnamefont {V.}~\bibnamefont
  {Shastry}}, \bibinfo {author} {\bibfnamefont {W.}~\bibnamefont {Broniowski}},
  \ and\ \bibinfo {author} {\bibfnamefont {E.}~\bibnamefont {Ruiz~Arriola}},\
  }\href {\doibase 10.1103/PhysRevD.106.114035} {\bibfield  {journal} {\bibinfo
   {journal} {Phys. Rev. D}\ }\textbf {\bibinfo {volume} {106}},\ \bibinfo
  {pages} {114035} (\bibinfo {year} {2022})},\ \Eprint
  {http://arxiv.org/abs/2209.02619} {arXiv:2209.02619 [hep-ph]} \BibitemShut
  {NoStop}%
\bibitem [{\citenamefont {Ma}\ \emph {et~al.}(2023)\citenamefont {Ma},
  \citenamefont {Pang}, \citenamefont {Zhang},\ and\ \citenamefont
  {Zhang}}]{Ma:2022gty}%
  \BibitemOpen
  \bibfield  {author} {\bibinfo {author} {\bibfnamefont {J.~P.}\ \bibnamefont
  {Ma}}, \bibinfo {author} {\bibfnamefont {Z.~Y.}\ \bibnamefont {Pang}},
  \bibinfo {author} {\bibfnamefont {C.~P.}\ \bibnamefont {Zhang}}, \ and\
  \bibinfo {author} {\bibfnamefont {G.~P.}\ \bibnamefont {Zhang}},\ }\href
  {\doibase 10.1007/JHEP04(2023)001} {\bibfield  {journal} {\bibinfo  {journal}
  {JHEP}\ }\textbf {\bibinfo {volume} {04}},\ \bibinfo {pages} {001} (\bibinfo
  {year} {2023})},\ \Eprint {http://arxiv.org/abs/2212.08238} {arXiv:2212.08238
  [hep-ph]} \BibitemShut {NoStop}%
\bibitem [{\citenamefont {Ji}\ \emph {et~al.}(2022)\citenamefont {Ji},
  \citenamefont {Yao},\ and\ \citenamefont {Zhang}}]{Ji:2022thb}%
  \BibitemOpen
  \bibfield  {author} {\bibinfo {author} {\bibfnamefont {Y.}~\bibnamefont
  {Ji}}, \bibinfo {author} {\bibfnamefont {F.}~\bibnamefont {Yao}}, \ and\
  \bibinfo {author} {\bibfnamefont {J.-H.}\ \bibnamefont {Zhang}},\ }\href@noop
  {} {\  (\bibinfo {year} {2022})},\ \Eprint {http://arxiv.org/abs/2212.14415}
  {arXiv:2212.14415 [hep-ph]} \BibitemShut {NoStop}%
\bibitem [{\citenamefont {Bhattacharya}\ \emph
  {et~al.}(2022{\natexlab{a}})\citenamefont {Bhattacharya}, \citenamefont
  {Cichy}, \citenamefont {Constantinou}, \citenamefont {Dodson}, \citenamefont
  {Gao}, \citenamefont {Metz}, \citenamefont {Mukherjee}, \citenamefont
  {Scapellato}, \citenamefont {Steffens},\ and\ \citenamefont
  {Zhao}}]{Bhattacharya:2022aob}%
  \BibitemOpen
  \bibfield  {author} {\bibinfo {author} {\bibfnamefont {S.}~\bibnamefont
  {Bhattacharya}}, \bibinfo {author} {\bibfnamefont {K.}~\bibnamefont {Cichy}},
  \bibinfo {author} {\bibfnamefont {M.}~\bibnamefont {Constantinou}}, \bibinfo
  {author} {\bibfnamefont {J.}~\bibnamefont {Dodson}}, \bibinfo {author}
  {\bibfnamefont {X.}~\bibnamefont {Gao}}, \bibinfo {author} {\bibfnamefont
  {A.}~\bibnamefont {Metz}}, \bibinfo {author} {\bibfnamefont {S.}~\bibnamefont
  {Mukherjee}}, \bibinfo {author} {\bibfnamefont {A.}~\bibnamefont
  {Scapellato}}, \bibinfo {author} {\bibfnamefont {F.}~\bibnamefont
  {Steffens}}, \ and\ \bibinfo {author} {\bibfnamefont {Y.}~\bibnamefont
  {Zhao}},\ }\href {\doibase 10.1103/PhysRevD.106.114512} {\bibfield  {journal}
  {\bibinfo  {journal} {Phys. Rev. D}\ }\textbf {\bibinfo {volume} {106}},\
  \bibinfo {pages} {114512} (\bibinfo {year} {2022}{\natexlab{a}})},\ \Eprint
  {http://arxiv.org/abs/2209.05373} {arXiv:2209.05373 [hep-lat]} \BibitemShut
  {NoStop}%
\bibitem [{\citenamefont {Bhattacharya}\ \emph
  {et~al.}(2023{\natexlab{a}})\citenamefont {Bhattacharya}, \citenamefont
  {Cichy}, \citenamefont {Constantinou}, \citenamefont {Dodson}, \citenamefont
  {Gao}, \citenamefont {Metz}, \citenamefont {Mukherjee}, \citenamefont
  {Scapellato}, \citenamefont {Steffens},\ and\ \citenamefont
  {Zhao}}]{Bhattacharya:2023tik}%
  \BibitemOpen
  \bibfield  {author} {\bibinfo {author} {\bibfnamefont {S.}~\bibnamefont
  {Bhattacharya}}, \bibinfo {author} {\bibfnamefont {K.}~\bibnamefont {Cichy}},
  \bibinfo {author} {\bibfnamefont {M.}~\bibnamefont {Constantinou}}, \bibinfo
  {author} {\bibfnamefont {J.}~\bibnamefont {Dodson}}, \bibinfo {author}
  {\bibfnamefont {X.}~\bibnamefont {Gao}}, \bibinfo {author} {\bibfnamefont
  {A.}~\bibnamefont {Metz}}, \bibinfo {author} {\bibfnamefont {S.}~\bibnamefont
  {Mukherjee}}, \bibinfo {author} {\bibfnamefont {A.}~\bibnamefont
  {Scapellato}}, \bibinfo {author} {\bibfnamefont {F.}~\bibnamefont
  {Steffens}}, \ and\ \bibinfo {author} {\bibfnamefont {Y.}~\bibnamefont
  {Zhao}},\ }\href {\doibase 10.22323/1.430.0095} {\bibfield  {journal}
  {\bibinfo  {journal} {PoS}\ }\textbf {\bibinfo {volume} {LATTICE2022}},\
  \bibinfo {pages} {095} (\bibinfo {year} {2023}{\natexlab{a}})},\ \Eprint
  {http://arxiv.org/abs/2301.03400} {arXiv:2301.03400 [hep-lat]} \BibitemShut
  {NoStop}%
\bibitem [{\citenamefont {Constantinou}\ \emph {et~al.}(2023)\citenamefont
  {Constantinou}, \citenamefont {Bhattacharya}, \citenamefont {Cichy},
  \citenamefont {Dodson}, \citenamefont {Gao}, \citenamefont {Metz},
  \citenamefont {Mukherjee}, \citenamefont {Scapellato}, \citenamefont
  {Steffens},\ and\ \citenamefont {Zhao}}]{Constantinou:2022fqt}%
  \BibitemOpen
  \bibfield  {author} {\bibinfo {author} {\bibfnamefont {M.}~\bibnamefont
  {Constantinou}}, \bibinfo {author} {\bibfnamefont {S.}~\bibnamefont
  {Bhattacharya}}, \bibinfo {author} {\bibfnamefont {K.}~\bibnamefont {Cichy}},
  \bibinfo {author} {\bibfnamefont {J.}~\bibnamefont {Dodson}}, \bibinfo
  {author} {\bibfnamefont {X.}~\bibnamefont {Gao}}, \bibinfo {author}
  {\bibfnamefont {A.}~\bibnamefont {Metz}}, \bibinfo {author} {\bibfnamefont
  {S.}~\bibnamefont {Mukherjee}}, \bibinfo {author} {\bibfnamefont
  {A.}~\bibnamefont {Scapellato}}, \bibinfo {author} {\bibfnamefont
  {F.}~\bibnamefont {Steffens}}, \ and\ \bibinfo {author} {\bibfnamefont
  {Y.}~\bibnamefont {Zhao}},\ }\href {\doibase 10.22323/1.430.0096} {\bibfield
  {journal} {\bibinfo  {journal} {PoS}\ }\textbf {\bibinfo {volume}
  {LATTICE2022}},\ \bibinfo {pages} {096} (\bibinfo {year} {2023})},\ \Eprint
  {http://arxiv.org/abs/2212.09818} {arXiv:2212.09818 [hep-lat]} \BibitemShut
  {NoStop}%
\bibitem [{\citenamefont {Cichy}\ \emph {et~al.}(2023)\citenamefont {Cichy}
  \emph {et~al.}}]{Cichy:2023dgk}%
  \BibitemOpen
  \bibfield  {author} {\bibinfo {author} {\bibfnamefont {K.}~\bibnamefont
  {Cichy}} \emph {et~al.},\ }in\ \href@noop {} {\emph {\bibinfo {booktitle}
  {{29th Cracow Epiphany Conference}}}}\ (\bibinfo {year} {2023})\ \Eprint
  {http://arxiv.org/abs/2304.14970} {arXiv:2304.14970 [hep-lat]} \BibitemShut
  {NoStop}%
\bibitem [{\citenamefont {Bhattacharya}\ \emph
  {et~al.}(2023{\natexlab{b}})\citenamefont {Bhattacharya}, \citenamefont
  {Cichy}, \citenamefont {Constantinou}, \citenamefont {Gao}, \citenamefont
  {Metz}, \citenamefont {Miller}, \citenamefont {Mukherjee}, \citenamefont
  {Petreczky}, \citenamefont {Steffens},\ and\ \citenamefont
  {Zhao}}]{Bhattacharya:2023ays}%
  \BibitemOpen
  \bibfield  {author} {\bibinfo {author} {\bibfnamefont {S.}~\bibnamefont
  {Bhattacharya}}, \bibinfo {author} {\bibfnamefont {K.}~\bibnamefont {Cichy}},
  \bibinfo {author} {\bibfnamefont {M.}~\bibnamefont {Constantinou}}, \bibinfo
  {author} {\bibfnamefont {X.}~\bibnamefont {Gao}}, \bibinfo {author}
  {\bibfnamefont {A.}~\bibnamefont {Metz}}, \bibinfo {author} {\bibfnamefont
  {J.}~\bibnamefont {Miller}}, \bibinfo {author} {\bibfnamefont
  {S.}~\bibnamefont {Mukherjee}}, \bibinfo {author} {\bibfnamefont
  {P.}~\bibnamefont {Petreczky}}, \bibinfo {author} {\bibfnamefont
  {F.}~\bibnamefont {Steffens}}, \ and\ \bibinfo {author} {\bibfnamefont
  {Y.}~\bibnamefont {Zhao}},\ }\href@noop {} {\  (\bibinfo {year}
  {2023}{\natexlab{b}})},\ \Eprint {http://arxiv.org/abs/2305.11117}
  {arXiv:2305.11117 [hep-lat]} \BibitemShut {NoStop}%
\bibitem [{\citenamefont {Bhattacharya}\ \emph
  {et~al.}(2020{\natexlab{b}})\citenamefont {Bhattacharya}, \citenamefont
  {Cichy}, \citenamefont {Constantinou}, \citenamefont {Metz}, \citenamefont
  {Scapellato},\ and\ \citenamefont {Steffens}}]{Bhattacharya:2020xlt}%
  \BibitemOpen
  \bibfield  {author} {\bibinfo {author} {\bibfnamefont {S.}~\bibnamefont
  {Bhattacharya}}, \bibinfo {author} {\bibfnamefont {K.}~\bibnamefont {Cichy}},
  \bibinfo {author} {\bibfnamefont {M.}~\bibnamefont {Constantinou}}, \bibinfo
  {author} {\bibfnamefont {A.}~\bibnamefont {Metz}}, \bibinfo {author}
  {\bibfnamefont {A.}~\bibnamefont {Scapellato}}, \ and\ \bibinfo {author}
  {\bibfnamefont {F.}~\bibnamefont {Steffens}},\ }\href {\doibase
  10.1103/PhysRevD.102.034005} {\bibfield  {journal} {\bibinfo  {journal}
  {Phys. Rev. D}\ }\textbf {\bibinfo {volume} {102}},\ \bibinfo {pages}
  {034005} (\bibinfo {year} {2020}{\natexlab{b}})},\ \Eprint
  {http://arxiv.org/abs/2005.10939} {arXiv:2005.10939 [hep-ph]} \BibitemShut
  {NoStop}%
\bibitem [{\citenamefont {Bhattacharya}\ \emph
  {et~al.}(2020{\natexlab{c}})\citenamefont {Bhattacharya}, \citenamefont
  {Cichy}, \citenamefont {Constantinou}, \citenamefont {Metz}, \citenamefont
  {Scapellato},\ and\ \citenamefont {Steffens}}]{Bhattacharya:2020jfj}%
  \BibitemOpen
  \bibfield  {author} {\bibinfo {author} {\bibfnamefont {S.}~\bibnamefont
  {Bhattacharya}}, \bibinfo {author} {\bibfnamefont {K.}~\bibnamefont {Cichy}},
  \bibinfo {author} {\bibfnamefont {M.}~\bibnamefont {Constantinou}}, \bibinfo
  {author} {\bibfnamefont {A.}~\bibnamefont {Metz}}, \bibinfo {author}
  {\bibfnamefont {A.}~\bibnamefont {Scapellato}}, \ and\ \bibinfo {author}
  {\bibfnamefont {F.}~\bibnamefont {Steffens}},\ }\href {\doibase
  10.1103/PhysRevD.102.114025} {\bibfield  {journal} {\bibinfo  {journal}
  {Phys. Rev. D}\ }\textbf {\bibinfo {volume} {102}},\ \bibinfo {pages}
  {114025} (\bibinfo {year} {2020}{\natexlab{c}})},\ \Eprint
  {http://arxiv.org/abs/2006.12347} {arXiv:2006.12347 [hep-ph]} \BibitemShut
  {NoStop}%
\bibitem [{\citenamefont {Braun}\ \emph
  {et~al.}(2021{\natexlab{a}})\citenamefont {Braun}, \citenamefont {Ji},\ and\
  \citenamefont {Vladimirov}}]{Braun:2021aon}%
  \BibitemOpen
  \bibfield  {author} {\bibinfo {author} {\bibfnamefont {V.~M.}\ \bibnamefont
  {Braun}}, \bibinfo {author} {\bibfnamefont {Y.}~\bibnamefont {Ji}}, \ and\
  \bibinfo {author} {\bibfnamefont {A.}~\bibnamefont {Vladimirov}},\ }\href
  {\doibase 10.1007/JHEP05(2021)086} {\bibfield  {journal} {\bibinfo  {journal}
  {JHEP}\ }\textbf {\bibinfo {volume} {05}},\ \bibinfo {pages} {086} (\bibinfo
  {year} {2021}{\natexlab{a}})},\ \Eprint {http://arxiv.org/abs/2103.12105}
  {arXiv:2103.12105 [hep-ph]} \BibitemShut {NoStop}%
\bibitem [{\citenamefont {Braun}\ \emph
  {et~al.}(2021{\natexlab{b}})\citenamefont {Braun}, \citenamefont {Ji},\ and\
  \citenamefont {Vladimirov}}]{Braun:2021gvv}%
  \BibitemOpen
  \bibfield  {author} {\bibinfo {author} {\bibfnamefont {V.~M.}\ \bibnamefont
  {Braun}}, \bibinfo {author} {\bibfnamefont {Y.}~\bibnamefont {Ji}}, \ and\
  \bibinfo {author} {\bibfnamefont {A.}~\bibnamefont {Vladimirov}},\ }\href
  {\doibase 10.1007/JHEP10(2021)087} {\bibfield  {journal} {\bibinfo  {journal}
  {JHEP}\ }\textbf {\bibinfo {volume} {10}},\ \bibinfo {pages} {087} (\bibinfo
  {year} {2021}{\natexlab{b}})},\ \Eprint {http://arxiv.org/abs/2108.03065}
  {arXiv:2108.03065 [hep-ph]} \BibitemShut {NoStop}%
\bibitem [{\citenamefont {Bhattacharya}\ \emph
  {et~al.}(2020{\natexlab{d}})\citenamefont {Bhattacharya}, \citenamefont
  {Cichy}, \citenamefont {Constantinou}, \citenamefont {Metz}, \citenamefont
  {Scapellato},\ and\ \citenamefont {Steffens}}]{Bhattacharya:2020cen}%
  \BibitemOpen
  \bibfield  {author} {\bibinfo {author} {\bibfnamefont {S.}~\bibnamefont
  {Bhattacharya}}, \bibinfo {author} {\bibfnamefont {K.}~\bibnamefont {Cichy}},
  \bibinfo {author} {\bibfnamefont {M.}~\bibnamefont {Constantinou}}, \bibinfo
  {author} {\bibfnamefont {A.}~\bibnamefont {Metz}}, \bibinfo {author}
  {\bibfnamefont {A.}~\bibnamefont {Scapellato}}, \ and\ \bibinfo {author}
  {\bibfnamefont {F.}~\bibnamefont {Steffens}},\ }\href {\doibase
  10.1103/PhysRevD.102.111501} {\bibfield  {journal} {\bibinfo  {journal}
  {Phys. Rev. D}\ }\textbf {\bibinfo {volume} {102}},\ \bibinfo {pages}
  {111501} (\bibinfo {year} {2020}{\natexlab{d}})},\ \Eprint
  {http://arxiv.org/abs/2004.04130} {arXiv:2004.04130 [hep-lat]} \BibitemShut
  {NoStop}%
\bibitem [{\citenamefont {Bhattacharya}\ \emph {et~al.}(2021)\citenamefont
  {Bhattacharya}, \citenamefont {Cichy}, \citenamefont {Constantinou},
  \citenamefont {Metz}, \citenamefont {Scapellato},\ and\ \citenamefont
  {Steffens}}]{Bhattacharya:2021moj}%
  \BibitemOpen
  \bibfield  {author} {\bibinfo {author} {\bibfnamefont {S.}~\bibnamefont
  {Bhattacharya}}, \bibinfo {author} {\bibfnamefont {K.}~\bibnamefont {Cichy}},
  \bibinfo {author} {\bibfnamefont {M.}~\bibnamefont {Constantinou}}, \bibinfo
  {author} {\bibfnamefont {A.}~\bibnamefont {Metz}}, \bibinfo {author}
  {\bibfnamefont {A.}~\bibnamefont {Scapellato}}, \ and\ \bibinfo {author}
  {\bibfnamefont {F.}~\bibnamefont {Steffens}},\ }\href {\doibase
  10.1103/PhysRevD.104.114510} {\bibfield  {journal} {\bibinfo  {journal}
  {Phys. Rev. D}\ }\textbf {\bibinfo {volume} {104}},\ \bibinfo {pages}
  {114510} (\bibinfo {year} {2021})},\ \Eprint
  {http://arxiv.org/abs/2107.02574} {arXiv:2107.02574 [hep-lat]} \BibitemShut
  {NoStop}%
\bibitem [{\citenamefont {Bhattacharya}\ \emph
  {et~al.}(2022{\natexlab{b}})\citenamefont {Bhattacharya}, \citenamefont
  {Cichy}, \citenamefont {Constantinou}, \citenamefont {Metz}, \citenamefont
  {Scapellato},\ and\ \citenamefont {Steffens}}]{Bhattacharya:2022fai}%
  \BibitemOpen
  \bibfield  {author} {\bibinfo {author} {\bibfnamefont {S.}~\bibnamefont
  {Bhattacharya}}, \bibinfo {author} {\bibfnamefont {K.}~\bibnamefont {Cichy}},
  \bibinfo {author} {\bibfnamefont {M.}~\bibnamefont {Constantinou}}, \bibinfo
  {author} {\bibfnamefont {A.}~\bibnamefont {Metz}}, \bibinfo {author}
  {\bibfnamefont {A.}~\bibnamefont {Scapellato}}, \ and\ \bibinfo {author}
  {\bibfnamefont {F.}~\bibnamefont {Steffens}},\ }\href {\doibase
  10.21468/SciPostPhysProc.8.056} {\bibfield  {journal} {\bibinfo  {journal}
  {SciPost Phys. Proc.}\ }\textbf {\bibinfo {volume} {8}},\ \bibinfo {pages}
  {056} (\bibinfo {year} {2022}{\natexlab{b}})}\BibitemShut {NoStop}%
\bibitem [{\citenamefont {Bhattacharya}\ \emph
  {et~al.}(2022{\natexlab{c}})\citenamefont {Bhattacharya}, \citenamefont
  {Cichy}, \citenamefont {Constantinou}, \citenamefont {Metz}, \citenamefont
  {Scapellato},\ and\ \citenamefont {Steffens}}]{Bhattacharya:2021rua}%
  \BibitemOpen
  \bibfield  {author} {\bibinfo {author} {\bibfnamefont {S.}~\bibnamefont
  {Bhattacharya}}, \bibinfo {author} {\bibfnamefont {K.}~\bibnamefont {Cichy}},
  \bibinfo {author} {\bibfnamefont {M.}~\bibnamefont {Constantinou}}, \bibinfo
  {author} {\bibfnamefont {A.}~\bibnamefont {Metz}}, \bibinfo {author}
  {\bibfnamefont {A.}~\bibnamefont {Scapellato}}, \ and\ \bibinfo {author}
  {\bibfnamefont {F.}~\bibnamefont {Steffens}},\ }\href {\doibase
  10.21468/SciPostPhysProc.8.057} {\bibfield  {journal} {\bibinfo  {journal}
  {SciPost Phys. Proc.}\ }\textbf {\bibinfo {volume} {8}},\ \bibinfo {pages}
  {057} (\bibinfo {year} {2022}{\natexlab{c}})},\ \Eprint
  {http://arxiv.org/abs/2107.12818} {arXiv:2107.12818 [hep-lat]} \BibitemShut
  {NoStop}%
\bibitem [{\citenamefont {Bhattacharya}\ \emph
  {et~al.}(2022{\natexlab{d}})\citenamefont {Bhattacharya}, \citenamefont
  {Cichy}, \citenamefont {Constantinou}, \citenamefont {Metz}, \citenamefont
  {Scapellato},\ and\ \citenamefont {Steffens}}]{Bhattacharya:2022gfu}%
  \BibitemOpen
  \bibfield  {author} {\bibinfo {author} {\bibfnamefont {S.}~\bibnamefont
  {Bhattacharya}}, \bibinfo {author} {\bibfnamefont {K.}~\bibnamefont {Cichy}},
  \bibinfo {author} {\bibfnamefont {M.}~\bibnamefont {Constantinou}}, \bibinfo
  {author} {\bibfnamefont {A.}~\bibnamefont {Metz}}, \bibinfo {author}
  {\bibfnamefont {A.}~\bibnamefont {Scapellato}}, \ and\ \bibinfo {author}
  {\bibfnamefont {F.}~\bibnamefont {Steffens}},\ }\href {\doibase
  10.22323/1.396.0105} {\bibfield  {journal} {\bibinfo  {journal} {PoS}\
  }\textbf {\bibinfo {volume} {LATTICE2021}},\ \bibinfo {pages} {105} (\bibinfo
  {year} {2022}{\natexlab{d}})}\BibitemShut {NoStop}%
\bibitem [{\citenamefont {Constantinou}\ \emph {et~al.}(2022)\citenamefont
  {Constantinou}, \citenamefont {Bhattacharya}, \citenamefont {Cichy},
  \citenamefont {Metz}, \citenamefont {Scapellato},\ and\ \citenamefont
  {Steffens}}]{Constantinou:2021nbn}%
  \BibitemOpen
  \bibfield  {author} {\bibinfo {author} {\bibfnamefont {M.}~\bibnamefont
  {Constantinou}}, \bibinfo {author} {\bibfnamefont {S.}~\bibnamefont
  {Bhattacharya}}, \bibinfo {author} {\bibfnamefont {K.}~\bibnamefont {Cichy}},
  \bibinfo {author} {\bibfnamefont {A.}~\bibnamefont {Metz}}, \bibinfo {author}
  {\bibfnamefont {A.}~\bibnamefont {Scapellato}}, \ and\ \bibinfo {author}
  {\bibfnamefont {F.}~\bibnamefont {Steffens}},\ }\href {\doibase
  10.22323/1.396.0391} {\bibfield  {journal} {\bibinfo  {journal} {PoS}\
  }\textbf {\bibinfo {volume} {LATTICE2021}},\ \bibinfo {pages} {391} (\bibinfo
  {year} {2022})},\ \Eprint {http://arxiv.org/abs/2111.01056} {arXiv:2111.01056
  [hep-lat]} \BibitemShut {NoStop}%
\bibitem [{\citenamefont {Kiptily}\ and\ \citenamefont
  {Polyakov}(2004)}]{Kiptily:2002nx}%
  \BibitemOpen
  \bibfield  {author} {\bibinfo {author} {\bibfnamefont {D.~V.}\ \bibnamefont
  {Kiptily}}\ and\ \bibinfo {author} {\bibfnamefont {M.~V.}\ \bibnamefont
  {Polyakov}},\ }\href {\doibase 10.1140/epjc/s2004-01957-3} {\bibfield
  {journal} {\bibinfo  {journal} {Eur. Phys. J. C}\ }\textbf {\bibinfo {volume}
  {37}},\ \bibinfo {pages} {105} (\bibinfo {year} {2004})},\ \Eprint
  {http://arxiv.org/abs/hep-ph/0212372} {arXiv:hep-ph/0212372} \BibitemShut
  {NoStop}%
\bibitem [{\citenamefont {Bhattacharya}(2021)}]{Bhattacharya:2021grn}%
  \BibitemOpen
  \bibfield  {author} {\bibinfo {author} {\bibfnamefont {S.}~\bibnamefont
  {Bhattacharya}},\ }\emph {\bibinfo {title} {{A Comprehensive Study of the
  Proton Structure: From PDFs to Wigner Functions}}},\ \href@noop {} {Ph.D.
  thesis},\ \bibinfo  {school} {Temple U.} (\bibinfo {year} {2021})\BibitemShut
  {NoStop}%
\bibitem [{\citenamefont {Aslan}\ \emph {et~al.}(2018)\citenamefont {Aslan},
  \citenamefont {Burkardt}, \citenamefont {Lorc\'e}, \citenamefont {Metz},\
  and\ \citenamefont {Pasquini}}]{Aslan:2018zzk}%
  \BibitemOpen
  \bibfield  {author} {\bibinfo {author} {\bibfnamefont {F.}~\bibnamefont
  {Aslan}}, \bibinfo {author} {\bibfnamefont {M.}~\bibnamefont {Burkardt}},
  \bibinfo {author} {\bibfnamefont {C.}~\bibnamefont {Lorc\'e}}, \bibinfo
  {author} {\bibfnamefont {A.}~\bibnamefont {Metz}}, \ and\ \bibinfo {author}
  {\bibfnamefont {B.}~\bibnamefont {Pasquini}},\ }\href {\doibase
  10.1103/PhysRevD.98.014038} {\bibfield  {journal} {\bibinfo  {journal} {Phys.
  Rev. D}\ }\textbf {\bibinfo {volume} {98}},\ \bibinfo {pages} {014038}
  (\bibinfo {year} {2018})},\ \Eprint {http://arxiv.org/abs/1802.06243}
  {arXiv:1802.06243 [hep-ph]} \BibitemShut {NoStop}%
\bibitem [{\citenamefont {Aslan}\ and\ \citenamefont
  {Burkardt}(2020)}]{Aslan:2018tff}%
  \BibitemOpen
  \bibfield  {author} {\bibinfo {author} {\bibfnamefont {F.}~\bibnamefont
  {Aslan}}\ and\ \bibinfo {author} {\bibfnamefont {M.}~\bibnamefont
  {Burkardt}},\ }\href {\doibase 10.1103/PhysRevD.101.016010} {\bibfield
  {journal} {\bibinfo  {journal} {Phys. Rev. D}\ }\textbf {\bibinfo {volume}
  {101}},\ \bibinfo {pages} {016010} (\bibinfo {year} {2020})},\ \Eprint
  {http://arxiv.org/abs/1811.00938} {arXiv:1811.00938 [nucl-th]} \BibitemShut
  {NoStop}%
\bibitem [{\citenamefont {Diehl}(2003)}]{Diehl:2003ny}%
  \BibitemOpen
  \bibfield  {author} {\bibinfo {author} {\bibfnamefont {M.}~\bibnamefont
  {Diehl}},\ }\href {\doibase 10.1016/j.physrep.2003.08.002,
  10.3204/DESY-THESIS-2003-018} {\bibfield  {journal} {\bibinfo  {journal}
  {Phys. Rept.}\ }\textbf {\bibinfo {volume} {388}},\ \bibinfo {pages} {41}
  (\bibinfo {year} {2003})},\ \Eprint {http://arxiv.org/abs/hep-ph/0307382}
  {arXiv:hep-ph/0307382 [hep-ph]} \BibitemShut {NoStop}%
\bibitem [{\citenamefont {Belitsky}\ \emph {et~al.}(2002)\citenamefont
  {Belitsky}, \citenamefont {Mueller},\ and\ \citenamefont
  {Kirchner}}]{Belitsky:2001ns}%
  \BibitemOpen
  \bibfield  {author} {\bibinfo {author} {\bibfnamefont {A.~V.}\ \bibnamefont
  {Belitsky}}, \bibinfo {author} {\bibfnamefont {D.}~\bibnamefont {Mueller}}, \
  and\ \bibinfo {author} {\bibfnamefont {A.}~\bibnamefont {Kirchner}},\ }\href
  {\doibase 10.1016/S0550-3213(02)00144-X} {\bibfield  {journal} {\bibinfo
  {journal} {Nucl. Phys. B}\ }\textbf {\bibinfo {volume} {629}},\ \bibinfo
  {pages} {323} (\bibinfo {year} {2002})},\ \Eprint
  {http://arxiv.org/abs/hep-ph/0112108} {arXiv:hep-ph/0112108} \BibitemShut
  {NoStop}%
\bibitem [{\citenamefont {Constantinou}\ and\ \citenamefont
  {Panagopoulos}(2017)}]{Constantinou:2017sej}%
  \BibitemOpen
  \bibfield  {author} {\bibinfo {author} {\bibfnamefont {M.}~\bibnamefont
  {Constantinou}}\ and\ \bibinfo {author} {\bibfnamefont {H.}~\bibnamefont
  {Panagopoulos}},\ }\href {\doibase 10.1103/PhysRevD.96.054506} {\bibfield
  {journal} {\bibinfo  {journal} {Phys. Rev.}\ }\textbf {\bibinfo {volume}
  {D96}},\ \bibinfo {pages} {054506} (\bibinfo {year} {2017})},\ \Eprint
  {http://arxiv.org/abs/1705.11193} {arXiv:1705.11193 [hep-lat]} \BibitemShut
  {NoStop}%
\bibitem [{\citenamefont {Bhattacharya}\ and\ \citenamefont
  {Metz}(2022)}]{Bhattacharya:2021boh}%
  \BibitemOpen
  \bibfield  {author} {\bibinfo {author} {\bibfnamefont {S.}~\bibnamefont
  {Bhattacharya}}\ and\ \bibinfo {author} {\bibfnamefont {A.}~\bibnamefont
  {Metz}},\ }\href {\doibase 10.1103/PhysRevD.105.054027} {\bibfield  {journal}
  {\bibinfo  {journal} {Phys. Rev. D}\ }\textbf {\bibinfo {volume} {105}},\
  \bibinfo {pages} {054027} (\bibinfo {year} {2022})},\ \Eprint
  {http://arxiv.org/abs/2105.07282} {arXiv:2105.07282 [hep-ph]} \BibitemShut
  {NoStop}%
\bibitem [{\citenamefont {Martinelli}\ \emph {et~al.}(1995)\citenamefont
  {Martinelli}, \citenamefont {Pittori}, \citenamefont {Sachrajda},
  \citenamefont {Testa},\ and\ \citenamefont {Vladikas}}]{Martinelli:1994ty}%
  \BibitemOpen
  \bibfield  {author} {\bibinfo {author} {\bibfnamefont {G.}~\bibnamefont
  {Martinelli}}, \bibinfo {author} {\bibfnamefont {C.}~\bibnamefont {Pittori}},
  \bibinfo {author} {\bibfnamefont {C.~T.}\ \bibnamefont {Sachrajda}}, \bibinfo
  {author} {\bibfnamefont {M.}~\bibnamefont {Testa}}, \ and\ \bibinfo {author}
  {\bibfnamefont {A.}~\bibnamefont {Vladikas}},\ }\href {\doibase
  10.1016/0550-3213(95)00126-D} {\bibfield  {journal} {\bibinfo  {journal}
  {Nucl. Phys.}\ }\textbf {\bibinfo {volume} {B445}},\ \bibinfo {pages} {81}
  (\bibinfo {year} {1995})},\ \Eprint {http://arxiv.org/abs/hep-lat/9411010}
  {arXiv:hep-lat/9411010 [hep-lat]} \BibitemShut {NoStop}%
\bibitem [{\citenamefont {Gockeler}\ \emph {et~al.}(1999)\citenamefont
  {Gockeler}, \citenamefont {Horsley}, \citenamefont {Oelrich}, \citenamefont
  {Perlt}, \citenamefont {Petters}, \citenamefont {Rakow}, \citenamefont
  {Schafer}, \citenamefont {Schierholz},\ and\ \citenamefont
  {Schiller}}]{Gockeler:1998ye}%
  \BibitemOpen
  \bibfield  {author} {\bibinfo {author} {\bibfnamefont {M.}~\bibnamefont
  {Gockeler}}, \bibinfo {author} {\bibfnamefont {R.}~\bibnamefont {Horsley}},
  \bibinfo {author} {\bibfnamefont {H.}~\bibnamefont {Oelrich}}, \bibinfo
  {author} {\bibfnamefont {H.}~\bibnamefont {Perlt}}, \bibinfo {author}
  {\bibfnamefont {D.}~\bibnamefont {Petters}}, \bibinfo {author} {\bibfnamefont
  {P.~E.~L.}\ \bibnamefont {Rakow}}, \bibinfo {author} {\bibfnamefont
  {A.}~\bibnamefont {Schafer}}, \bibinfo {author} {\bibfnamefont
  {G.}~\bibnamefont {Schierholz}}, \ and\ \bibinfo {author} {\bibfnamefont
  {A.}~\bibnamefont {Schiller}},\ }\href {\doibase
  10.1016/S0550-3213(99)00036-X} {\bibfield  {journal} {\bibinfo  {journal}
  {Nucl. Phys.}\ }\textbf {\bibinfo {volume} {B544}},\ \bibinfo {pages} {699}
  (\bibinfo {year} {1999})},\ \Eprint {http://arxiv.org/abs/hep-lat/9807044}
  {arXiv:hep-lat/9807044 [hep-lat]} \BibitemShut {NoStop}%
\bibitem [{\citenamefont {Alexandrou}\ \emph {et~al.}(2017)\citenamefont
  {Alexandrou}, \citenamefont {Constantinou},\ and\ \citenamefont
  {Panagopoulos}}]{Alexandrou:2015sea}%
  \BibitemOpen
  \bibfield  {author} {\bibinfo {author} {\bibfnamefont {C.}~\bibnamefont
  {Alexandrou}}, \bibinfo {author} {\bibfnamefont {M.}~\bibnamefont
  {Constantinou}}, \ and\ \bibinfo {author} {\bibfnamefont {H.}~\bibnamefont
  {Panagopoulos}} (\bibinfo {collaboration} {ETM}),\ }\href {\doibase
  10.1103/PhysRevD.95.034505} {\bibfield  {journal} {\bibinfo  {journal} {Phys.
  Rev.}\ }\textbf {\bibinfo {volume} {D95}},\ \bibinfo {pages} {034505}
  (\bibinfo {year} {2017})},\ \Eprint {http://arxiv.org/abs/1509.00213}
  {arXiv:1509.00213 [hep-lat]} \BibitemShut {NoStop}%
\bibitem [{\citenamefont {Alexandrou}\ \emph {et~al.}(2019)\citenamefont
  {Alexandrou}, \citenamefont {Cichy}, \citenamefont {Constantinou},
  \citenamefont {Hadjiyiannakou}, \citenamefont {Jansen}, \citenamefont
  {Scapellato},\ and\ \citenamefont {Steffens}}]{Alexandrou:2019lfo}%
  \BibitemOpen
  \bibfield  {author} {\bibinfo {author} {\bibfnamefont {C.}~\bibnamefont
  {Alexandrou}}, \bibinfo {author} {\bibfnamefont {K.}~\bibnamefont {Cichy}},
  \bibinfo {author} {\bibfnamefont {M.}~\bibnamefont {Constantinou}}, \bibinfo
  {author} {\bibfnamefont {K.}~\bibnamefont {Hadjiyiannakou}}, \bibinfo
  {author} {\bibfnamefont {K.}~\bibnamefont {Jansen}}, \bibinfo {author}
  {\bibfnamefont {A.}~\bibnamefont {Scapellato}}, \ and\ \bibinfo {author}
  {\bibfnamefont {F.}~\bibnamefont {Steffens}},\ }\href {\doibase
  10.1103/PhysRevD.99.114504} {\bibfield  {journal} {\bibinfo  {journal} {Phys.
  Rev.}\ }\textbf {\bibinfo {volume} {D99}},\ \bibinfo {pages} {114504}
  (\bibinfo {year} {2019})},\ \Eprint {http://arxiv.org/abs/1902.00587}
  {arXiv:1902.00587 [hep-lat]} \BibitemShut {NoStop}%
\bibitem [{\citenamefont {Karpie}\ \emph {et~al.}(2019)\citenamefont {Karpie},
  \citenamefont {Orginos}, \citenamefont {Rothkopf},\ and\ \citenamefont
  {Zafeiropoulos}}]{Karpie:2019eiq}%
  \BibitemOpen
  \bibfield  {author} {\bibinfo {author} {\bibfnamefont {J.}~\bibnamefont
  {Karpie}}, \bibinfo {author} {\bibfnamefont {K.}~\bibnamefont {Orginos}},
  \bibinfo {author} {\bibfnamefont {A.}~\bibnamefont {Rothkopf}}, \ and\
  \bibinfo {author} {\bibfnamefont {S.}~\bibnamefont {Zafeiropoulos}},\ }\href
  {\doibase 10.1007/JHEP04(2019)057} {\bibfield  {journal} {\bibinfo  {journal}
  {JHEP}\ }\textbf {\bibinfo {volume} {04}},\ \bibinfo {pages} {057} (\bibinfo
  {year} {2019})},\ \Eprint {http://arxiv.org/abs/1901.05408} {arXiv:1901.05408
  [hep-lat]} \BibitemShut {NoStop}%
\bibitem [{\citenamefont {Backus}\ and\ \citenamefont
  {Gilbert}(1968)}]{BackusGilbert}%
  \BibitemOpen
  \bibfield  {author} {\bibinfo {author} {\bibfnamefont {G.}~\bibnamefont
  {Backus}}\ and\ \bibinfo {author} {\bibfnamefont {F.}~\bibnamefont
  {Gilbert}},\ }\href {\doibase 10.1111/j.1365-246X.1968.tb00216.x} {\bibfield
  {journal} {\bibinfo  {journal} {Geophysical Journal International}\ }\textbf
  {\bibinfo {volume} {16}},\ \bibinfo {pages} {169} (\bibinfo {year}
  {1968})}\BibitemShut {NoStop}%
\bibitem [{\citenamefont {Alexandrou}\ \emph
  {et~al.}(2021{\natexlab{a}})\citenamefont {Alexandrou} \emph
  {et~al.}}]{Alexandrou:2021gqw}%
  \BibitemOpen
  \bibfield  {author} {\bibinfo {author} {\bibfnamefont {C.}~\bibnamefont
  {Alexandrou}} \emph {et~al.} (\bibinfo {collaboration} {Extended Twisted
  Mass}),\ }\href {\doibase 10.1103/PhysRevD.104.074515} {\bibfield  {journal}
  {\bibinfo  {journal} {Phys. Rev. D}\ }\textbf {\bibinfo {volume} {104}},\
  \bibinfo {pages} {074515} (\bibinfo {year} {2021}{\natexlab{a}})},\ \Eprint
  {http://arxiv.org/abs/2104.13408} {arXiv:2104.13408 [hep-lat]} \BibitemShut
  {NoStop}%
\bibitem [{\citenamefont {Iwasaki}(1985)}]{Iwasaki:1985we}%
  \BibitemOpen
  \bibfield  {author} {\bibinfo {author} {\bibfnamefont {Y.}~\bibnamefont
  {Iwasaki}},\ }\href {\doibase 10.1016/0550-3213(85)90606-6} {\bibfield
  {journal} {\bibinfo  {journal} {Nucl. Phys. B}\ }\textbf {\bibinfo {volume}
  {258}},\ \bibinfo {pages} {141} (\bibinfo {year} {1985})}\BibitemShut
  {NoStop}%
\bibitem [{\citenamefont {Sheikholeslami}\ and\ \citenamefont
  {Wohlert}(1985)}]{Sheikholeslami:1985ij}%
  \BibitemOpen
  \bibfield  {author} {\bibinfo {author} {\bibfnamefont {B.}~\bibnamefont
  {Sheikholeslami}}\ and\ \bibinfo {author} {\bibfnamefont {R.}~\bibnamefont
  {Wohlert}},\ }\href {\doibase 10.1016/0550-3213(85)90002-1} {\bibfield
  {journal} {\bibinfo  {journal} {Nucl. Phys.}\ }\textbf {\bibinfo {volume}
  {B259}},\ \bibinfo {pages} {572} (\bibinfo {year} {1985})}\BibitemShut
  {NoStop}%
\bibitem [{\citenamefont {Constantinou}\ \emph {et~al.}(2010)\citenamefont
  {Constantinou} \emph {et~al.}}]{Constantinou:2010gr}%
  \BibitemOpen
  \bibfield  {author} {\bibinfo {author} {\bibfnamefont {M.}~\bibnamefont
  {Constantinou}} \emph {et~al.} (\bibinfo {collaboration} {ETM}),\ }\href
  {\doibase 10.1007/JHEP08(2010)068} {\bibfield  {journal} {\bibinfo  {journal}
  {JHEP}\ }\textbf {\bibinfo {volume} {08}},\ \bibinfo {pages} {068} (\bibinfo
  {year} {2010})},\ \Eprint {http://arxiv.org/abs/1004.1115} {arXiv:1004.1115
  [hep-lat]} \BibitemShut {NoStop}%
\bibitem [{\citenamefont {Alexandrou}\ \emph
  {et~al.}(2021{\natexlab{b}})\citenamefont {Alexandrou} \emph
  {et~al.}}]{Alexandrou:2020okk}%
  \BibitemOpen
  \bibfield  {author} {\bibinfo {author} {\bibfnamefont {C.}~\bibnamefont
  {Alexandrou}} \emph {et~al.},\ }\href {\doibase 10.1103/PhysRevD.103.034509}
  {\bibfield  {journal} {\bibinfo  {journal} {Phys. Rev. D}\ }\textbf {\bibinfo
  {volume} {103}},\ \bibinfo {pages} {034509} (\bibinfo {year}
  {2021}{\natexlab{b}})},\ \Eprint {http://arxiv.org/abs/2011.13342}
  {arXiv:2011.13342 [hep-lat]} \BibitemShut {NoStop}%
\bibitem [{\citenamefont {Efremov}\ \emph {et~al.}(1997)\citenamefont
  {Efremov}, \citenamefont {Teryaev},\ and\ \citenamefont
  {Leader}}]{Efremov:1996hd}%
  \BibitemOpen
  \bibfield  {author} {\bibinfo {author} {\bibfnamefont {A.~V.}\ \bibnamefont
  {Efremov}}, \bibinfo {author} {\bibfnamefont {O.~V.}\ \bibnamefont
  {Teryaev}}, \ and\ \bibinfo {author} {\bibfnamefont {E.}~\bibnamefont
  {Leader}},\ }\href {\doibase 10.1103/PhysRevD.55.4307} {\bibfield  {journal}
  {\bibinfo  {journal} {Phys. Rev. D}\ }\textbf {\bibinfo {volume} {55}},\
  \bibinfo {pages} {4307} (\bibinfo {year} {1997})},\ \Eprint
  {http://arxiv.org/abs/hep-ph/9607217} {arXiv:hep-ph/9607217} \BibitemShut
  {NoStop}%
\bibitem [{\citenamefont {Penttinen}\ \emph
  {et~al.}(2000{\natexlab{b}})\citenamefont {Penttinen}, \citenamefont
  {Polyakov},\ and\ \citenamefont {Goeke}}]{Penttinen:1999th}%
  \BibitemOpen
  \bibfield  {author} {\bibinfo {author} {\bibfnamefont {M.}~\bibnamefont
  {Penttinen}}, \bibinfo {author} {\bibfnamefont {M.~V.}\ \bibnamefont
  {Polyakov}}, \ and\ \bibinfo {author} {\bibfnamefont {K.}~\bibnamefont
  {Goeke}},\ }\href {\doibase 10.1103/PhysRevD.62.014024} {\bibfield  {journal}
  {\bibinfo  {journal} {Phys. Rev. D}\ }\textbf {\bibinfo {volume} {62}},\
  \bibinfo {pages} {014024} (\bibinfo {year} {2000}{\natexlab{b}})},\ \Eprint
  {http://arxiv.org/abs/hep-ph/9909489} {arXiv:hep-ph/9909489} \BibitemShut
  {NoStop}%
\bibitem [{\citenamefont {Frommer}\ \emph {et~al.}(2014)\citenamefont
  {Frommer}, \citenamefont {Kahl}, \citenamefont {Krieg}, \citenamefont
  {Leder},\ and\ \citenamefont {Rottmann}}]{Frommer:2013fsa}%
  \BibitemOpen
  \bibfield  {author} {\bibinfo {author} {\bibfnamefont {A.}~\bibnamefont
  {Frommer}}, \bibinfo {author} {\bibfnamefont {K.}~\bibnamefont {Kahl}},
  \bibinfo {author} {\bibfnamefont {S.}~\bibnamefont {Krieg}}, \bibinfo
  {author} {\bibfnamefont {B.}~\bibnamefont {Leder}}, \ and\ \bibinfo {author}
  {\bibfnamefont {M.}~\bibnamefont {Rottmann}},\ }\href {\doibase
  10.1137/130919507} {\bibfield  {journal} {\bibinfo  {journal} {SIAM J. Sci.
  Comput.}\ }\textbf {\bibinfo {volume} {36}},\ \bibinfo {pages} {A1581}
  (\bibinfo {year} {2014})},\ \Eprint {http://arxiv.org/abs/1303.1377}
  {arXiv:1303.1377 [hep-lat]} \BibitemShut {NoStop}%
\bibitem [{\citenamefont {Alexandrou}\ \emph {et~al.}(2016)\citenamefont
  {Alexandrou}, \citenamefont {Bacchio}, \citenamefont {Finkenrath},
  \citenamefont {Frommer}, \citenamefont {Kahl},\ and\ \citenamefont
  {Rottmann}}]{Alexandrou:2016izb}%
  \BibitemOpen
  \bibfield  {author} {\bibinfo {author} {\bibfnamefont {C.}~\bibnamefont
  {Alexandrou}}, \bibinfo {author} {\bibfnamefont {S.}~\bibnamefont {Bacchio}},
  \bibinfo {author} {\bibfnamefont {J.}~\bibnamefont {Finkenrath}}, \bibinfo
  {author} {\bibfnamefont {A.}~\bibnamefont {Frommer}}, \bibinfo {author}
  {\bibfnamefont {K.}~\bibnamefont {Kahl}}, \ and\ \bibinfo {author}
  {\bibfnamefont {M.}~\bibnamefont {Rottmann}},\ }\href {\doibase
  10.1103/PhysRevD.94.114509} {\bibfield  {journal} {\bibinfo  {journal} {Phys.
  Rev. D}\ }\textbf {\bibinfo {volume} {94}},\ \bibinfo {pages} {114509}
  (\bibinfo {year} {2016})},\ \Eprint {http://arxiv.org/abs/1610.02370}
  {arXiv:1610.02370 [hep-lat]} \BibitemShut {NoStop}%
\end{thebibliography}%

\end{document}